\documentclass[11pt,a4paper]{article}
\pdfoutput=1

\usepackage[colorlinks=true, linkcolor=black!50!blue, urlcolor=blue, citecolor=blue, anchorcolor=blue]{hyperref}
\usepackage[font=small,labelfont=bf,margin=0mm,labelsep=period,tableposition=top]{caption}
\usepackage[a4paper,top=1.5cm,bottom=2cm,left=1.5cm,right=1.5cm,bindingoffset=0mm]{geometry}

\usepackage{placeins,setup,cite}
\usepackage{graphicx}
\usepackage{float}
\usepackage{afterpage}
\usepackage{epsfig}
\usepackage{amssymb}
\usepackage{amsmath}
\usepackage{bm}
\usepackage{multirow}
\usepackage{url}
\usepackage{xcolor}
\usepackage{ulem}
\usepackage{url}
\usepackage{booktabs,multirow}
\usepackage{textcomp}
\usepackage{fancyvrb}
\graphicspath{{../}{./figures/}}

\makeatletter
\def\thickhline{%
             \noalign{\ifnum0 =`}\fi\hrule \@height \thickarrayrulewidth \futurelet
             \reserved@a\@xthickhline}
\def\@xthickhline{\ifx\reserved@a\thickhline
                \vskip\doublerulesep
                \vskip -\thickarrayrulewidth
                \fi
                \ifnum0 =`{\fi}}
\makeatother
\newlength{\thickarrayrulewidth}
\usepackage{tikz}
\usetikzlibrary{positioning}
\usetikzlibrary{shapes.geometric, arrows}
\usetikzlibrary{arrows.meta}
\usepackage{varwidth}
\usepackage{xcolor}
\definecolor{mtplotlib1}{HTML}{1f77b4}
\definecolor{mtplotlib2}{HTML}{ff7f0e}
\definecolor{mtplotlib3}{HTML}{2ca02c}
\definecolor{mtplotlib4}{HTML}{d62728}

\tikzset{%
  >={Latex[width=2mm,length=2mm]},
            base/.style = {rectangle, rounded corners, draw=black,
                           minimum width=4cm, minimum height=1cm,
                           text centered}, 
            mystyle/.style={rectangle, rounded corners, draw=black,
            minimum width=12cm, minimum height=1cm,
            text centered}, 
    col0/.style = {base, fill=white!30},
    col1/.style = {base, fill=mtplotlib1!30},
    col11/.style = {mystyle, fill=mtplotlib1!30},
    col2/.style = {base, fill=mtplotlib2!30},
    col3/.style = {base, fill=mtplotlib3!30},
    col4/.style = {base, minimum width=2.5cm, fill=mtplotlib4!15,}
}

\newcommand{\be}{\begin{equation}}
\newcommand{\ee}{\end{equation}}
\newcommand{\bea}{\begin{eqnarray}}
\newcommand{\eea}{\end{eqnarray}}
\newcommand{\bi}{\begin{itemize}}
\newcommand{\ei}{\end{itemize}}
\newcommand{\ben}{\begin{enumerate}}
\newcommand{\een}{\end{enumerate}}

\newcommand{\lc}{\left[}
\newcommand{\rc}{\right]}
\newcommand{\lp}{\left(}
\newcommand{\rp}{\right)}

\def\frac#1#2{{{#1}\over {#2}}}
\def\gsim{\mathrel{\rlap{\lower4pt\hbox{\hskip1pt$\sim$}}
    \raise1pt\hbox{$>$}}}       
\def\lsim{\mathrel{\rlap{\lower4pt\hbox{\hskip1pt$\sim$}}
    \raise1pt\hbox{$<$}}}

\newcommand{\rep}{\mathrm{rep}}

\newcommand{\draft}[1]{}

\def\beq{\begin{equation}}
\def\eeq{\end{equation}}

\numberwithin{equation}{section}
\numberwithin{figure}{section}
\numberwithin{table}{section}

\usepackage{tabularx}
\newcolumntype{C}[1]{>{\centering\arraybackslash}p{#1}}

\definecolor{darkblue}{rgb}{0.0,0,0.5}
\definecolor{darkgreen}{rgb}{0.0,0.3,0.0}
\definecolor{redish}{rgb}{0.675,0,0.2}
\definecolor{red}{rgb}{0.8,0,0}
\definecolor{green}{rgb}{0,0.6,0}
\definecolor{bluish}{rgb}{0.2,0.2,0.675}
\definecolor{mygrey}{rgb}{0.6,0.6,0.6}

\usepackage{tikz} 
\usetikzlibrary{shapes,arrows,positioning,automata,backgrounds,calc,er,patterns,arrows.meta}
\usepackage{tikz-feynman}
\tikzfeynmanset{compat=1.0.0}
\usepackage{varwidth}
\usepackage{xcolor}
\definecolor{mtplotlib1}{HTML}{1f77b4}
\definecolor{mtplotlib2}{HTML}{ff7f0e}
\definecolor{mtplotlib3}{HTML}{2ca02c}
\definecolor{mtplotlib4}{HTML}{d62728}

\tikzset{%
  >={Latex[width=2mm,length=2mm]},
            base/.style = {rectangle, rounded corners, draw=black,
                           minimum width=4cm, minimum height=1cm,
                           text centered}, 
            mystyle/.style={rectangle, rounded corners, draw=black,
            minimum width=12cm, minimum height=1cm,
            text centered}, 
    col0/.style = {base, fill=white!30},
    col1/.style = {base, fill=mtplotlib1!30},
    col11/.style = {mystyle, fill=mtplotlib1!30},
    col2/.style = {base, fill=mtplotlib2!30},
    col3/.style = {base, fill=mtplotlib3!30},
    col4/.style = {base, minimum width=2.5cm, fill=mtplotlib4!15,}
}

\usepackage{tabularx}
\usepackage{subcaption}
\newcolumntype{C}[1]{>{\centering\arraybackslash}p{#1}}

\renewcommand{\arraystretch}{1.3}
\def\lra#1{\overset{\text{\scriptsize$\leftrightarrow$}}{#1}}

\begin{document}
\newgeometry{top=1.5cm,bottom=1.5cm,left=1.5cm,right=1.5cm,bindingoffset=0mm}

\vspace{-2.0cm}
\begin{flushright}
Nikhef-2022-015\\
\end{flushright}
\vspace{0.3cm}

\begin{center}
  {\Large \bf Unbinned multivariate observables for global SMEFT analyses \\[0.2cm] from machine learning}\\
  \vspace{1.1cm}
  {\small
    Raquel Gomez Ambrosio,$^{1,2}$
    Jaco ter Hoeve,$^{3,4}$
    Maeve Madigan,$^{5}$
    Juan Rojo,$^{3,4}$
    and Veronica Sanz$^{6,7}$
  }\\
  
\vspace{0.7cm}

{\it \small
   ~$^1$ Dipartimento di Fisica ``G. Occhialini'', Universita degli Studi di Milano-Bicocca, \\[0.1cm] 
and INFN, Sezione di Milano Bicocca, Piazza della Scienza 3, I – 20126 Milano, Italy\\[0.1cm]
	~$^2$ Dipartimento di Fisica, Universit\'a di Torino, and INFN, Sezione di Torino,
	Via P. Giuria 1, 10125 Torino, Italy \\[0.1cm]
  ~$^3$Department of Physics and Astronomy, VU Amsterdam, 1081HV Amsterdam,
  The Netherlands\\[0.1cm]
  ~$^4$Nikhef Theory Group, Science Park 105, 1098 XG Amsterdam,
  The Netherlands\\[0.1cm]
  ~$^5$DAMTP, University of Cambridge, Wilberforce Road, Cambridge CB3 0WA, UK\\[0.1cm]
  ~$^{6}$ Instituto de Física Corpuscular (IFIC), Universidad de Valencia-CSIC, E-46980 Valencia, Spain\\[0.07cm]
  ~$^7$ Department of Physics and Astronomy, University of Sussex, Brighton BN1 9QH, UK
}
\vspace{1.0cm}

{\bf \large Abstract}

\end{center}

Theoretical interpretations of particle physics data,
such as the determination of the  Wilson coefficients of the Standard Model Effective Field Theory (SMEFT),
often involve the inference of multiple parameters from a global dataset.
Optimizing such interpretations requires the identification of  observables that exhibit
the highest possible sensitivity to the underlying theory parameters.
In this work we develop a flexible open source framework, {\sc\small ML4EFT},
enabling the integration of unbinned multivariate observables into global SMEFT fits.
As compared to traditional measurements,
such observables enhance the sensitivity to the theory parameters by preventing
the information loss incurred when binning in a subset of
final-state kinematic variables.
Our strategy combines machine learning regression and classification techniques
to parameterize high-dimensional likelihood ratios, using
the Monte Carlo replica method to estimate and propagate methodological uncertainties.
As a proof of concept we construct unbinned multivariate observables for
top-quark pair and Higgs+$Z$ production  at the LHC,
demonstrate their impact on the SMEFT parameter space as
compared to  binned measurements, and study
the improved constraints associated to multivariate inputs.
Since the number of neural networks to be trained scales quadratically
with the number of parameters and can be fully parallelized,
the {\sc\small ML4EFT} framework is well-suited to construct unbinned
multivariate observables which depend on up to
tens of EFT coefficients, as required in global fits.

\clearpage

\tableofcontents
\section{Introduction}
\label{sec:intro}

The extensive characterization of the  Higgs boson properties achieved
at the LHC~\cite{ATLAS:2022vkf,CMS:2022dwd,Dawson:2018dcd} following the tenth anniversary of its
discovery~\cite{CMS:2012qbp,ATLAS:2012yve} represents a 
powerful example of the unique potential that precision measurements
have in unveiling hypothetical signals of beyond the Standard Model (BSM) physics in high-energy
collisions.
This potential motivates ongoing efforts within the theory and experimental
 communities to  develop novel frameworks, tools, and analysis techniques
that enhance the sensitivity of precision LHC measurements to BSM signals in comparison with
more traditional approaches.

Since the early days of quantum field theory, effective field theories (EFTs)
have proven
a robust framework to describe the low-energy limits of theories whose ultraviolet completions
are either unknown or with which the computation of predictions is too challenging.
Of particular relevance for the  model-independent interpretation of LHC measurements is the
Standard Model effective field theory (SMEFT)~\cite{Weinberg:1979sa,Buchmuller:1985jz,Grzadkowski:2010es}
(see also the reviews in~\cite{Manohar:2018aog,Alonso:2013hga,Boggia:2017hyq,Brivio:2017vri,Biekoetter:2018ypq}),
which extends the SM while preserving its (exact) symmetries and its field content.
In order to maximize the constraining power of this framework and explore the broadest
possible region in the parameter space, it is advantageous to integrate the information
contained in different types of processes within a coherent global SMEFT analysis.
Several groups have presented  combined SMEFT interpretations of LHC data from the Higgs, top-quark,
and electroweak sectors, eventually complemented with the information from low-energy
electroweak precision observables (EWPOs) and/or flavor data from $B$-meson decays,
e.g.~\cite{DeBlas:2019ehy,Ethier:2021bye,Ethier:2021ydt, Ellis:2020unq,Bissmann:2020mfi,Bruggisser:2021duo,deBlas:2021wap}.
These analyses rely on unfolded binned distributions  provided by the experiments, that is,
they are based on reinterpreting ``SM measurements'' within the SMEFT framework.

In addition to such a combination of multiple datasets and processes, another avenue towards
improved SMEFT analyses is provided by the  design
of tailored observables characterized by
enhanced, or even maximal, sensitivity to the underlying Wilson coefficients
for a given process.
Optimal observables are able to maximally exploit the kinematic
information contained within a given measurement, event by event,
to carry out parameter inference by comparing with the corresponding theoretical predictions.
The low-multiplicity final states present in electron-positron collisions make them particularly
amenable to this strategy, and optimal observables have been used  in the
context of parameter fitting at
LEP, e.g.~\cite{DELPHI:2010ykq,Diehl:1993br}, and for future lepton collider studies~\cite{Durieux:2018tev}.
Constructing optimal observables is instead more difficult
in hadron collisions, where the higher complexity and
multiplicity of the final state, the significant
QCD shower and non-perturbative effects, and the need to
account for detector simulation make difficult the evaluation of the event-by-event likelihood.
This is one of the reasons why
most LHC measurements are presented as unfolded binned cross-sections,
with the exact statistical model~\cite{Cranmer:2021urp} replaced by a multi-Gaussian approximation.

Despite technical challenges associated to their definition and their
presentation~\cite{Arratia:2021otl},
there is growing evidence that at the LHC unbinned multivariate observables accounting for
the full event-by-event
kinematic information are advantageous to constrain the SMEFT parameters.
As an illustration, the most stringent limits on top quark EFT operators from CMS
data are those arising from unbinned detector-level observables~\cite{CMS:2022hjj,CMS:2021aly}.
As compared to traditional measurements, unbinned observables
enhance the sensitivity to EFT coefficients
by preventing the information loss incurred when adopting a specific binning or
when restricting the analysis to a subset of the possible final-state kinematic variables.
Constructing such observables for hadronic collisions can be achieved
with the analytical evaluation of the event likelihood using e.g. the Matrix Element
Method (MEM)~\cite{Campbell:2012cz,Artoisenet:2010cn,Gainer:2013iya,Fiedler:2010sg,Martini:2015fsa}
or numerically by means of Monte Carlo (MC) simulations.
In the latter case, Machine Learning (ML) techniques
provide a powerful toolbox to efficiently construct high-sensitivity
observables for EFT studies~\cite{Chen:2020mev,DAgnolo:2019vbw,DAgnolo:2018cun,Brehmer:2019xox,Brehmer:2018kdj,Brehmer:2018eca,Letizia:2022xbe,Chatterjee:2022oco,Chatterjee:2021nms,Brehmer:2019gmn,Bortolato:2020zcg,Butter:2021rvz,Arganda:2022qzy,Arganda:2022cl,Arganda:2022zbs,Gritsan:2020pib}, see
also~\cite{DeCastro:2018psv,Brehmer:2018hga,Wunsch:2020iuh,dAgnolo:2021aun,Coccaro:2019lgs} for related work.
Such optimal observables are relevant in other contexts beyond
EFTs such
as PDF fits~\cite{Gao:2017yyd,Kovarik:2019xvh}, see~\cite{Aggarwal:2022cki}
for a recent example.

In this work we develop a general framework
enabling the integration of tailored unbinned multivariate observables
from LHC processes within global SMEFT fits.
Our strategy, implemented in the {\sc\small python} open source package
 {\sc\small ML4EFT}, combines machine learning regression and classification techniques to
parameterize high-dimensional likelihood ratios for an arbitrary number of kinematic inputs
and EFT coefficients.
Once the likelihood ratio is parametrized in terms of
neural networks trained on MC simulations,
the posterior probability distributions
in the EFT coefficients  can be inferred by means of Nested Sampling.
The Monte Carlo replica method is used
to estimate methodological uncertainties, such
as those associated to the finite number of training events, and to propagate
them to the inferred confidence level intervals.
A key feature of {\sc\small ML4EFT}
is that the number of networks to be trained, which scales quadratically with the number of
EFT parameters, can be fully parallelized.
While previous studies of ML-assisted optimized observables for EFT fits
consider relatively small operator bases, our
framework is hence well-suited to construct general unbinned multivariate observables which
depend on up to tens of EFT coefficients  as required in global fits.

As a proof of concept of the  {\sc\small ML4EFT} framework,
we construct unbinned multivariate observables for
two processes relevant for global EFT interpretations of LHC data:
inclusive top-quark pair production
and Higgs boson production in associated with a $Z$ boson,
in the $ b \ell^+  \nu_{\ell} \bar{b}  \ell^-  \bar{\nu}_{\ell}$ (dilepton)
and $b\bar{b}\ell^+\ell^-$ final states respectively.
We consider fiducial regions where these
measurements are statistically-limited and therefore systematic errors can be neglected.
Whenever possible, we compare the results based on the ML parametrization
with those provided by the analytical evaluation of the exact event-by-event likelihood.
We demonstrate the improved constraints that these unbinned multivariate observables
provide on the SMEFT parameter space as compared to their binned counterparts,
and study the information gain associated to the inclusion of multiple kinematic inputs.
Our analysis motivates and defines a possible roadmap towards the measurement (and delivery)
of unbinned observables tailored to SMEFT parameters at the LHC.

The outline of this paper is as follows.
First of all, Sect.~\ref{sec:statistical_framework} introduces
the statistical framework which is adopted to construct unbinned multivariate observables.
Then Sect.~\ref{sec:trainingMethodology} discusses how
this general framework applies to the SMEFT and how
machine learning is deployed to parametrize high-dimensional likelihood functions.
Sect.~\ref{sec:pseudodata} describes our pipeline for
the MC simulation of LHC events in the SMEFT and the settings
of the pseudo-data generation.
Our results are presented in Sect.~\ref{sec:results}, which
quantifies the constraints on the EFT parameter space
provided by unbinned observables in $t\bar{t}$
and in $hZ$ production.
Finally, in Sect.~\ref{sec:summary} we summarize and discuss possible
future avenues.
App.~\ref{app:code} presents the main features of the open source {\sc\small ML4EFT}
framework, while
App.~\ref{app:continous_asimov} discusses the Asimov dataset
in the case of unbinned observables.

\section{From binned to unbinned likelihoods}
\label{sec:statistical_framework}

We begin by presenting the statistical framework that will be adopted in this work
in order to construct unbinned observables in the context of global EFT analyses.
While we focus on applications to the SMEFT, we emphasize that this formalism
is fully general and can be deployed to also construct unbinned observables relevant
{\it e.g.} to the determination of SM parameters such as the parton distribution functions.

\subsection{Binned likelihoods}
\label{sec:binned_likelihoods}
Let us consider a dataset $\mathcal{D}$.
The corresponding theory prediction $\mathcal{T}$ will in general depend
on $n_{\rm p}$ model parameters, denoted by $\boldsymbol{c} = \{ c_1, c_2, \ldots, c_{n_{\rm p}}\}$,
and hence we write these predictions as $\mathcal{T}(\boldsymbol{c})$.
The likelihood function is defined as the
probability to observe the dataset $\mathcal{D}$ assuming that the corresponding
underlying law is described by the theory predictions $\mathcal{T}(\boldsymbol{c})$
associated to the specific set of parameters $\boldsymbol{c}$,
\be
\label{eq:likelihood}
\mathcal{L}( \boldsymbol{c}) = P\lp \mathcal{D} | \mathcal{T}(\boldsymbol{c}) \rp \, .
\ee
This likelihood function makes it possible to discriminate between different theory
hypotheses and to determine, within a given theory hypothesis $\mathcal{T}(\boldsymbol{c})$,
the preferred values and confidence level (CL) intervals for
a given set of model parameters.
In particular,
the best-fit values of the parameters $\boldsymbol{\hat{c}}$ are then
determined from the maximization of the likelihood function $\mathcal{L}( \boldsymbol{c})$,
with contours of fixed likelihood determining their CL intervals.

The most common manner of presenting the information contained in the dataset $\mathcal{D}$
is by binning the data in terms of specific values of selected variables characteristic of each
event, such as the final state kinematics.
In this case, the individual events are combined into $N_{b}$ bins.
Let us denote by $n_i$ the number of observed events in the $i$-th bin
and by $\nu_i(\boldsymbol{c})$ the corresponding theory prediction for the
model parameters $\boldsymbol{c}$.
For a sufficiently large number of events $n_i$ per bin (typically taken to be $n_i\gsim 30$)
one can rely on the Gaussian approximation.  Hence, the likelihood
to observe $\boldsymbol{n} = (n_1, \dots, n_{N_{b}})$ events in each bin, given the
theory predictions $\boldsymbol{\nu}(\boldsymbol{c})$,
is given by
\begin{equation}
  \label{eq:binned_gaussian_likelihood}
	\mathcal{L}(\boldsymbol{n}; \boldsymbol{\nu}(\boldsymbol{c})) = \prod_{i=1}^{N_b} \exp\left[-\frac{1}{2}\frac{(n_i-\nu_i(\boldsymbol{c}))^2}{\nu_i(\boldsymbol{c})}\right] \, ,
\end{equation}
where we consider only statistical uncertainties and neglect possible sources of
correlated systematic errors in the measurement (uncorrelated systematic errors can be
accounted for in the same manner as the statistical counterparts).
This approximation is justified since in this work we focus on
statistically-limited  observables, e.g. the high-energy tails
of differential distributions.
The binned Gaussian likelihood Eq.~(\ref{eq:binned_gaussian_likelihood}) can also be expressed
as
\begin{equation}
   \label{eq:binned_gaussian_likelihood_chi2}
   -2\log  \mathcal{L}(\boldsymbol{n}; \boldsymbol{\nu}(\boldsymbol{c})) =
    \sum_{i=1}^{N_b} \frac{(n_i-\nu_i(\boldsymbol{c}))^2}{\nu_i(\boldsymbol{c})} \equiv \chi^2(\boldsymbol{c})\, ,
\end{equation}
that is, as the usual $\chi^2$ corresponding to Gaussianly distributed binned measurements.
The most likely values of the parameters $\boldsymbol{\hat{c}}$ given the theory hypothesis
$\mathcal{T}(\boldsymbol{c})$ and the measured dataset $\mathcal{D}$ are obtained
from the minimization of Eq.~(\ref{eq:binned_gaussian_likelihood_chi2}).

The Gaussian binned likelihood, Eq.~(\ref{eq:binned_gaussian_likelihood_chi2}), is not appropriate
whenever the number of events in some bins becomes too small.
%
Denoting by $n_{\rm tot}$ the total number of observed events and $\nu_{\rm tot}(\boldsymbol{c})$
the corresponding theory prediction, the corresponding likelihood is the product
of Poisson and multinomial distributions:
\begin{equation}
  \label{eq:likelihood_binned_poisson}
  \mathcal{L}(\boldsymbol{n}; \boldsymbol{\nu}(\boldsymbol{c}))  = \frac{\lp \nu_{\mathrm{tot}}(\boldsymbol{c})\rp^{n_\mathrm{tot}}e^{-\nu_{\mathrm{tot}}(\boldsymbol{c})}}{n_\mathrm{tot}!}\frac{n_\mathrm{tot}!}{n_1!\dots n_{N_b}!}\prod_{i=1}^{N_b}
\left(\frac{\nu_i (\boldsymbol{c})}{\nu_{\mathrm{tot}}(\boldsymbol{c}))}\right)^{n_i} \, ,
\end{equation}
where the total number of observed events (and the corresponding theory prediction)
is equivalent to the sum over all bins,
\be
\nu_{\mathrm{tot}}(\boldsymbol{c}) = \sum_{i=1}^{N_b} \nu_i(\boldsymbol{c})\, , \qquad n_{\mathrm{tot}} = \sum_{i=1}^{N_b} n_i \, .
\ee
When imposing these constraints, Eq.~(\ref{eq:likelihood_binned_poisson}) simplifies to
\begin{equation}
\mathcal{L}(\boldsymbol{n}; \boldsymbol{\nu}(\boldsymbol{c})) = \prod_{i=1}^{N_b}\frac{\nu_i(\boldsymbol{c})^{n_i}}{n_i!}e^{-\nu_i(\boldsymbol{c})} \, ,
\label{eq:joint_pdf_simp}
\end{equation}
which is equivalent to the likelihood of a binned measurement  in which the number of events $n_i$ in each bin follows an independent Poisson distribution with mean $\nu_i(\boldsymbol{c})$.
As in the Gaussian case, Eq.~(\ref{eq:binned_gaussian_likelihood_chi2}), one often
considers the negative log-likelihood,
and for the Poissonian likelihood of Eq.~(\ref{eq:joint_pdf_simp})
this translates into
\begin{equation}
  -2 \log \mathcal{L}(\boldsymbol{n}; \boldsymbol{\nu}(\boldsymbol{c}))  = -2 \sum_{i=1}^{N_b} \lp 
  n_i\log \nu_i(\boldsymbol{c})-\nu_i(\boldsymbol{c}) \rp,
	\label{eq:likelihood_binned_2}
\end{equation}
where we have dropped the $c$-independent terms.
In the limit of large number of events per bin, $n_i\gg 1$, it can be shown
that the Poisson log-likelihood, Eq.~(\ref{eq:likelihood_binned_2}), reduces
to its Gaussian counterpart, Eq.~(\ref{eq:binned_gaussian_likelihood_chi2}).
Again, the most likely values of the model parameters, $\boldsymbol{\hat{c}}$,
    are those obtained from the minimization of Eq.~(\ref{eq:likelihood_binned_2}).

    \paragraph{Confidence level intervals.}
 In order to determine confidence level intervals associated to the
 model parameters for both the Gaussian and the Poisson likelihood
 one can adopt, rather than the likelihood $\mathcal{L}(\boldsymbol{c})$,
 the profile likelihood ratio (PLR) as test statistic of choice.
 The PLR is defined as
\begin{equation}
	q_{\boldsymbol{c}} \equiv -2 \log \frac{\mathcal{L}(\boldsymbol{c})}{\mathcal{L}(\boldsymbol{\hat{c}})} \, ,
	\label{eq:plr}
\end{equation} 
where as mentioned above $\boldsymbol{\hat{c}}$ denotes the maximum likelihood estimator of the theory parameters $\boldsymbol{c}$ given the observed dataset $\mathcal{D}$.
By construction, the PLR $q_{\boldsymbol{c}}$ is semi-positive definite for any value of the theory
parameters.
Larger values of $q_{\boldsymbol{c}}$  indicate increasing incompatibility between
theory predictions $\mathcal{T}(\boldsymbol{c})$
and observed data $\mathcal{D}$, while lower values (down to $q_{\boldsymbol{c}}=0$)
correspond to improved compatibility.
One important difference between the absolute likelihood
$\mathcal{L}( \boldsymbol{c})$ and  the profile likelihood ratio $q_{\boldsymbol{c}}$
is that the latter can only be constructed after having determined
$\boldsymbol{\hat{c}}$.

Adopting the  profile likelihood ratio as test statistic is advantageous, particularly
in light of the powerful result due to Wilks~\cite{10.1214/aoms/1177732360}
stating that $q_{\boldsymbol{c}}$ is distributed according
to a $\chi^2$ distribution under
the null hypothesis, that is, under data $\mathcal{D}$ whose underlying law
is described by the theory predictions $\mathcal{T}(\boldsymbol{c})$.
Furthermore, 
in the large sample limit and assuming specific regularity conditions, the profile
likelihood ratio Eq.~(\ref{eq:plr}) follows a $\chi^2$ distribution with
$n_{\rm p}$ degrees of freedom, $\chi^2_{n_{\rm p}}$.
The main benefit of the profile likelihood ratio  is hence that it allows
for  an efficient limit setting procedure given that one has direct access to the asymptotic probability
distribution.
For instance, for the Gaussian likelihood we can determine
the endpoints of the $100(1-\alpha)\%$ confidence level (CL) intervals
by imposing the condition
\begin{equation}
  \label{eq:confidence_level}
	p_{\boldsymbol{c}} \equiv 1 - F_{\chi^2_{n_{\rm p}}}(q_{\boldsymbol{c}}) = \alpha
\end{equation} 
on the p-value $p_{c}$, where $F_{\chi^2_k}(y)$ is the cumulative distribution function of the $\chi^2_k$
distribution with $k$ degrees of freedom,
\be
F_{\chi^2_k}(q_{\boldsymbol{c}} ) = P\lp \chi^2_k \le q_{\boldsymbol{c}}\rp \, ,
\ee
and recall that both $\chi^2_k$ and $q_{\boldsymbol{c}}$ are semi-positive-definite quantities.
For instance, $\alpha=0.05$ corresponds to the calculation of the 95\% CL intervals
of the theory parameters $\boldsymbol{c}$.
Isolating $q_{\boldsymbol{c}}$ from Eq.~(\ref{eq:confidence_level}) gives
\begin{equation}
  \label{eq:solving_c_from_q}
	q_{\boldsymbol{c}} = F^{-1}_{\chi^2_{n_p}}(1-\alpha) \, ,
\end{equation}
and hence the resulting confidence level intervals satisfy
\begin{equation}
  \label{eq:plr_3}
\chi^2(\boldsymbol{c} ) =  \chi^2(\boldsymbol{\hat{c}}) + F^{-1}_{\chi^2_{n_p}}(1-\alpha) \, .
\end{equation}
The determination of the CL contours on the theory
parameter space for the binned
Gaussian likelihood is obtained by solving Eq.~(\ref{eq:plr_3}) for $\boldsymbol{c}$.
Working with $q_c$ directly, the same result is obtained by demanding
Eq.~(\ref{eq:solving_c_from_q}).
A similar derivation can be used to determine CL intervals in the case
of the Poisson likelihood, Eq.~(\ref{eq:likelihood_binned_2}).

\subsection{Unbinned likelihood}
\label{sec:unbinned_likelihood}

The previous discussion applies to binned observables, and leads
to the standard Gaussian and Poisson likelihoods, Eqns.~(\ref{eq:binned_gaussian_likelihood})
and~(\ref{eq:joint_pdf_simp}) respectively, in the case of statistically-dominated
measurements.
Any binned measurement entails some information loss by construction,
since the information provided by individual events falling into the same bin
is being averaged out.
To eliminate the effects of this information loss, one can
construct unbinned likelihoods that reduce to their binned
counterparts Eqns.~(\ref{eq:binned_gaussian_likelihood})
and~(\ref{eq:joint_pdf_simp}) in the appropriate limits.

Instead of collecting the $N_{\rm ev}$ measured events into $N_b$ bins,
when constructing unbinned observables one treats  each event individually.
We denote now the dataset under consideration as
\be
\label{eq:dataset_unbinned_definition}
\mathcal{D}=\{ \boldsymbol{x}_i\}\, \qquad
\boldsymbol{x}_i= \lp x_{i,1}, x_{i,2},\ldots, x_{i,n_{k}} \rp \, ,\qquad  i=1, \ldots, N_{\rm ev} \, ,
\ee
with $\boldsymbol{x}_i$ denoting the
array indicating
the values of the $n_{k}$ final-state variables that are being measured.
Typically the array $\boldsymbol{x}_i$ will contain the values of the transverse
momenta, rapidities, and azimuthal angles of the measured final state particles, but
could also be composed of higher-level variables such as in jet substructure measurements.
Furthermore, the same approach can be applied to detector-level quantities,
in which case the array $\boldsymbol{x}_i$ contains information
such as energy deposits in the calorimeter cells.

As in the binned case, we assume that this process is described
by a theoretical framework $\mathcal{T}(\boldsymbol{c})$
depending on the $n_p$ model parameters  $\boldsymbol{c} = \{ c_1, c_2, \ldots, c_{n_{\rm p}}\}$.
The kinematic variables of the events constituting the
dataset Eq.~(\ref{eq:dataset_unbinned_definition})
are independent and identically distributed random variables following a given distribution,
which we denote by $f_\sigma\lp \boldsymbol{x}, \boldsymbol{c} \rp $, where the notation
reflects that this probability density will be given, in the cases we are interested in,
by the differential cross-section evaluated using the null hypothesis (theory
$\mathcal{T}(\boldsymbol{c})$ in this case).
For such an unbinned measurement, the likelihood factorizes into the contributions
from individual events such that
\begin{equation}
\mathcal{L}( \boldsymbol{c}) = \prod_{i=1}^{N_{\rm ev}} f_\sigma\lp \boldsymbol{x}_i, \boldsymbol{c} \rp \, .
\label{eq:likelihood_unbinned_fixednev}
\end{equation}
It is worth noting that in Eq.~(\ref{eq:likelihood_unbinned_fixednev}) the
data enters as the experimentally observed values of the kinematic variables
$\boldsymbol{x}_i$ for each event, while the theory predictions
enter at the level of the model adopted $f_\sigma\lp \boldsymbol{x}, \boldsymbol{c} \rp $
for the underlying probability density.

By analogy with the binned Poissonian case, the likelihood can be generalized
to the more realistic case where the measured number of events $N_{\rm ev}$ is not
fixed but rather distributed according to a Poisson with mean $\nu_{\rm tot}(\boldsymbol{c})$,
namely the total number of events predicted by the theory $\mathcal{T}(\boldsymbol{c})$,
see also Eq.~(\ref{eq:likelihood_binned_poisson}).
The likelihood Eq.~(\ref{eq:likelihood_unbinned_fixednev}) then receives an
extra contribution to account for the random size of the dataset $\mathcal{D}$ which reads
\begin{equation}
\mathcal{L}( \boldsymbol{c}) = \frac{\nu_{\rm tot}(\boldsymbol{c})^{N_{\rm ev}}}{N_{\rm ev}!}e^{-\nu_{\rm tot}(\boldsymbol{c})}\prod_{i=1}^{N_{\rm ev}} f_\sigma\lp \boldsymbol{x}_i, \boldsymbol{c} \rp \, .
\label{eq:likelihood_extended}
\end{equation}
Eq.~(\ref{eq:likelihood_extended}) defines the extended unbinned likelihood,
with corresponding log-likelihood given by
\begin{equation}
  \log \mathcal{L}( \boldsymbol{c}) = -\nu_{\rm tot}(\boldsymbol{c}) + N_{\rm ev}\log \nu_{\rm tot}(\boldsymbol{c}) +
  \sum_{i=1}^{N_{\rm ev}} \log f_\sigma\lp \boldsymbol{x}_i, \boldsymbol{c} \rp
  =  -\nu_{\rm tot}(\boldsymbol{c}) +
  \sum_{i=1}^{N_{\rm ev}} \log  \lp  \nu_{\rm tot}(\boldsymbol{c})  f_\sigma\lp \boldsymbol{x}_i, \boldsymbol{c} \rp \rp \, ,
\label{eq:log_likelihood_extended}
\end{equation}
where again we have  dropped all terms that do not depend on the theory
parameters $\boldsymbol{c} $ since these are not relevant to determine
the maximum likelihood estimators and confidence level intervals.
The unbinned log-likelihood Eq.~(\ref{eq:log_likelihood_extended}) can also be obtained from the Poissonian
binned likelihood Eq.~(\ref{eq:likelihood_binned_2}) in the infinitely narrow bin limit,
that is, when taking $n_i\rightarrow 1\,( \forall i)$ and $N_{b}\to N_{\rm ev}$, where $\nu_i(\boldsymbol{c})  \rightarrow
\nu_{\rm tot}(\boldsymbol{c})f_\sigma\lp \boldsymbol{x}_i, \boldsymbol{c} \rp $.
Indeed, in this limit one has that
\bea
 && \log \mathcal{L}(\boldsymbol{n}; \boldsymbol{\nu}(\boldsymbol{c}))\Big|_{\rm binned}  = \sum_{i=1}^{N_b} \lp 
  n_i\log \nu_i(\boldsymbol{c})-\nu_i(\boldsymbol{c})   \rp
  \label{eq:likelihood_binned_2_limit}\\
  && \to  \sum_{i=1}^{N_{\rm ev}} \lc 
  \log \lp \nu_{\rm tot}(\boldsymbol{c})f_\sigma\lp \boldsymbol{x}_i, \boldsymbol{c} \rp\rp -\nu_{\rm tot}(\boldsymbol{c})f_\sigma\lp \boldsymbol{x}_i, \boldsymbol{c} \rp   \rc = -\nu_{\rm tot}(\boldsymbol{c}) +
  \sum_{i=1}^{N_{\rm ev}} 
  \log \lp \nu_{\rm tot}(\boldsymbol{c})f_\sigma\lp \boldsymbol{x}_i, \boldsymbol{c} \rp\rp   \nonumber \, ,
  \eea
  as expected, 
  where we have used the normalization condition for the probability density
  \be
   \sum_{i=1}^{N_{\rm ev}}f_\sigma\lp \boldsymbol{x}_i, \boldsymbol{c} \rp  = 1 \, .
   \ee
   Hence one can smoothly interpolate between the (Poissonian) binned and unbinned
   likelihoods by reducing the bin size until there is at most one event per bin.
   Again, we ignore correlated systematic errors in this derivation.

As mentioned above, the probability density associated to the events
that constitute the dataset Eq.~(\ref{eq:dataset_unbinned_definition}) and enter
the corresponding likelihood Eq.~(\ref{eq:log_likelihood_extended}) is, in the case
of high-energy collisions, given by the  normalized
differential cross-section
\begin{equation}
f_\sigma\lp \boldsymbol{x}, \boldsymbol{c} \rp= \frac{1}{\sigma_{\rm fid}( \boldsymbol{c})}\frac{d\sigma(\boldsymbol{x}, \boldsymbol{c})}{d\boldsymbol{x}} \, ,
\label{eq:dsigma_dx}
\end{equation}
with $\sigma_{\rm fid}(\boldsymbol{c})$ indicating the total fiducial cross-section corresponding
to the phase space region in which the kinematic variables $\boldsymbol{x}$ that describe
the event are being measured.
By construction, Eq.~(\ref{eq:dsigma_dx}) is normalized as should be the case
for a probability density.
We can now use Eq.~(\ref{eq:log_likelihood_extended}) together with Eq.~(\ref{eq:dsigma_dx})
in order to evaluate the unbinned profile likelihood ratio, Eq.~(\ref{eq:plr}):
\begin{equation}
  q_{\boldsymbol{c}} =  -2 \log \frac{\mathcal{L}(\boldsymbol{c})}{\mathcal{L}(\boldsymbol{\hat{c}})}=
  2\left[\nu_{\rm tot}(\boldsymbol{c}) - \nu_{\rm tot}(\boldsymbol{\hat{c}}) -\sum_{i=1}^{N_{\rm ev}} \log \lp 
	  \frac{d\sigma(\boldsymbol{x}_i, \boldsymbol{c})}{d\boldsymbol{x}} \Bigg{/}\frac{d\sigma(\boldsymbol{x}_i, \boldsymbol{\hat{c}})}{d\boldsymbol{x}} \rp \right].
	\label{eq:plr_4}
\end{equation}
For convenience of notation, let us we define
\begin{equation}
  r_\sigma( \boldsymbol{x}_i, \boldsymbol{c}, \boldsymbol{\hat{c}}) \equiv
  \frac{d\sigma(\boldsymbol{x}_i, \boldsymbol{c})}{d\boldsymbol{x}} \Bigg{/}\frac{d\sigma(\boldsymbol{x}_i, \boldsymbol{\hat{c}})}{d\boldsymbol{x}}  \qquad \text{and}\qquad  r_\sigma( \boldsymbol{x}_i, \boldsymbol{c})\equiv  r_\sigma( \boldsymbol{x}_i, \boldsymbol{c}, \boldsymbol{0}) \, .
  \label{eq:ratio_rsigma_definition}
\end{equation}
The latter is especially useful in cases such as the SMEFT, where
the alternative
hypotheses corresponds to the vanishing of all the theory
parameters (the EFT Wilson coefficients), and $\mathcal{T}$ reduces
to the SM.
In terms of this notation, we can then express the profile likelihood ratio for the unbinned
observables Eq.~(\ref{eq:plr_4}) as
\begin{equation}
q_{\boldsymbol{c}}= 2\left[\nu_{\rm tot}(\boldsymbol{c})-\sum_{i=1}^{N_{\rm ev}} \log  r_\sigma( \boldsymbol{x}_i, \boldsymbol{c})\right] -  2\left[\nu_{\rm tot}(\boldsymbol{\hat{c}})-\sum_{i=1}^{N_{\rm ev}} \log  r_\sigma( \boldsymbol{x}_i, \boldsymbol{\hat{c}})\right] \, .
	\label{eq:plr_5}
\end{equation}
One can then use either Eq.~(\ref{eq:plr_4}) or Eq.~(\ref{eq:plr_5}) to derive confidence level intervals associated
to the theory parameters $\boldsymbol{c}$ in the same manner as in the binned case, namely
by imposing Eq.~(\ref{eq:solving_c_from_q}) for a given choice of the CL range.

Provided double counting is avoided, binned and unbinned observables can simultaneously be used
in the context of parameter limit setting.
In this general case one assembles a joint likelihood which
accounts for the contribution of all available types of observables,
namely
\be
\mathcal{L}( \boldsymbol{c}) = \prod_{k=1}^{N_{\mathcal{D}}}\mathcal{L}_k( \boldsymbol{c}) =
\prod_{k=1}^{N_{\mathcal{D}}^{\rm (ub)}}\mathcal{L}_k^{\rm (ub)}( \boldsymbol{c})
\prod_{j=1}^{N_{\mathcal{D}}^{\rm (bp)}}\mathcal{L}_j^{\rm (bp)}( \boldsymbol{c})
\prod_{\ell=1}^{N_{\mathcal{D}}^{\rm (bg)}}\mathcal{L}_\ell^{\rm (bg)}( \boldsymbol{c}) \, ,
\ee
where we have $N_{\mathcal{D}} = N_{\mathcal{D}}^{\rm (ub)}+N_{\mathcal{D}}^{\rm (bp)} + N_{\mathcal{D}}^{\rm (bg)}$
datasets classified into unbinned (ub), binned Poissonian (bp), and binned Gaussian (bg)
datasets, where the corresponding likelihoods are given by Eq.~(\ref{eq:likelihood_extended})
for unbinned, Eq.~(\ref{eq:joint_pdf_simp}) for binned Poissonian, and
Eq.~(\ref{eq:binned_gaussian_likelihood}) for binned Gaussian observables.
The associated log-likelihood function is then
\be
\label{eq:log_likelihood_global}
\log \mathcal{L}( \boldsymbol{c}) = \sum_{k=1}^{N_{\mathcal{D}}} \log \mathcal{L}_k( \boldsymbol{c}) =
\sum_{k=1}^{N_{\mathcal{D}}^{\rm (ub)}}\log \mathcal{L}_k^{\rm (ub)}( \boldsymbol{c}) +
\sum_{j=1}^{N_{\mathcal{D}}^{\rm (bp)}}\log \mathcal{L}_j^{\rm (bp)}( \boldsymbol{c}) + 
\sum_{\ell=1}^{N_{\mathcal{D}}^{\rm (bg)}} \log \mathcal{L}_\ell^{\rm (bg)}( \boldsymbol{c}) \, ,
\ee
which can then be used to construct the profile likelihood ratio Eq.~(\ref{eq:plr})
in order to test the null hypothesis and determine confidence level intervals
in the theory parameters $\boldsymbol{c}$. 

The main challenge for the integration of unbinned observables in global fits
using the framework summarized by Eq.~(\ref{eq:log_likelihood_global}) is that the evaluation of $\mathcal{L}_k^{\rm (ub)}( \boldsymbol{c})$
is in general costly, since the underlying probability density is not known in closed
form and hence needs to be computed numerically using Monte Carlo methods.
In this next section we discuss how to bypass this problem by adopting machine learning
techniques to parametrize this probability density (the differential cross-section)
and hence assemble unbinned observables which are  fast and efficient to evaluate,
as required for their integration into a global SMEFT analysis.

\section{Unbinned observables from machine learning}
\label{sec:trainingMethodology}

In this section we describe our approach
 to construct unbinned multivariate
observables tailored for global EFT analyses by means
of supervised machine learning.
We discuss how neural networks are deployed as universal unbiased interpolants in order
to parametrize likelihood ratios associated to the theoretical models
of the SM and EFT differential cross-sections, making possible the efficient
 evaluation of the likelihood functions for arbitrary values
of the Wilson coefficients as required for parameter inference.
We emphasize the scalability and robustness of our approach with respect
to the number of coefficients and to the dimensionality of the final state
kinematics, and validate the performance of the neural network training.

\subsection{Differential cross-sections}
\label{sec:param_eft_xsec}

Following the notation of Sect.~\ref{sec:unbinned_likelihood},
we consider a given process
whose associated measurement $\mathcal{D}$ consists of
$ N_{\rm ev}$ events, each of them characterized by $n_{k}$ final state
variables,
\be
\mathcal{D}=\{ \boldsymbol{x}_i\}\, \qquad
\boldsymbol{x}_i= \lp x_{i,1}, x_{i,2},\ldots, x_{i,n_{k}} \rp \, ,\qquad  i=1, \ldots, N_{\rm ev} \, .
\ee
The kinematic variables (features) $\boldsymbol{x}$ under consideration depend on the type
of measurement that is being carried out.
For instance, for a top quark measurement at the parton level, one would have
that the $\boldsymbol{x}_i$ are the transverse momenta and rapidities of the top quark,
while for the corresponding particle-level measurement, one would use instead
$b$-jet and leptonic kinematic variables.
Likewise,  $\boldsymbol{x}_i$ could also correspond
to detector-level kinematic variables for measurements
carried out without unfolding.
The inclusive cross-section case corresponds to $n_k=0$
when one integrates over all final state kinematics subject to fiducial cuts.
The probability distribution associated to the events constituting $\mathcal{D}$
is given by the differential cross-section
\begin{equation}
  \label{eq:prob_density_sect3}
f_\sigma\lp \boldsymbol{x}, \boldsymbol{c} \rp= \frac{1}{\sigma_{\rm fid}(\boldsymbol{c}) }\frac{d\sigma(\boldsymbol{x}, \boldsymbol{c})}{d\boldsymbol{x}} \, ,
\end{equation}
in terms of the model parameters $\boldsymbol{c}$.

In general not all $n_k$ kinematic variables that one can consider for a given process
will be independent.
For example, $2\to 2$ processes with on-shell
particles (like $pp\to t\bar{t}$
before decay) are fully described by three independent final-state variables.
For more exclusive measurements, $n_k$ grows rapidly yet the final-state variables
remain partially correlated to each other.
The best choice of $\boldsymbol{x}$ and $n_k$ should in this respect
be studied separately from the impact associated
to the use of unbinned observables as compared to their binned
counterparts.
Furthermore, in the same manner that one expects that the constraints provided by a binned observable
tend to those from unbinned ones in the narrow bin limit,
these constraints will also saturate once $n_k$ becomes large enough
that adding more variables does not provide independent information.

In the specific case of the SMEFT,
the parameters of the theory framework $\mathcal{T}\lp \boldsymbol{c}\rp$
are the Wilson coefficients associated
to the $n_{\rm eft}$
higher dimensional operators that enter the description of the processes
under consideration for a given set of flavor assumptions.
Given that a differential cross-section in the dimension-six SMEFT
exhibits at most a quadratic dependence with the Wilson coefficients,
one can write the differential probability density in
Eq.~(\ref{eq:prob_density_sect3}) as\footnote{We adopt a notation
  where the cutoff scale $\Lambda$ is being reabsorbed into a redefinition
  of the Wilson coefficients.
  Therefore, the coefficients ${\boldsymbol c}$ in
  Eq.~(\ref{eq:EFT_structure})
  are to be understood as 
  $\tilde{c}_i/\Lambda^2$ in terms of the dimensionless coefficients
   $\tilde{c}_i$
  entering the SMEFT Lagrangian. }
\be
\label{eq:EFT_structure}
f_\sigma(\boldsymbol{x},\boldsymbol{c}) =
f_\sigma(\boldsymbol{x},\boldsymbol{0}) + \sum_{j=1}^{n_{\rm{eft}}}
f^{(j)}_\sigma(\boldsymbol{x})c_j
+\sum_{j=1}^{n_{\rm{eft}}}\sum_{k \ge j}^{n_{\rm{eft}}}
f^{(j,k)}_\sigma(\boldsymbol{x})c_j c_k \, ,
\ee
where $f_\sigma(\boldsymbol{x},\boldsymbol{0}) $ corresponds to the SM
cross-section, $f^{(j)}_\sigma(\boldsymbol{x})$ indicates the linear EFT corrections
arising from the interference with the SM amplitude,
and $f^{(j,k)}_\sigma(\boldsymbol{x})$ corresponds to the quadratic corrections
associated to the square of the EFT amplitude.
We note that while $f_\sigma(\boldsymbol{x},\boldsymbol{0}) $
and $f^{(j,k)}_\sigma(\boldsymbol{x})$ arise from squared amplitudes
and hence are positive-definite, this is not necessarily the case
for the interference cross-section $f^{(j)}_\sigma(\boldsymbol{x})$.

The SM and EFT
cross-sections $f_\sigma(\boldsymbol{x},\boldsymbol{0}), f^{(j)}_\sigma(\boldsymbol{x})$,
and $f^{(j,k)}_\sigma(\boldsymbol{x})$ can be evaluated in perturbation theory, and one can account
for different types of effects such as parton shower, hadronization, or detector simulation,
depending on the observable under consideration.
The SM cross-sections $f_\sigma(\boldsymbol{x},\boldsymbol{0})$ can be
computed at NNLO QCD (eventually matched
to parton showers) for most of the LHC processes relevant for global EFT fits, while
for the EFT linear and quadratic corrections the accuracy frontier is NLO QCD~\cite{Degrande:2020evl}.
The settings of the calculation should be chosen to reproduce
as close as possible those of the corresponding
experimental measurement, while aiming to minimize
the associated theoretical uncertainties.
In this work we evaluate the differential cross-sections $f_\sigma(\boldsymbol{x},\boldsymbol{c})$
numerically, cross-checking with analytic calculations  whenever possible.

In order to construct unbinned observables in an efficient manner it
is advantageous to work in terms of the ratio between EFT and SM
cross-sections, Eq.~(\ref{eq:ratio_rsigma_definition}), which,
accounting for the quadratic structure of the EFT cross-sections in Eq.~(\ref{eq:EFT_structure}),
can be expressed as
\be
\label{eq:EFT_structure_v2}
r_{\sigma}(\boldsymbol{x}, \boldsymbol{c}) \equiv \frac{f_\sigma(\boldsymbol{x},\boldsymbol{c})}{f_\sigma(\boldsymbol{x},\boldsymbol{0})} =
1 + \sum_{j=1}^{n_{\rm eft}}
r^{(j)}_\sigma(\boldsymbol{x})c_j
+\sum_{j=1}^{n_{\rm eft}}\sum_{k \ge j}^{n_{\rm eft}}
r^{(j,k)}_\sigma(\boldsymbol{x})c_j c_k \, ,
\ee
where we have defined the linear and quadratic ratios to the SM cross-section as
\be
r^{(j)}_\sigma(\boldsymbol{x}) = \frac{f^{(j)}_\sigma(\boldsymbol{x})}{
  f_\sigma(\boldsymbol{x},\boldsymbol{0})} \, ,\qquad
r^{(j,k)}_\sigma(\boldsymbol{x}) = \frac{f^{(j,k)}_\sigma(\boldsymbol{x})}{
f_\sigma(\boldsymbol{x},\boldsymbol{0})} \, .
\ee
Parameterizing the ratios between the EFT and SM cross-sections, Eq.~(\ref{eq:EFT_structure_v2}), is
beneficial as compared to directly
parameterizing the absolute cross-sections since in general EFT effects
represent a moderate distortion of the SM baseline prediction.

As indicated by Eq.~(\ref{eq:plr_5}), the profile likelihood ratio used to derive limits on the EFT coefficients
can be expressed in terms of the ratio Eq.~(\ref{eq:EFT_structure_v2}).
Indeed, in the case of the dimension-six SMEFT
the PLR reads
\bea
  q_{\boldsymbol{c}}&=& 2\left[\nu_{\rm tot}(\boldsymbol{c})-\sum_{i=1}^{N_{\rm ev}} \log\lp
    1 + \sum_{j=1}^{n_{\rm eft}}
r^{(j)}_\sigma(\boldsymbol{x}_i)c_j
+\sum_{j=1}^{n_{\rm eft}}\sum_{k \ge j}^{n_{\rm eft}}
r^{(j,k)}_\sigma(\boldsymbol{x}_i)c_j c_k\rp
\right] \nonumber \\
 &-& 2\left[\nu_{\rm tot}(\boldsymbol{\hat{c}})-\sum_{i=1}^{N_{\rm ev}} \log
\lp 1 + \sum_{j=1}^{n_{\rm eft}}
r^{(j)}_\sigma(\boldsymbol{x}_i)\hat{c}_j
+\sum_{j=1}^{n_{\rm eft}}\sum_{k \ge j}^{n_{\rm eft}}
r^{(j,k)}_\sigma(\boldsymbol{x}_i)\hat{c}_j \hat{c}_k\rp
    \right] \, .
\label{eq:plr_6}
\eea
where the  $\boldsymbol{\hat{c}}$
denotes the maximum likelihood estimator of the Wilson coefficients.
We emphasize that in this derivation the SM serves as a natural reference hypothesis in the EFT parameter space - ratios expressed with respect to another reference point, say $\boldsymbol{c'}$, are trivially equivalent according to the following identity
\be
\label{eq:EFT_structure_v3}
r_{\sigma}(\boldsymbol{x}, \boldsymbol{c}) = \frac{f_\sigma(\boldsymbol{x},\boldsymbol{c})}{f_\sigma(\boldsymbol{x},\boldsymbol{0})} =
\frac{f_\sigma(\boldsymbol{x},\boldsymbol{c})}{f_\sigma(\boldsymbol{x},\boldsymbol{c'})} \frac{f_\sigma(\boldsymbol{x},\boldsymbol{c'})}{f_\sigma(\boldsymbol{x},\boldsymbol{0})} \, .
\ee

The main challenge in applying limit setting to unbinned observables
by means of the profile likelihood ratio of Eq.~(\ref{eq:plr_6}) is that the evaluation
of the EFT cross-section ratios $r^{(j)}_\sigma(\boldsymbol{x})$ and $r^{(j,k)}_\sigma(\boldsymbol{x})$ is
computationally intensive, and in many cases intractable, specifically for high-multiplicity
observables and  when the number of events considered $N_{\rm ev}$ is large.
As we explain next, in this work we bypass this challenge by parameterizing the EFT cross-section ratios in terms
of feed-forward neural networks, with the kinematic variables $\boldsymbol{x}$ as inputs,
trained on the outcome of Monte Carlo simulations.

\subsection{Cross-section parametrization}
\label{sec:methodology_xsec_param}

As first introduced in Sect.~\ref{sec:statistical_framework}, the profile likelihood
ratio provides an optimal test statistic
in the sense that no statistical power is lost in the process of mapping the high-dimensional feature vector $\boldsymbol{x}$
onto the scalar ratio $r_\sigma(\boldsymbol{x}, \boldsymbol{c})$.
Performing inference on the Wilson coefficients
using the profile likelihood ratio from Eq.~(\ref{eq:plr_6})
requires a precise knowledge about the differential cross section ratio $r_\sigma(\boldsymbol{x}, \boldsymbol{c})$ for arbitrary values
of $\boldsymbol{c}$.
However, in general one does not have direct access to $r_\sigma(\boldsymbol{x}, \boldsymbol{c})$
whenever MC event generators can only be run in the forward mode, i.e. used to generate samples.
The inverse problem, namely statistical inference, is often rendered intractable due to the many paths in parameter space that lead from the theory parameters $\boldsymbol{c}$ to the final measurement in the detector.
In the machine learning literature this intermediate (hidden) space is known as the latent space.

Feed-forward neural networks are suitable in this context
as model-independent unbiased interpolants to construct a surrogate of the true profile likelihood ratio.
Consider two balanced
datasets $\mathcal{D}_{\rm eft}(\boldsymbol{c})$ and $\mathcal{D}_{\rm sm}$ generated
based on the theory hypotheses $\mathcal{T}(\boldsymbol{c})$ and $\mathcal{T}(\boldsymbol{0})$ respectively,
where by balanced we mean that the same number of unweighted Monte Carlo events are generated in both cases.
We would like to determine the decision boundary
function $g(\boldsymbol{x,c})$ which can be used to classify an event
$\boldsymbol{x}$ into either $\mathcal{T}(\boldsymbol{0})$, the Standard Model,
or $\mathcal{T}(\boldsymbol{c})$, the SMEFT hypothesis for point $\boldsymbol{c}$
in parameter space.
We can determine this decision boundary
by using the balanced datasets $\mathcal{D}_{\rm eft}(\boldsymbol{c})$ and $\mathcal{D}_{\rm sm}$
to train a binary classifier by means of the cross-entropy loss-functional, defined as
\be 
L[g(\boldsymbol{x},\boldsymbol{c})] = -\int d\boldsymbol{x}\frac{d\sigma(\boldsymbol{x},\boldsymbol{c})}{d\boldsymbol{x}}\log(1-g(\boldsymbol{x},\boldsymbol{c})) - \int d\boldsymbol{x}\frac{d\sigma(\boldsymbol{x},\boldsymbol{0})}{d\boldsymbol{x}}\log g(\boldsymbol{x},\boldsymbol{c}) \, .
\label{eq:loss_CE}
\ee
In practice, the integrations required in the evaluation of the cross-entropy loss Eq.~(\ref{eq:loss_CE}) are carried out numerically from the generated Monte Carlo events, such that
\be 
L[g(\boldsymbol{x},\boldsymbol{c})] = - \sigma_{\rm fid}(\boldsymbol{c})\sum_{i=1}^{N_{\rm ev}}\log(1-g(\boldsymbol{x}_i,\boldsymbol{c})) - \sigma_{\rm fid}(\boldsymbol{0}) \sum_{j=1}^{N_{\rm ev}} \log g(\boldsymbol{x}_j,\boldsymbol{c}) \, ,
\label{eq:loss_CE_numeric}
\ee
where $\sigma_{\rm fid}(\boldsymbol{c})$ and $\sigma_{\rm fid}(\boldsymbol{0})$
represent the integrated fiducial cross-sections in the SMEFT and the SM
respectively.
Recall that we have two independent sets of $N_{\rm ev}$ events each generated under
$\mathcal{T}(\boldsymbol{0})$ and $\mathcal{T}(\boldsymbol{c})$ respectively,
and hence in Eq.~(\ref{eq:loss_CE_numeric}) the first (second) term
in the RHS involves the sum over the $N_{\rm ev}$ events generated according to $\mathcal{T}(\boldsymbol{c})$
($\mathcal{T}(\boldsymbol{0})$).

It is also possible to adopt other loss functions for the binary classifier Eq.~(\ref{eq:loss_CE}),
such as the quadratic loss used in~\cite{Chen:2020mev}.
The outcome of the classification
should be stable with respect to alternative choices of the loss function, and indeed
we find that both methods lead to consistent results, while the cross entropy formulation
benefits from a faster convergence due to presence of stronger gradients as compared
to the quadratic loss.

In the limit of an infinitely large training dataset
and sufficiently flexible parametrization, one can take the functional derivative of $L$ with respect to the
decision boundary function $g(\boldsymbol{x}, \boldsymbol{c})$ to determine that it is given by
\be 
\frac{\delta L}{\delta g} = 0 \implies g(\boldsymbol{x}, \boldsymbol{c}) = \lp 1 +\frac{d\sigma(\boldsymbol{x},\boldsymbol{c})}{d\boldsymbol{x}}
\Bigg/ \frac{d\sigma(\boldsymbol{x},\boldsymbol{0})}{d\boldsymbol{x}} \rp^{-1} = \frac{1}{1 + r_\sigma(\boldsymbol{x}, \boldsymbol{c})} \, ,
\label{eq:optimal_f}
\ee
and hence in this limit
the solution of the classification problem defined by the cross-entropy
function Eq.~(\ref{eq:loss_CE}) is given by the EFT ratios $r_\sigma(\boldsymbol{x}, \boldsymbol{c})$
that need to be evaluated
in order to determine the associated profile likelihood ratio.
Hence our strategy will be to parametrize $r_\sigma(\boldsymbol{x}, \boldsymbol{c})$ with neural networks,
benefiting from the characteristic quadratic structure of the EFT cross-sections, and then
training these machine learning classifiers by minimizing the loss function Eq.~(\ref{eq:loss_CE}).

In practice, one can only expect to obtain a reasonably good estimator $\hat{g}$ of the true result due to finite size effects
in the Monte Carlo training data $\mathcal{D}_{\rm eft}$ and $\mathcal{D}_{\rm sm}$ and in the neural network
architecture.
Since EFT and SM predictions largely overlap in a significant region of the phase space,
it is crucial to obtain a decision boundary trained with as much precision as possible
in order to have a reliable test statistic to carry out inference.
The situation is in this respect different from  usual classification problems, for which an imperfect decision boundary
parameterized by $g$ can still achieve high performances whenever most features are disjoint, and hence
a slight modification of  $g$ does not lead to a significant performance drop.
In order to estimate the uncertainties associated to the fact that
the actual estimator $\hat{g}$ differs from the true result $g(\boldsymbol{x}, \boldsymbol{c})$,
in this work we use the Monte Carlo replica method
described in Sect.~\ref{sec:nntraining}.

Given the quadratic structure of the EFT cross-sections and their ratios to the SM prediction, Eqns.~(\ref{eq:EFT_structure})
and~(\ref{eq:EFT_structure_v2}) respectively, once the  linear and quadratic ratios $r_\sigma^{(j)}({\boldsymbol{x}})$ and
$r_\sigma^{(j, k)}({\boldsymbol{x}})$ are determined throughout the entire phase space one can
straightforwardly evaluate the EFT
differential cross sections (and their ratios to the SM) for
any point in the EFT parameter space.
Here we exploit this property during the neural network
training by decoupling the learning problem of the linear cross section ratios from that
of the quadratic ones.
This allows one to extract $r_\sigma^{(j)}$ and $r_\sigma^{(j, k)}$ independently from each other, meaning that the neural network classifiers can be trained in parallel and also that the
training scales at most quadratically with the number of EFT operators considered
$n_{\rm eft}$. 

To be specific, at the linear level we determine the EFT cross-section ratios
$r_\sigma^{(j)}({\boldsymbol{x}})$ by training the binary classifier from the cross-entropy loss
Eq.~(\ref{eq:loss_CE}) on a reference dataset $\mathcal{D}_{\rm sm}$ and an EFT dataset defined by
\be
\label{eq:eft_dataset_onecoeff}
\mathcal{D}_{\rm eft}(\boldsymbol{c}=(0,\ldots,0,c^{\rm (tr)}_j,0,\ldots,0)) \, ,
\ee
and generated at linear order,  $\mathcal{O}\left(\Lambda^{-2}\right)$, in the EFT expansion with all Wilson coefficients
set to zero except for the $j$-th one, which we denote by $c^{\rm (tr)}_j$.
For such model configuration, the EFT cross-section ratio can be parametrized as
\be 
r_\sigma(\boldsymbol{x}, c^{\rm (tr)}_j) = 1 + c^{\rm (tr)}_j \mathrm{NN}^{(j)}(\boldsymbol{x}) \, ,
\label{eq:nn_param_lin}
\ee 
where only the individual coefficient $c^{\rm (tr)}_j$ has survived
the sum in Eq.~(\ref{eq:EFT_structure_v2}) since all other EFT parameters
are switched off by construction.
Comparing Eq.~(\ref{eq:nn_param_lin}) and Eq.~(\ref{eq:EFT_structure_v2}) we see that in the large sample limit
\be 
\mathrm{NN}^{(j)}(\boldsymbol{x}) \rightarrow r_{\sigma}^{(j)}(\boldsymbol{x}) \, .
\label{eq:trained_lin_coeff}
\ee 
In practice, this relation will only be met with a certain finite accuracy due to statistical fluctuations in the finite training sets.
This limitation is especially relevant in phase space regions where the cross-section is suppressed, such as in the tails of invariant mass distributions,
and indicates that it is important to account for these methodological uncertainties associated to the training procedure.
By means of the Monte Carlo replica method one can estimate and propagate these uncertainties
first to the parametrization of the EFT ratio $r_{\sigma}$ and then to the associated limits on the
Wilson coefficients.

Concerning the training of the EFT quadratic cross-section ratios $r_\sigma^{(j, k)}$, we follow the same strategy as in the linear case, except that now we construct the EFT dataset at quadratic order without any linear contributions.
By omitting the linear term, we reduce the learning problem at the quadratic level to a linear one.
Specifically, we generate events at pure $\mathcal{O}\left(\Lambda^{-4}\right)$ level, without the
interference (linear) contributions,
in the EFT by switching off all Wilson coefficients except two of them,
denoted by $c^{\rm (tr)}_j$ and $c^{\rm (tr)}_k$, 
\be
\label{eq:eft_dataset_twocoeff}
\mathcal{D}_{\rm eft}(\boldsymbol{c}=(0,\ldots,0,c^{\rm (tr)}_j,0,\ldots,0, c^{\rm (tr)}_k, 0, \ldots )) \, ,
\ee
and parametrize the cross-section ratio as
\be 
r_\sigma(\boldsymbol{x}, c^{\rm (tr)}_j, c^{\rm (tr)}_k) = 1 + c^{\rm (tr)}_j c^{\rm (tr)}_k \mathrm{NN}^{(j, k)}(\boldsymbol{x}) \, ,
\label{eq:nn_param_quad}
\ee 
where only purely quadratic terms with both $c^{\rm (tr)}_j$ and $c^{\rm (tr)}_k$ have survived the sum.
Note that when $j \neq k$, this parametrization of the cross-section ratio $r_\sigma(\boldsymbol{x}, c^{\rm (tr)}_j, c^{\rm (tr)}_k)$ depends only on the product $c_{j} c_{k}$,
whereas when $j=k$ it depends only on terms proportional to $c_{j}^{2}$.
The cross-section ratio is parametrized in this way
to facilitiate separate training of the $c_{j}^{2}$, $c_{k}^{2}$ and $c_{j} c_{k}$ terms, and we make use of training data
in which the contributions from each of these terms has been separately generated, as discussed in more detail in Sect.~\ref{sec:pipeline}. 
By the same reasoning as above, in the large sample limit we will have that
\be 
\mathrm{NN}^{(j, k)}(\boldsymbol{x}) \rightarrow r_{\sigma}^{(j, k)}(\boldsymbol{x}) \, .
\label{eq:trained_quad_coeff}
\ee
We note that in the case that the Monte Carlo generator used to evaluate
the theory predictions $\mathcal{T}(\boldsymbol{c})$ does not allow
the separate evaluation of the EFT quadratic terms, one can always subtract the linear
contribution numerically by means of the outcome of Eq.~(\ref{eq:trained_lin_coeff}).

By repeating this procedure $n_{\rm eft}$ times for
the linear terms and $n_{\rm eft}(n_{\rm eft}+1)/2$ times for the quadratic terms, one ends up with the
set of functions that parametrize the EFT cross-section ratio Eq.~(\ref{eq:EFT_structure_v2}),
\be
\label{eq:set_of_trained_NNs}
\{ \mathrm{NN}^{(j)}(\boldsymbol{x}) \} \quad {\rm and}\quad \{ \mathrm{NN}^{(j,k)}(\boldsymbol{x}) \} \, ,
\quad  j, k=1,\ldots,n_{\rm eft}\,, \quad k\ge j \, .
\ee
The similar structure that is shared between Eq.~(\ref{eq:nn_param_lin}) and Eq.~(\ref{eq:nn_param_quad}) implies that parameterizing the quadratic EFT contributions
in this manner is ultimately a linear problem, i.e. redefining the product $c^{\rm (tr)}_j c^{\rm (tr)}_k$ as  $\tilde{c}^{\rm (tr)}_{j,k}$ maps the quadratic learning problem back to a linear one:
\be 
r_\sigma(\boldsymbol{x}, \tilde{c}^{\rm (tr)}_{j,k}) = 1 + \tilde{c}^{\rm (tr)}_{j,k}\mathrm{NN}^{(j, k)}(\boldsymbol{x}) \, .
\label{eq:nn_param_quad_redef}
\ee
Eq.~(\ref{eq:set_of_trained_NNs}) represents the final outcome of the training procedure,
namely an approximate parametrization  $\hat{r}_{\sigma}(\boldsymbol{x}, \boldsymbol{c})$ of the true
EFT cross-section ratio  $r_\sigma(\boldsymbol{x}, \boldsymbol{c})$,
\be
\label{eq:EFT_structure_v4}
\hat{r}_{\sigma}(\boldsymbol{x}, \boldsymbol{c}) =
1 + \sum_{j=1}^{n_{\rm eft}}
 \mathrm{NN}^{(j)}(\boldsymbol{x}) c_j
+\sum_{j=1}^{n_{\rm eft}}\sum_{k \ge j}^{n_{\rm eft}}
\mathrm{NN}^{(j, k)}(\boldsymbol{x}) c_j c_k \, ,
\ee
valid for any point in the model parameter $\boldsymbol{c}$,
as 
required to evaluate the profile likelihood ratio in Eq.~(\ref{eq:plr_5})
and to
perform inference on the Wilson coefficients.
Below we provide technical details about how the neural network training
is carried out and how uncertainties are estimated by means of the replica method.

\paragraph{Cross-section positivity during training}
While the differential cross-section $f_\sigma\lp \boldsymbol{x}, \boldsymbol{c}\rp$ (and its ratio
to the SM) is positive-definite, this is not necessarily the case for the linear
(interference) EFT term, and hence in principle Eq.~(\ref{eq:nn_param_lin}) is unbounded from below.

At the level of the training pseudo-data, we avoid the issue of negative cross-sections by generating our
pseudo-data at fixed values of the Wilson coefficients, specifically chosen such that the differential cross sections 
are always positive.  For example, in the case of negative interference between the EFT and the SM, we generate our training pseudo-data
assuming a negative Wilson coefficient such that the net effect of the EFT is an enhancement relative to the SM.
The choices of Wilson coefficients used in our study will be further discussed in Sect.~\ref{sec:inputs_traiing} and in Table~\ref{tab:settings_nntrain}.

It only then remains to ensure  
that the physical requirement of cross-section positivity
is satisfied at the level
of neural network training, and hence that the parameter space region leading
to negative cross-sections is avoided.
Cross-section positivity can be implemented at the training level
by means
of adding a penalty term to the loss function
whenever the likelihood ratio becomes negative through a Lagrange multiplier.
That is, the loss function is extended as
\be
\label{eq:cross-section-pos}
L[g] \rightarrow L[g] + \lambda \, \mathrm{ReLU}\left(\frac{g(\boldsymbol{x},\boldsymbol{c})-1}{g(\boldsymbol{x},\boldsymbol{c})}\right) = L[g] + \lambda \, \mathrm{ReLU}\left(
- r_{\sigma}(\boldsymbol{x},\boldsymbol{c})\right)  \, ,
\ee
where $\mathrm{ReLU}$ stands for the Rectified Linear Unit activation function.
Such a Lagrange multiplier penalizes configurations where the likelihood ratio
becomes negative,  with the penalty increasing the more negative
$r_{\sigma}$ becomes. The value of the hyperparameter $\lambda$
should be chosen such that the training in the physically
allowed region is not distorted.
This is the same method used in the NNPDF4.0 
analysis to implement PDF positivity and integrability~\cite{NNPDF:2021uiq,NNPDF:2021njg}
at the training level without having to impose these constraints in at the parametrization level.

However, the Lagrange multiplier method defined by Eq.~(\ref{eq:cross-section-pos}) is not compatible with the cross-entropy loss function of Eq.~(\ref{eq:loss_CE}), given that this loss function is only well defined for $0 < g(\boldsymbol{x}, \boldsymbol{c}) < 1$ corresponding to positive likelihood ratios.
We note that this is not the case for other loss-functions for which configurations
with $r_\sigma(\boldsymbol{x}, \boldsymbol{c}) < 0$ are allowed, such as the quadratic loss-function used by~\cite{Chen:2020mev}, making them in principle compatible with the Lagrange multiplier method
to ensure cross-section positivity.

Instead of using the Lagrange multiplier method, in this work we introduce an alternative parameterization of the cross section ratio $r_\sigma$ such that cross-section positivity is guaranteed by construction.
Specifically, we modify  Eq.~(\ref{eq:nn_param_lin}) to enforce positivity, namely
the condition
\be 
r_{\sigma}(\boldsymbol{x}, \boldsymbol{c}) = \lp 1 + c^{(\rm tr)}_{j}\cdot \mathrm{NN}^{(j)}(\boldsymbol{x}) \rp > 0 \, ,
\ee
for any value of $\boldsymbol{x}$ and $ \boldsymbol{c}$,
by transforming the outcome of the neural network $\mathrm{NN}^{(j)}(\boldsymbol{x})$
as follows
\be 
\mathrm{NN}^{(j)}(\boldsymbol{x}) \rightarrow \widetilde{\mathrm{NN}}^{(j)}(\boldsymbol{x}; c^{(\rm tr)}_{j}) =
\begin{cases}
      \mathrm{ReLU}(\mathrm{NN}^{(j)}(\boldsymbol{x})) - 1/c^{(\rm tr)}_{j} + \epsilon,  & \text{if}\ c^{(\rm tr)}_{j}>0 \\
      - \mathrm{ReLU}(\mathrm{NN}^{(j)}(\boldsymbol{x})) - 1/c^{(\rm tr)}_{j} - \epsilon, & \text{if}\ c^{(\rm tr)}_{j}<0
    \end{cases},
\label{eq:xsec_pve_2}
\ee 
where $\epsilon$ is an infinitesimal positive constant to ensure $r_\sigma(\boldsymbol{x}, \boldsymbol{c}) > 0 $ when the linear contribution becomes negative, $\mathrm{NN}^{(j)}(\boldsymbol{x})<0$.
The transformation of Eq.~(\ref{eq:xsec_pve_2}) can be thought of as adding a custom activation function at the end of the network such that the cross-entropy loss is well-defined throughout the entire training procedure.
We stress that it is the transformed neural network $\widetilde{\mathrm{NN}}^{(j)}$ which is subject to training and not the original $\mathrm{NN}^{(j)}$.
Regarding imposing cross-section positivity at the quadratic level, we note that the transformation of Eq.~(\ref{eq:xsec_pve_2}) applies just as well in the quadratic case by virtue of Eq.~(\ref{eq:nn_param_quad_redef}),
and therefore the same approach can be taken there.
The main advantage of Eq.~(\ref{eq:xsec_pve_2}) as compared to the Lagrange multiplier
method is that we always work with a positive-definite likelihood ratio
as required by the cross-entropy loss function.

\subsection{Neural network training}
\label{sec:nntraining}

Here we describe the settings of the neural network training leading to
the parametrization of Eq.~(\ref{eq:EFT_structure_v3}).
We consider in turn the choice of neural network architecture, minimizer, and other
related hyperparameters; how the input data is preprocessed;
the settings of the stopping criterion used to avoid overfitting;  the estimate
of methodological uncertainties by means of the Monte Carlo replica method;
the scaling
of the ML training with respect to the number of EFT parameters;
and finally the validation procedure where the machine learning model
is compared to the analytic calculation of the likelihood ratio.

\paragraph{Architecture, optimizer, and hyperparameters.}
Table~\ref{tab:training_settings} specifies the training settings that are
adopted for each process, e.g. the features that were trained on, the architecture of the hidden layers, the
learning rate $\eta$ and the number of mini-batches.
Given a process for which the SMEFT parameter space is spanned by $n_{\rm eft}$ Wilson coefficients,
there are a maximum of $N_{\rm nn}=(n_{\rm eft}^2 + 3n_{\rm eft})/2$
independent neural networks to be trained.
In practice, this number can be smaller due to vanishing contributions,
in which case we will mention this explicitly. 
We have verified that we select redundant architectures, meaning that
training results are stable in the event that a somewhat less flexible architecture were
to be adopted.
For every choice of  $n_k$ kinematic features, these $N_{\rm nn}$ neural
networks share the same hyperparameters listed there.
The last column of Table~\ref{tab:training_settings}
indicates the average training time per replica and the corresponding standard deviation,
evaluated over the $N_{\rm nn}\times N_{\rm rep}$ networks to be trained
for a given process.
In future work one can consider an automated process to optimize
the choice of the hyperparameters listed in Table~\ref{tab:training_settings} along the lines of 
the strategy adopted for the NNPDF4.0 analysis~\cite{NNPDF:2021njg,Carrazza:2019mzf}.

\begin{table}[t]
  \centering
  \renewcommand{\arraystretch}{1.5}
    \begin{tabular}{c|c|c|c|c|c}
      \toprule
      \small
        process & $\quad$ features $\quad$ &$\quad$ hidden layers $\quad$& learning rate & $n_{\mathrm{batch}}$ & time (min) \\
        \midrule
     \multirow{2}{*}{$pp \to t\bar{t}$}& $m_{t\bar{t}}$ & $25\times 25\times 25 $ & $10^{-3}$& 5 &$17.3\pm13.9$\\
     & $m_{t\bar{t}}, y_{{t\bar{t}}}$ & $25\times 25\times 25$ & $10^{-3}$& 5 &$16.4\pm12.7$\\
     \midrule
         \multirow{3}{*}{ $pp\to t\bar{t}\to b\bar{b}\ell^+\ell^- \nu_\ell \bar{\nu}_{\ell}$} & $p_T^{\ell\bar{\ell}}$ & $25\times 25\times 25 $& $10^{-3}$&1&$46.8\pm35.0$\\
          & $p_T^{\ell\bar{\ell}}, \eta_\ell$ & $ 25\times 25\times 25 $& $10^{-3}$&1&$53.7\pm 29.9$\\
     & $18$ & $100\times 100 \times 100 $& $10^{-4}$&$50$&$5.4 \pm 2.7$\\
     \midrule
         \multirow{2}{*}{ $pp\to hZ\to b\bar{b}\ell^+\ell^-$} & $p_T^Z$ &$100 \times 100 \times 100$&$10^{-3}$ &$100$ &$9.4\pm9.0$\\
         & 7 &$ 100 \times 100\times 100$ &$10^{-4}$ &$50$ &$14.1\pm8.7$\\        
         \bottomrule
    \end{tabular}
    \vspace{0.3cm}
    \caption{Overview of the settings for the neural network trainings.
      For each of the processes to be described in Sect.~\ref{sec:pseudodata}, we specify the $n_k$ kinematic features $\boldsymbol{x}$ used
      for the likelihood ratio parametrization, the architecture, the learning rate, the number of mini-batches, and the training time per network averaged over all replicas.
      As indicated by Eq.~(\ref{eq:EFT_structure_v4}),
      given a process for which the parameter space is spanned by $n_{\rm eft}$ Wilson coefficients,
      there are  $(n_{\rm eft}^2 + 3n_{\rm eft})/2$
      independent neural networks to be trained.
      For each choice of  $n_k$ kinematic features, these neural
      networks share the settings listed here.
    }
    \label{tab:training_settings}
\end{table}

We train these neural networks by performing (mini)-batch gradient descent on the cross-entropy loss function Eq.~(\ref{eq:loss_CE_numeric}) using the \verb|AdamW|~\cite{Loshchilov2017} optimizer. Training was implemented in \verb|PyTorch|~\cite{Paszke2019} and run on \verb|AMD Rome| with $19.17$ HS06 per CPU core. We point the interested reader
to App.~\ref{app:code} and the corresponding online
documentation where the main features of the {\sc\small ML4EFT} software
framework are highlighted. 

\paragraph{Data preprocessing.}
The kinematic features $\boldsymbol{x}$ that enter
the evaluation of the likelihood function $f_\sigma(\boldsymbol{x}, \boldsymbol{c})$ and
its ratio $r_\sigma(\boldsymbol{x}, \boldsymbol{c})$
cannot be used directly as an input to the neural network training algorithm and should be preprocessed
first to ensure that the input information is provided to the neural nets
in their region of maximal sensitivity.
For instance, considering parton-level top quark pair production at the LHC,
the typical invariant masses $m_{t\bar{t}}$ to be used for the training
cover the range between 350 GeV and 3000 GeV,
while the rapidities are dimensionless and restricted to the range $y_{t\bar{t}}\in [-2.5,2.5]$.
To ensure a homogeneous and well-balanced training, especially for high-multiplicity
observables, all features should be transformed 
to a common range and their distribution in this range should be reasonably
similar.

A common data preprocessing
method for Gaussianly distributed variables is to standardize all features to zero mean and unit variance.
However, for typical LHC process the kinematic distributions are highly non-Gaussian,
in particular invariant mass and $p_T$ distributions are very skewed.
In such cases,  one instead should perform a rescaling based on a suitable interquartile range,
such as the $68\%$ CL interval.
This method is particularly interesting for our application because of its robustness to outliers at high invariant masses and transverse momenta, in the same way that the median is less sensitive to them than the sample mean.
In our approach we use a robust feature scaler which
subtracts the median and scales to an inter-quantile range,
resulting into input feature distributions peaked around zero with their bulk well contained within the
$\lc -1,1\rc$ region, which is not necessarily the case for the standardized Gaussian scaler.
Further justification of this choice will be provided in Sect.~\ref{sec:pseudodata}.
See also~\cite{Carrazza:2021yrg} for a recent application of feature scaling to the training
of neural networks in the context of PDF fits.

\paragraph{Stopping and regularization.}
The high degree of flexibility of neural networks in supervised learning applications
has an associated risk of overlearning, whereby the model ends up learning the statistical
fluctuations present in the data rather than the actual underlying law.
This implies that for a sufficiently flexible architecture a training with a fixed number of epochs will result
in either underlearning or overfitting, and hence that the optimal
number of epochs should be determined separately for each individual training by means
of a stopping criterion.

Here the optimal stopping point is determined separately
for each trained neural network by means of a variant
of the cross-validation dynamical stopping algorithm introduced in~\cite{NNPDF:2014otw}.
Within this approach, one splits up randomly each of the input datasets
$\mathcal{D}_{\rm sm}$ and $\mathcal{D}_{\rm eft}$ into two
disjoint sets known as the training set and the validation set, in a $80\%/20\%$ ratio.
The points in the validation subset are excluded from the optimization procedure,
and the loss function evaluated on them, denoted by $L_{\rm val}$, is used as a diagnosis tool
to prevent overfitting.
The minimization of the training loss function $L_{\rm tr}$ is carried out while
monitoring the value of $L_{\rm val}$.
One continues training until $L_{\rm val}$ has stopped decreasing following $n_{\rm p}$ (patience) epochs with respect to its last recorded local minimum.
The optimal network parameters, those with the smallest generalization error,
then correspond to those at which $L_{\rm val}$
exhibits this global minimum within the patience threshold.

\begin{figure}[t]
    \centering
    \includegraphics[width=\textwidth]{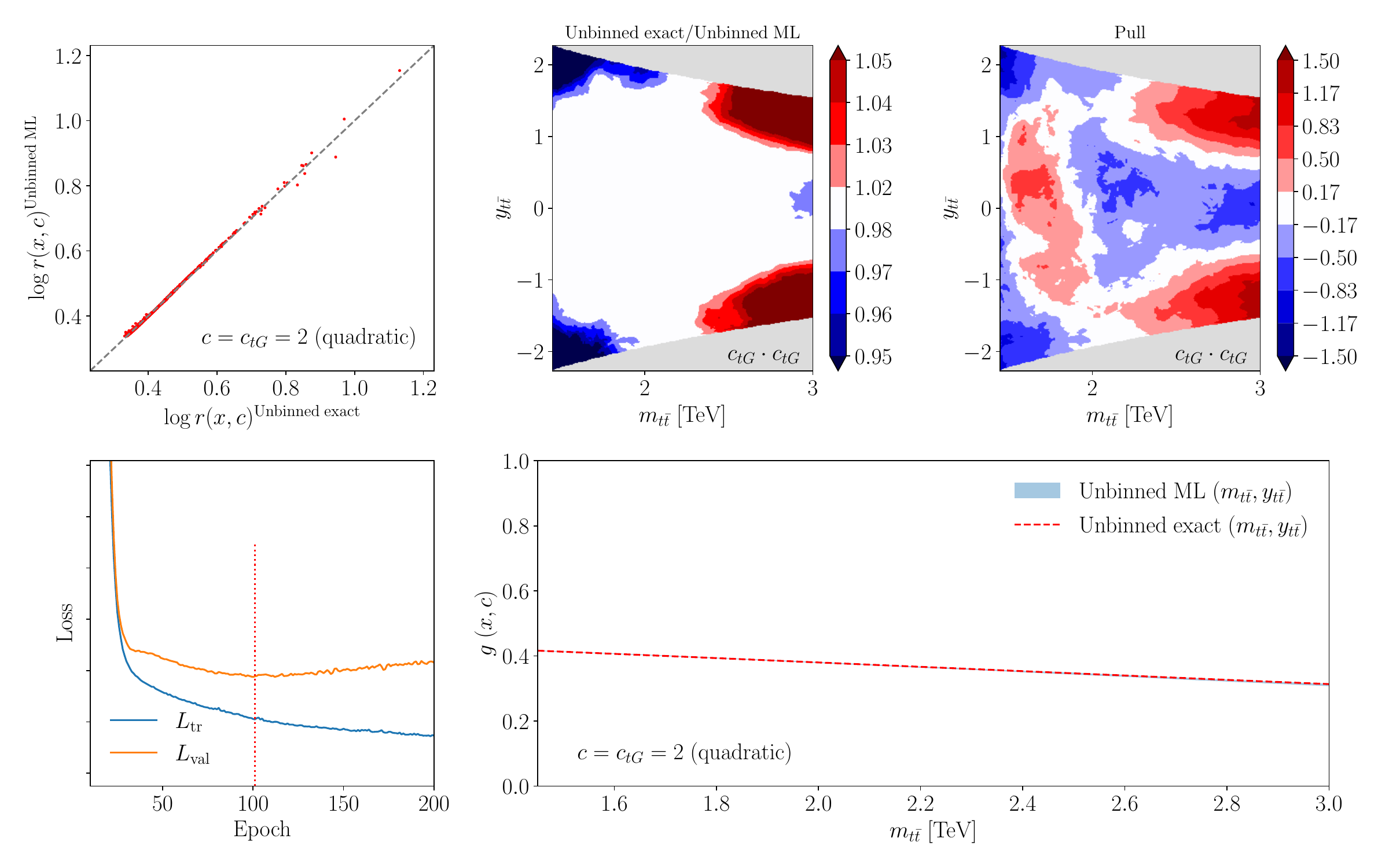}
    \caption{Validation of the machine learning parametrization
      of the EFT cross-section ratios
      when applied to  the case of parton-level top quark pair production to be
      described in Sect.~\ref{sec:pseudodata}.
      The results shown here correspond to the training of
      the quadratic neural network
      $\mathrm{NN}^{(j, j)}(m_{t\bar{t}},y_{t\bar{t}})$ in Eq.(\ref{eq:EFT_structure_v4})
      with $j$ indicating the chromomagnetic operator $c_{tG}$.
      From left to right and top to bottom we display a point-by-point comparison of the log-likelihood ratio in the ML model and the corresponding analytical calculation; the median
      of the ratio  between the ML model
      and the analytical calculation and the associated pull in the
      $(m_{t\bar{t}},y_{t\bar{t}})$ feature space; the evolution of the loss function split in training and validation sets for a representative replica as a function of the number of training epochs; and the resultant decision boundary $g(\boldsymbol{x,c})$ for $c_{tG}=2$ including MC replica uncertainties at the end of the training procedure.}
    \label{fig:nn_overview_perf}
\end{figure}

The bottom-left plot of Fig.~\ref{fig:nn_overview_perf} illustrates the dependence of
the training and validation loss functions in a representative training.
While $L_{\rm tr}$ continues to decrease as the number of epochs increases,
at some point  $L_{\rm val}$ exhibits a global minimum
and does not decrease further during $n_{\mathrm{p}}$ epochs.
The position of this global minimum is indicated with a vertical dashed line,
corresponding to the optimal stopping point.
The parameters of the trained network are stored for each iteration,
and once the optimal stopping point has been identified the final parameters
are assigned to be those of the epoch where $L_{\rm val}$ has its global minimum.

\paragraph{Uncertainty estimate from the replica method.}
In general the ML parametrization  $\hat{r}_{\sigma}(\boldsymbol{x}, \boldsymbol{c})$ will
differ from the true
EFT cross-section ratio $r_\sigma(\boldsymbol{x}, \boldsymbol{c})$ for two main reasons:
first, because of the finite statistics of the MC event samples used for the
neural network training, leading to a functional uncertainty in the ML model,
and second, due to residual inefficiencies of the optimization and stopping algorithms.
In order to quantify these sources of methodological uncertainty and their impact
on the subsequent EFT parameter inference procedure, we adopt the
neural network replica method developed in the
context of PDF determinations~\cite{Giele:2001mr,Giele:1998gw,Ball:2008by,DelDebbio:2004xtd}.

The basic idea is to generate $N_{\mathrm{rep}}$
replicas of the MC training dataset,
each of them statistically independent, and then train separate sets of neural networks
on each of these replicas.
As explained in Sect.~\ref{sec:methodology_xsec_param}, we train
the decision boundary $g({\boldsymbol{x}},{\boldsymbol{c}})$
from a balanced sample of SM and EFT events.
If we aim to carry out
the training of $\hat{r}_{\sigma}$ on a sample of $N_{\rm ev}$ events
(balanced between the EFT and SM hypotheses), one generates
a total of  $N_{\rm ev}\times N_{\rm rep}$ events and divides them into $N_{\mathrm{rep}}$
replicas, each of them containing the same amount of information on the
underlying EFT cross-section $r_\sigma$.
Subsequently, one trains the full set of neural networks required
to parametrize $\hat{r}_{\sigma}$ separately for each of these replicas,
using in each case different random seeds for the initialization of the network
parameters and other settings of the optimization algorithm.

In this manner, at the end of the training procedure, one ends up 
instead of Eq.~(\ref{eq:EFT_structure_v4}) with  an ensemble of  $N_{\rm rep}$
 replicas of the cross-section ratio parametrization,
\be
\label{eq:EFT_structure_v5}
\hat{r}^{(i)}_{\sigma}(\boldsymbol{x}, \boldsymbol{c}) \equiv
1 + \sum_{j=1}^{n_{\rm eft}}
 \mathrm{NN}^{(j)}_i(\boldsymbol{x}) c_j
+\sum_{j=1}^{n_{\rm eft}}\sum_{k \ge j}^{n_{\rm eft}}
\mathrm{NN}^{(j, k)}_i(\boldsymbol{x}) c_j c_k \, , \qquad i=1,\ldots,N_{\rm rep}
\ee
which estimates the methodological uncertainties associated to the parametrization
and training.
Confidence level intervals associated with these uncertainties can then be determined in the usual way, for instance by taking suitable lower and upper quantiles.
In other words, the replica ensemble given by Eq.~(\ref{eq:EFT_structure_v5}) provides
a suitable representation of the probability density in the space of NN models,
which can be used to quantify the impact of methodological
uncertainties at the level of EFT parameter inference.
For the processes considered in this work we find that values of  $N_{\rm rep}$ between 25 and 50
are sufficient to estimate the impact of these procedural uncertainties at the level
of EFT parameter inference.

\paragraph{Scaling with number of EFT parameters.}
If unbinned observables such as those constructed here are to be integrated
into global SMEFT fits, their scaling with the number of EFT operators $n_{\rm eft}$ considered
should be not too computationally costly, given that typical fits involve up
to $n_{\rm eft}\sim 50$ independent degrees of freedom.
In this respect, exploiting the polynomial structure of  EFT  cross-sections as done in this work
allows for an efficient scaling of the neural network training and makes complete paralellisation possible.
We note that most related approaches in the literature, such as e.g.~\cite{Cowan:2010js}, are limited to a small
number of EFT parameters and hence not amenable to global fits.
In other approaches, e.g.~~\cite{Chen:2020mev}, the proposed ML parametrization
is such that the coefficients of the linear and the quadratic terms mix, and in such case  no separation between linear and quadratic terms
and between different Wilson coefficients is possible.
This implies that in such approaches all
 neural networks
 parameterizing the likelihood functions
 need to be trained simultaneously and hence that parallelization is not possible.

 Within our framework,
 assembling the parametrization of the cross-section ratio Eq.~(\ref{eq:EFT_structure_v3}) involves
 $n_{\rm eft}$ independent trainings for the linear contributions followed by
$n_{\rm eft}(n_{\rm eft}+1)/2$ ones for the quadratic terms.
Hence the total number of independent neural network trainings required will be given by
\be
 N_{\rm nn} = \frac{n_{\rm eft}^2 + 3n_{\rm eft}}{2} \, ,
\ee
which scales polynomially ($n_{\rm eft}^2$) for a large number of EFT parameters.
Furthermore, since each neural net is trained independently,
the procedure is fully parallelizable and the total computing time
required scales rather as $n_{\rm eft}^2/n_{\rm proc}$
with $n_{\rm proc}$ being the number of available processors.
Thanks to this property, even for the case in which $n_{\rm eft}\sim 40$ in a typical cluster with $\sim 10^3$
 nodes the computational effort required to construct Eq.~(\ref{eq:EFT_structure_v3}) is only 50\%
larger as compared to the case with $n_{\rm eft}=1$.
This means that our method is well suited
for the large parameter spaces considered in global EFT analyses.

Furthermore, for each unbinned multivariate
observable that is constructed we  repeat the training of the
neural networks $N_{\rm rep}$ times  to estimate methodological
uncertainties.
Hence the maximal number of neural network trainings involved
will be given by
\be
\#\,{\rm trainings} = N_{\rm rep}\times N_{\rm nn} = \frac{N_{\rep}\lp n_{\rm eft}^2 + 3n_{\rm eft}\rp}{2} \, .
\ee
For example, for $hZ$ production with quadratic EFT corrections
we will have $n_{\rm eft}=7$ coefficients and $N_{\rm rep}=50$ replicas,
resulting into a maximum of 1750 neural networks to be trained.\footnote{In this
  case the actual number of trainings is smaller, $\#\,{\rm trainings}=1500$, given
that some quadratic cross-terms vanish.}
While this number may appear daunting, these trainings are parallelizable
and the total integrated computing requirements end up being not too different
from those of the single-network training.

\paragraph{Validation with analytical likelihood.}
As will be explained in Sect.~\ref{sec:pseudodata},
for relatively simple processes one can evaluate the cross-section ratios
Eq.~(\ref{eq:EFT_structure_v3}) also in a purely analytic manner.
In such cases, the PLR and the associated parameter inference can be evaluated exactly
without the need to resort to numerical simulations.
The availability of such analytical calculations offers the possibility
to independently validate its machine learning counterpart, Eq.~(\ref{eq:EFT_structure_v4}),
at various levels during the training process.

Fig.~\ref{fig:nn_overview_perf} presents an overview of representative validation checks
of our procedure
that we carry out whenever the analytical cross-sections are available.
In this case the process under consideration is parton-level top quark pair production, to be
described in Sect.~\ref{sec:pseudodata}, where the kinematic features
are the top quark pair invariant mass $m_{t\bar{t}}$  and rapidity $y_{t\bar{t}}$, that is,
the feature array is given by
$\boldsymbol{x} = \lp m_{t\bar{t}},y_{t\bar{t}} \rp$.
The neural
network training shown corresponds to the quadratic term
$\mathrm{NN}^{(j, j)}$ with $j$ being the chromomagnetic operator $c_{tG}$.

First, we display a point-by-point comparison of the log-likelihood ratio in the ML model and the
corresponding analytical calculation, namely comparing Eqns.~(\ref{eq:EFT_structure_v4})
and~(\ref{eq:EFT_structure_v3}) evaluated on the kinematics of the Monte Carlo
events generated for the training in the specific case of  $c_{tG}=2$.
One obtains excellent agreement within the full phase space considered.
Then we show the median value (over replicas)
of the ratio between the analytical and machine learning calculations of $\mathrm{NN}^{(j, j)}$ 
evaluated in the $\lp m_{t\bar{t}},y_{t\bar{t}} \rp$ kinematic feature space, with $j$ again being the chromomagnetic operator $c_{tG}$.
We also show the pull between the analytical and numerical calculations in units
of the Monte Carlo replica uncertainty.
From the median plot we see that the parametrized ratio $\hat{r}_{\sigma}$ reproduces
the exact result within a few percent except for low-statistics regions (large  $|y_{t\bar{t}}|$
and $ m_{t\bar{t}},y_{t\bar{t}}$ tails), and that these differences are in general well contained
within the one-sigma MC replica uncertainty band. 

The bottom right plot of Fig.~\ref{fig:nn_overview_perf}
displays the resultant decision boundary $g(\boldsymbol{x,c})$
for $y_{t\bar{t}}=0$ as a function of the invariant mass $m_{t\bar{t}}$
in the training of
the quadratic cross-section ratio proportional to $c_{tG}^2$ in
the specific case of also for $c_{tG}=2$.
The band in the ML model is evaluated as the 68\% CL interval
over the trained MC replicas, and is the largest at high $m_{t\bar{t}}$ values
where statistics are the smallest.
Again we find that the ML parametrization is in agreement within uncertainties
when compared to the exact analytical calculation, further validating the procedure.
Similar good agreement is observed for other EFT operators
both for the linear and for the quadratic cross-sections.

\section{Theoretical modeling}
\label{sec:pseudodata}

We describe here the settings adopted
for the theoretical modeling and simulation of unbinned observables at the LHC
and their subsequent SMEFT interpretation.
We consider two representative processes relevant
for global EFT fits, namely top-quark pair production
and Higgs boson production in association with a $Z$-boson.
We describe the calculational setups used for the SM and EFT cross-sections
at both the parton and the particle level, justify the 
choice of EFT operator basis, motivate the selection
and acceptance cuts applied to final-state particles, present
the validation
of our numerical simulations with  analytical calculations whenever possible,
and summarize the inputs to the neural network training.

\subsection{Benchmark processes and simulation pipeline}
\label{sec:pipeline}

We apply the methodology developed in
Sect.~\ref{sec:trainingMethodology} to construct
unbinned observables for inclusive top-quark pair production
and Higgs boson production in association with a $Z$-boson in proton-proton collisions.
We evaluate theoretical predictions in the SM and in the SMEFT
for both processes at leading order (LO), which suffices
in this context given that we are considering pseudo-data.
For particle-level event generation we consider the fully leptonic decay channel of top quark
pair production,
\be
\label{eq:tt_particlelevel}
{\rm p}+{\rm p} \to t + \bar{t} \to b + \ell^+ + \nu_{\ell} + \bar{b} + \ell^- + \bar{\nu}_{\ell} \, ,
\ee
and that of the Higgs decaying to a pair of bottom quarks and with the $Z$-boson decaying leptonically,
\be
{\rm p}+{\rm p} \to h + Z \to b + \bar{b} + \ell^+ + \ell^- \, .
\ee
The evaluation of the SM and SMEFT cross-sections at LO is carried out
with \verb|MadGraph5_aMC@NLO|~\cite{Alwall:2014hca}
interfaced to {\sc\small SMEFTsim}~\cite{Brivio:2017btx,Brivio:2020onw}
with  \verb|NNPDF31_nnlo_as_0118| as the input PDF set~\cite{Ball:2012cx}. 
\VerbatimFootnotes
As discussed in Sect.~\ref{sec:methodology_xsec_param}, at the quadratic level in the EFT we parametrize 
the cross-section ratio such that we have separate neural networks for terms proportional to $c_{j}^{2}$ and for the quadratic mixed terms $c_{j} c_{k}$ \:\footnote{
To generate the corresponding training sample we make use of the \verb|MadGraph5_aMC@NLO| syntax which allows for the evaluation
of cross sections dependent only on the product $c_{j} c_{k}$, for example
$$
\verb|p p > t t~  NP<=1 NP^2==2 NPc[j]^2==1 NPc[k]^2==1| \, .
$$
}.
In addition to Eq.~(\ref{eq:tt_particlelevel}),
we also carry out $t\bar{t}$ simulations at the undecayed
parton level with the goal of comparing with the corresponding exact analytical calculation of
the likelihood ratio for benchmarking purposes.
Such analytical evaluation becomes more difficult (or impossible) for 
realistic unbinned multivariate measurements  presented
in terms of  particle-level or detector-level observables.

\begin{figure}[t]
    \centering
    \includegraphics[width=0.90\textwidth]{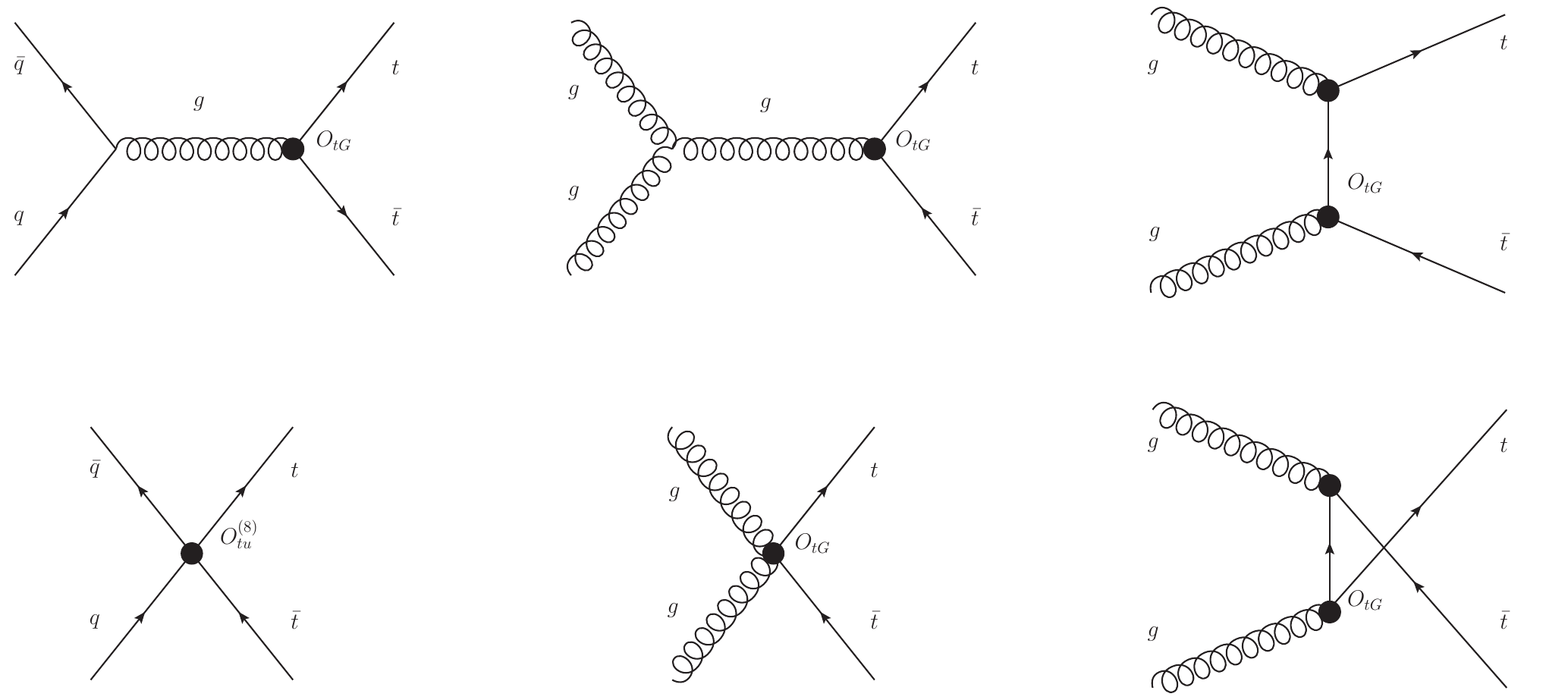}
    \caption{Representative Feynman diagrams
      for parton-level top-quark pair production
      in the SMEFT, indicating some of the corrections induced
      by the $O_{tG}$ and $O_{tu}^{(8)}$ operators.
      These operators both modify existing SM interaction vertices, such as $gt\bar{t}$,
      and induce new ones, such as $ggt\bar{t}$.
      The operators considered do not modify
      top quark decays.     
    }
    \label{fig:tt_production}
    \end{figure}

\tikzstyle{startstop} = [rectangle, rounded corners, minimum width=3cm, minimum height=0.5cm,text centered, draw=black, fill=red!30]
\tikzstyle{io} = [trapezium, trapezium left angle=70, trapezium right angle=110, minimum width=3cm, minimum height=1cm, text centered, draw=black, fill=blue!30]
\tikzstyle{decision} = [diamond, minimum width=3cm, minimum height=1cm, text centered, draw=black, fill=green!30]
\tikzstyle{arrow} = [thick,->,>=stealth]
\tikzstyle{process} = [rectangle, minimum width=4cm, minimum height=1cm, text centered, draw=black, fill=orange!30, rounded corners, text width=4cm]

\begin{figure}[t]
\small
    \centering
    \begin{tikzpicture}[node distance=2cm]

\node (in1) [io] {\verb|smeftsim_main.fr|};
\node (pro1) [process, below of=in1] {\verb|SMEFTsim| \verb|Flavor = topU3l|, \verb|Scheme=MwScheme|};
\node (out1) [io, below left=of pro1] {\verb|FeynArts .mod file|};
\node (out2) [io, below right=of pro1] {\verb|FeynArts .gen file|};
\node (pro2a) [process, below of=pro1, node distance=5cm] {\verb|FeynArts|: draw topologies and insert fields};
\node (pro2b) [process, below of=pro2a] {\verb|FormCalc|: compute $\left|\mathcal{M}_{\mathrm{sm}} + \mathcal{M}_{\mathrm{eft}}(\boldsymbol{c})\right|^2$};
\node (pro2c) [process, below of=pro2b] {\verb|Mathematica|: compute $d\sigma(\boldsymbol{c})$};
\node (out3) [io, below of=pro2c] {Analytical $d\sigma(\boldsymbol{c})$};

\draw [arrow] (in1) -- (pro1);

\draw [arrow] (pro1) -| (out1);
\draw [arrow] (pro1) -| (out2);
\draw [arrow] (out1) |- (pro2a);
\draw [arrow] (out2) |- (pro2a);
\draw [arrow] (pro2a) -- (pro2b);
\draw [arrow] (pro2b) -- (pro2c);
\draw [arrow] (pro2c) -- (out3);

\end{tikzpicture}
    \caption{Flow diagram displaying the pipeline adopted
    to evaluate at LO analytical expressions
    of differential distributions in the SMEFT for
    the LHC processes considered.
    These analytical cross-sections provide access to the
    exact likelihood ratio to benchmark the numerical MC simulations
    and the performance of the ML algorithm.
    }
    \label{fig:pipeline_flowchart}
\end{figure}
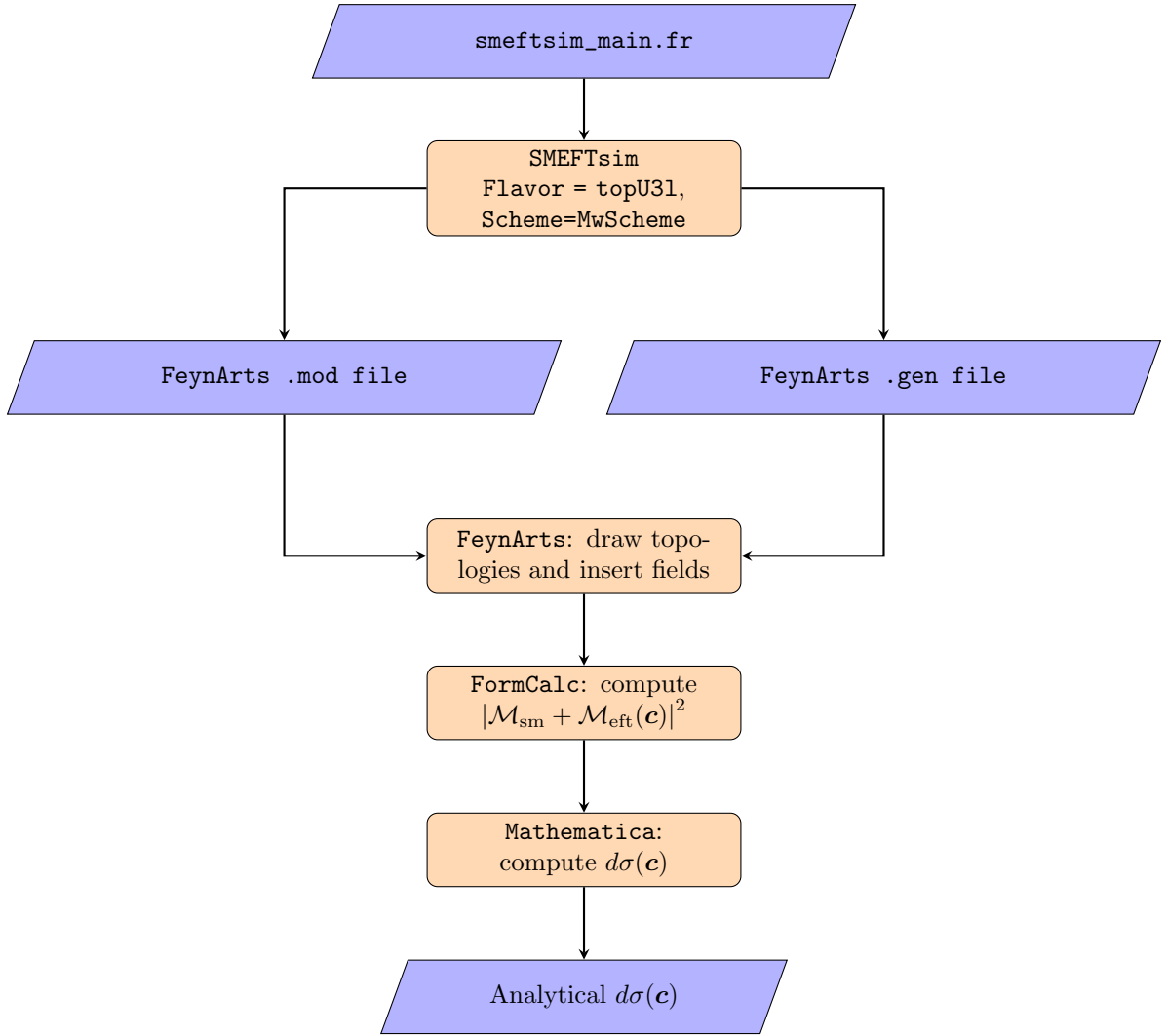

The flow chart in Fig.~\ref{fig:pipeline_flowchart}
describes the pipeline adopted
to evaluate the analytical expressions for the parton level analysis
at LO in the QCD expansion. 
First, we generate a \verb|FeynArts| \cite{Hahn:2000kx} model file from the \verb|SMEFTsim top U3l|
UFO model in the $\{m_W, m_Z, G_F\}$ input scheme~\cite{Brivio:2020onw}.
This amounts to a $U(3)_l \times U(3)_e$ flavor symmetry in the leptonic sector and $U(2)_q\times U(3)_d\times U(2)_d$
in the quark sector, consistent with the flavor assumptions made in the
{\sc\small SMEFiT}  analysis~\cite{Ethier:2021bye}.
Then we use \verb|FeynArts| to construct the diagrams associated to a given production process before passing pass them on to  \verb|FormCalc_v9.9|~\cite{Hahn:1998yk} interfaced to \verb|Mathematica|, that ultimately produces the analytical differential cross section in the SMEFT.

\begin{figure}[t]
    \centering
    \includegraphics[width=1.0\textwidth]{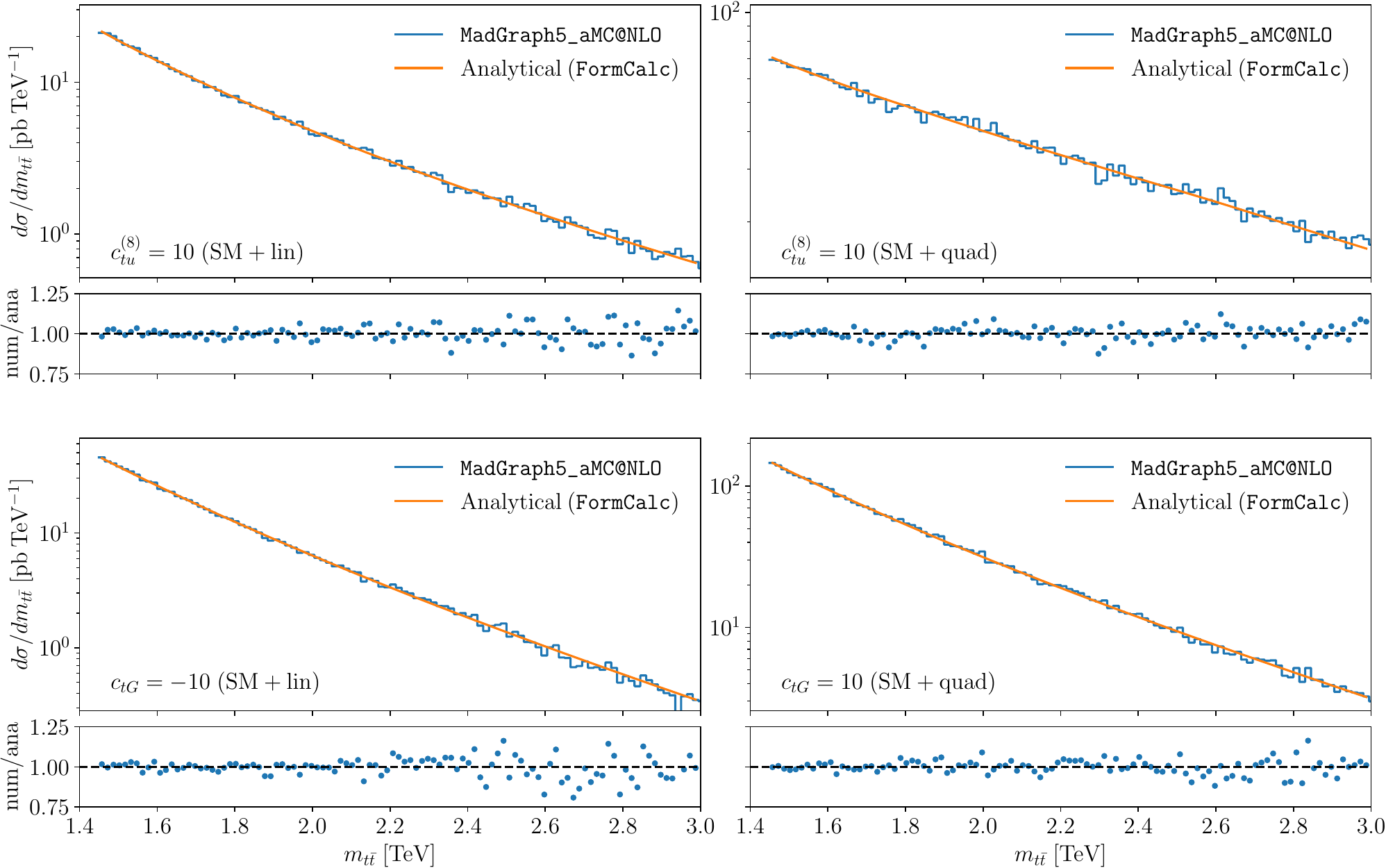}
    \caption{Parton level $m_{t\bar{t}}$ distribution in
    the analytical calculation
    of top-quark pair production at LO based on the pipeline of
    Fig.~\ref{fig:pipeline_flowchart} compared with the corresponding
    numerical simulations based on
    {\sc\small MadGraph5\_aMC@NLO}.
    We display the SM predictions and those of the SMEFT
    for non-zero values of the $c_{tG}$ and $c_{tu}^{(8)}$ coefficients
    in the linear-only and quadratic-only cases.
    The bottom panels display the ratio between the numerical and analytical calculations.
    \label{fig:tt_benchmark_parton}
    }
\end{figure}
   
Satisfactory agreement is found between the analytical
SM and SMEFT calculations and the outcome of the
corresponding \verb|MadGraph5_aMC@NLO| simulations both at the linear
and quadratic EFT level for all processes considered here. 
 This agreement is illustrated by Fig.~\ref{fig:tt_benchmark_parton},
comparing numerical and analytical SMEFT  predictions for the $m_{t\bar{t}}$
distribution in top-quark pair production for some of the values of
the  $c_{tG}$ and $c_{tu}^{(8)}$ coefficients used for the neural network training.
Similar agreement is found for other distributions and other points
in the EFT parameter space.
These analytical
calculations also make possible validating the accuracy of the neural
network training, as exemplified in
Fig.~\ref{fig:nn_overview_perf}, and indeed the agreement persists at the level
of the training of the decision boundary  $g(\boldsymbol{x},\boldsymbol{c})$.

\paragraph{Dominance of statistical uncertainties}
As discussed in Sect.~\ref{sec:statistical_framework}, we restrict our analysis
to measurements dominated by statistical uncertainties for which correlated systematic uncertainties
can be neglected.
This condition can be enforced by
restricting the fiducial phase space
such that the number of events per bin satisfies
\be
\label{eq:condition_luminosity}
\frac{\delta \sigma_i^{(\rm stat)}}{\sigma_i(\boldsymbol{0})} = \frac{1}{\sqrt{\nu_i(\boldsymbol{0})}} \ge \delta_{\rm min}^{(\rm stat)} \, ,
\qquad i=1,\ldots,N_{b} \, ,
\ee
where $\nu_i(\boldsymbol{0})$ is the number of expected events in bin $i$ according to the
SM hypothesis  after applying selection, acceptance, and efficiency cuts.
The threshold parameter $\delta_{\rm min}^{(\rm stat)}$
is set to $\delta_{\rm min}^{(\rm stat)}=0.02$ for our baseline analysis.
We have verified that our qualitative findings are not modified upon moderate variations of its value.
Since Eq.~(\ref{eq:condition_luminosity}) must apply for all possible binning choices, it should also hold
for $N_{b}=1$, namely for the total fiducial cross-section.
Therefore, we require that the  selection and acceptance cuts applied lead
to a fiducial region satisfying
$(\delta \sigma_{\rm fid}^{(\rm stat)}/\sigma_{\rm fid}) \ge \delta_{\rm min}^{(\rm stat)}$.
This condition implies that the requirement of Eq.~(\ref{eq:condition_luminosity}) will also be satisfied
for any particular choice of binning, including the narrow bin limit, i.e.~the unbinned case.

Within our approach
there are two options by which the condition Eq.~(\ref{eq:condition_luminosity}) can be enforced
when applied to the fiducial cross-section, given by
\be
\label{eq:nu_tot_lumi_discussion}
\nu_{\rm tot}(\boldsymbol{0}) = \mathcal{L}_{\rm int} \times \sigma_{\rm fid}(\boldsymbol{0}) \, .
\ee
The first option is adjusting the integrated luminosity $\mathcal{L}_{\rm int} $
corresponding to this measurement.
In this work we will take a fixed baseline luminosity $\mathcal{L}_{\rm int}=300$ fb$^{-1}$, corresponding to the
integrated luminosity accumulated at the end of Run III.
The second option is to adjust the fiducial region
such that Eq.~(\ref{eq:condition_luminosity}) is satisfied.
Taking into account Eqns.~(\ref{eq:condition_luminosity}) and~(\ref{eq:nu_tot_lumi_discussion}),
for a given luminosity $ \mathcal{L}_{\rm int}$ the fiducial (SM) cross-section should satisfy
\be
\sigma_{\rm fid}(\boldsymbol{0}) \ge \lc \lp  \delta_{\rm min}^{(\rm stat)}  \rp^2\mathcal{L}_{\rm int} \rc^{-1} \, .
\ee
In this work we take the second option,
imposing kinematic cuts restricting the events
to the high-energy, low-yield tails of distributions, such as by means of a strong $m_{t\bar{t}}$ cut
in the case of top quark pair production, see Table~\ref{tab:tt_cuts}.
It is then possible to generalize the
results presented in this work for  $\mathcal{L}_{\rm int}=300$ fb$^{-1}$
to higher integrated luminosities by  making
the cuts that define the fiducial region more stringent.

\subsection{Top-quark pair production: parton level}
\label{subsec:ttbar_theorysim_parton}

For inclusive top quark pair production with stable tops
 the  {\sc\small MadGraph5\_aMC@NLO}
calculation is accompanied by and benchmarked against the analytical
evaluation of the likelihood, see also Fig.~\ref{fig:tt_benchmark_parton}.
We consider the effects
of two representative dimension-six SMEFT operators modifying inclusive top quark pair
production, namely the chromomagnetic dipole operator $\mathcal{O}_{tG}$ and the two-light-two-heavy
four-fermion color-octet
operator $\mathcal{O}_{tu}^{(8)}$  defined as in~\cite{Brivio:2020onw} in the \verb|topU3l| flavor scheme:
\begin{align}
    \nonumber \mathcal{O}_{tu}^{(8)} &= (\bar{t}\gamma_\mu T^A t)(\bar{u}\gamma^\mu T^A u) \, ,\\
    \mathcal{O}_{tG} &= (\bar{Q}\sigma^{\mu\nu}T^a t)\Tilde{H}G_{\mu\nu}^a \, .
    \label{eq:tt_part_operators}
\end{align}
Representative Feynman diagrams displaying SMEFT corrections associated to these operators in
top-quark pair production
are shown in
Figs.~\ref{fig:tt_production}.
The dipole operator $\mathcal{O}_{tG}$ modifies the $gt\bar{t}$ coupling
as well as induces new four-body $ggt\bar{t}$ interactions, while
the four-fermion octet operator $\mathcal{O}_{tu}^{(8)}$ leads to a new $\bar{q}q\bar{t}t$ vertex.

At the  level of undecayed tops, a $2\rightarrow 2$ process  such as 
top quark pair production is uniquely determined by specifying three independent
kinematic variables, since the four-momenta $p_t^\mu$ and $p_{\bar{t}}^\mu$
satisfy the mass-shell conditions and transverse momentum conservation.
We refer to these kinematic variables 
as features in the context of ML classification problems.
We choose the three independent features to be the transverse
momentum of the top quark, $p_T^t$, and the invariant mass and rapidity
of the top quark pair, $m_{t\bar{t}}$ and $y_{t\bar{t}}$ respectively.
It can be verified how considering additional variables does not improve
the sensitivity to the EFT parameters given the redundancy of the extra features.
No fiducial cuts are imposed in the {\sc\small MadGraph5\_aMC@NLO} calculation
to facilitate the comparison with the analytical result.

Concerning the event generation settings, for each point in the
EFT parameter space ${\boldsymbol c}$ that enters the neural network
training
 we generate $N_{\rm rep}=50$ independent sets of events (replicas) containing
 $\widetilde{N}_{\rm ev}=10^5$ events each, for a total of $5 \times 10^6$ events, see also the overview in
 Table~\ref{tab:settings_nntrain}.
 Note that we adopt the convention whereby $\widetilde{N}_{\rm ev}$ denotes Monte Carlo events generated
   to train the machine learning classifier while $N_{\rm ev}=\nu_{\rm tot}$ indicates the physical events
   that enter the EFT parameter inference. The former can be made as large as one wants, while the latter
   is fixed by the assumed integrated luminosity and the value of the fiducial cross-section,
   Eq.~(\ref{eq:nu_tot_lumi_discussion}).
In addition, for each replica we generate an independent set of $10^5$ SM events.
This is required so that the two terms of the cross-entropy loss
function Eq.~(\ref{eq:loss_CE_numeric}) are properly balanced,
and results in a training set of $2\times 10^5$ events per replica.
Similar settings are used by the particle level processes described next,
and we have verified that the size of this Monte Carlo dataset is sufficient
to ensure a stable and accurate parametrization of the likelihood ratio.

\subsection{Top-quark pair production: particle level}
\label{subsec:ttbar_theorysim}

\begin{figure}[t]
    \centering
    \includegraphics[width=\textwidth]{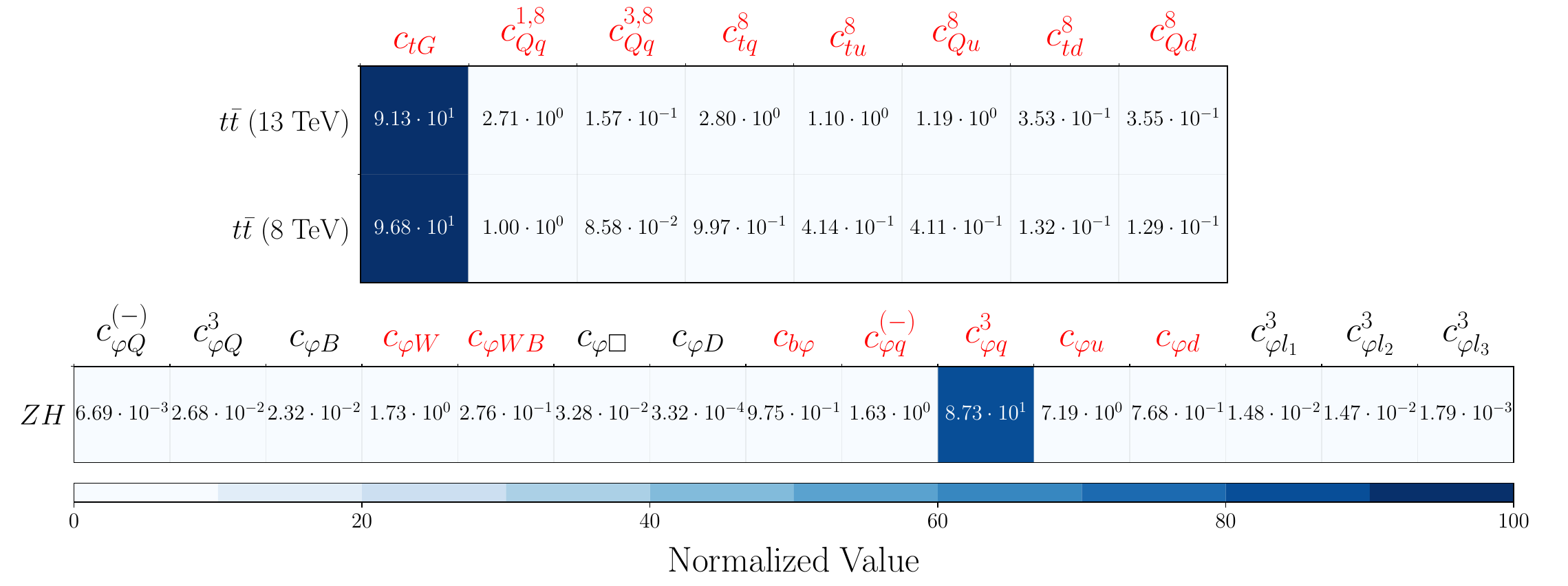}
    \caption{The diagonal entries of the Fisher matrix corresponding
    to the operators that enter the calculation
    of the LHC  $t\bar{t}$ production
    at 8 TeV and 13 TeV (top) and of the $hZ$ production  at 13 TeV (bottom)~\cite{Ethier:2021bye},
    where each row is normalized to 100.
    The dark blue entry indicates the operator that dominates the Fisher information,
    and those marked in red are selected to construct the unbinned observables.
}
    \label{fig:fisher_heatmap}
\end{figure}

In the particle-level case, where the top-quark events generated
from the diagrams in Fig.~\ref{fig:tt_production} are decayed into the
$ b \ell^+  \nu_{\ell} \bar{b}  \ell^-  \bar{\nu}_{\ell}$ final state,
one considers a broader set of kinematic features.
As in the parton level case, SM and EFT events are simulated
with {\sc\small MadGraph5\_aMC@NLO} at LO in the QCD expansion,
though now the analytical calculation
is not available as a cross-check.
In order to select the relevant EFT operators,
we adopt the following strategy.
Since we consider a single process, it
is only possible to constrain a subset of operators,
which are taken to be the  $n_{\rm eft}$ Wilson coefficients with the highest Fisher information value,
namely those that can be better determined from the fit.
In the upper part of Fig.~\ref{fig:fisher_heatmap} we display
the diagonal entries of the Fisher information matrix corresponding
 to the operators that enter the calculation
 of the LHC  $t\bar{t}$ production
 measurements at 8 TeV and 13 TeV from~\cite{Ethier:2021bye}, where
each row is normalized to 100.
The dark blue entry indicates the dominating operator,
while
the operators listed in red are those with the highest Fisher information and
selected to construct the unbinned observables.
Constraining additional Wilson coefficients would require extending
the analysis to consider unbinned observables for processes such as $t\bar{t}V$
which span complementary directions in the parameter space.

In Table~\ref{tab:op_defn_tt} we indicate the SMEFT operators 
entering inclusive top-quark pair production and listed in
Fig.~\ref{fig:fisher_heatmap}.
For each operator we provide its definition in terms of the SM fields
and the notation used to refer to the corresponding
Wilson coefficients in {\sc\small SMEFTsim} (in the {\tt topU3l} flavor scheme),
{\sc\small SMEFiT}, and {\sc\small SMEFT@NLO}~\cite{Degrande:2020evl}.
These operator definitions are consistent with those used in the  {\sc\small SMEFiT}
global analyses~\cite{Hartland:2019bjb,Ethier:2021bye} as required
for the eventual integration of the unbinned observables there.

\begin{table}[t]
  \centering
  \small
  \renewcommand{\arraystretch}{1.5}
    \begin{tabular}{c|c|c|c|c}
    \toprule
        $\qquad$ operator $\qquad$& $\qquad$\verb|SMEFiT| $\qquad$ & $\qquad$\verb|SMEFTsim|$\qquad$&\verb|SMEFT@NLO|&Definition  \\
         \midrule
         $\mathcal{O}_{tG}$ & \verb|ctG|& $-\verb|ctGRe|/\verb|gs|$ &\verb|ctG|&$ig_s(\bar{Q}\tau^{\mu\nu}T_A t)\Tilde{\varphi}G_{\mu\nu}^A + \mathrm{h.c.}$ \\
         
         $\mathcal{O}_{Qq}^{1,8}$&\verb|c81qq|& \verb|cQj18| &\verb|cQq18|&$\sum\limits_{i=1,2}c_{qq}^{1(i33i)} + 3c_{qq}^{3(i33i)}$\\
         
         $\mathcal{O}_{Qq}^{3,8}$&\verb|c83qq|& \verb|cQj38| &\verb|cQq38|&$\sum\limits_{i=1,2}c_{qq}^{1(i33i)}-c_{qq}^{3(i33i)}$ \\
         
         $\mathcal{O}_{tq}^{8}$& \verb|c8qt|& \verb|ctj8| &\verb|ctq8|&$\sum\limits_{i=1,2}c_{qu}^{8(ii33)}$ \\
         
         $\mathcal{O}_{tu}^{8}$ &\verb|c8ut|&
         \verb|ctu8|&\verb|ctu8|&$\sum\limits_{i=1,2}2c_{uu}^{(i33i)}$\\
         
         $\mathcal{O}_{Qu}^8$ &\verb|c8qu|&
         \verb|cQu8|&\verb|cQu8|&$\sum\limits_{i=1,2}c_{qu}^{8(33ii)}$\\
         
         $\mathcal{O}_{td}^8$ &\verb|c8dt|&
         \verb|ctd8|\:\verb|=|\:\verb|ctb8|&\verb|ctd8|&$\sum\limits_{j=1,2,3}c_{ud}^{8(33jj)}$\\
         
         $\mathcal{O}_{Qd}^8$ &\verb|c8qd|&
         \verb|cQd8|\:\verb|=|\:\verb|cQb8|&\verb|cQd8|&$\sum\limits_{j=1,2,3}c_{qd}^{8(33jj)}$\\
         
         \bottomrule
    \end{tabular}
    \vspace{0.3cm}
    \caption{The SMEFT operators entering inclusive top-quark pair production.
    We indicate their definition in terms of the SM fields
    and the notation used for corresponding
    Wilson coefficients in {\sc\small SMEFTsim} (in the {\tt topU3l} flavor scheme),
    {\sc\small SMEFiT}, and {\sc\small SMEFT@NLO}.
The {\sc\small =} sign indicates that two coefficients
are fixed to  the same value~\cite{Brivio:2020onw}.
}
    \label{tab:op_defn_tt}
\end{table}

\begin{table}[t]
    \centering
    \small
    \begin{tabular}{c|c}
        \toprule
        $\qquad$  $\qquad$ kinematic feature  $\qquad$ $\qquad$ & $\qquad$ cut$\qquad$ \\
         \midrule
         $m_{t\bar{t}}$&$>1.45$ TeV\\
         $p_T^\ell\:\rm{~(leading)}$&$>25$ GeV\\
         $p_T^\ell\:\rm{~(trailing)}$&$>20$ GeV\\
         $p_T^j$&$>30$ GeV\\
         $|\eta^j|$&$< 2.5$\\
         $m_{\ell\bar{\ell}}$&$m_{\ell\bar{\ell}} > 106$ GeV, or $m_{\ell\bar{\ell}} < 76$ GeV and $m_{\ell\bar{\ell}}>20$ GeV\\
         $\Delta R (j, \ell)$&$>0.4$\\
         $p_T^{\rm{miss}}$&$>40$ GeV\\
              \bottomrule
                  \end{tabular}
    \vspace{0.3cm}
    \caption{Selection and acceptance cuts
    imposed on the final-state particles of the $t\bar{t}\to b\bar{b}\ell^+\ell^-\nu_\ell\bar{\nu}_\ell$ process.}
    \label{tab:tt_cuts}
\end{table}

The selection and acceptance cuts
imposed on the final-state particles of
the $t\bar{t}\to b\bar{b}\ell^+\ell^-\nu_\ell\bar{\nu}_\ell$ process
are adapted from the Run II dilepton CMS analysis~\cite{CMS:2018adi}
and listed in Table~\ref{tab:tt_cuts}.
Concerning the  array of kinematic features ${\boldsymbol{x}}$,
it is composed of $n_k=18$ features:
$p_T$ of the lepton $p_T^\ell$,
$p_T$ of the antilepton $p_T^{\bar{\ell}}$,
leading $p_T^\ell$,
trailing $p_T^\ell$,
lepton pseudorapidity  $\eta_\ell$,
antilepton pseudorapidity  $\eta_{\bar{\ell}}$,
leading $\eta_\ell$,
trailing $\eta_\ell$,
$p_T$ of the dilepton system $p_T^{\ell\bar{\ell}}$,
invariant mass of the dilepton system $m_{\ell\bar{\ell}}$,
absolute difference in azimuthal angle $|\Delta \phi(\ell, \bar{\ell})|$,
difference in absolute rapidity $\Delta \eta(\ell, \bar{\ell})$,
leading $p_T$ of the $b$-jet, 
trailing $p_T$ of the $b$-jet,
pseudorapidity of the leading $b$-jet $\eta_b$,
pseudorapidity of the trailing $b$-jet $\eta_b$,
$p_T$ of the $b\bar{b}$ system $p_T^{b\bar{b}}$, and
invariant mass of the $b\bar{b}$ system $m_{b\bar{b}}$.
These features are partially correlated among them, and hence
maximal sensitivity of the unbinned
observables to constrain the EFT coefficients will be achieved
for $n_k<18$.

Since no parton shower  or hadronization effects are included, the $b$-quarks can be reconstructed
without the need of jet clustering and assuming  perfect tagging efficiency.
These simulation settings are not suited to describe
actual data but suffice for the present analysis based on pseudo-data,
whose goal is the consistent comparison of the impact
on the EFT parameter space of  unbinned multivariate ML observables
with
their binned counterparts.

\subsection{Higgs associated production with a $Z$-boson}
\label{subsec:hz_theorysim}

\begin{figure}[t]
    \centering
    \includegraphics[width=\textwidth]{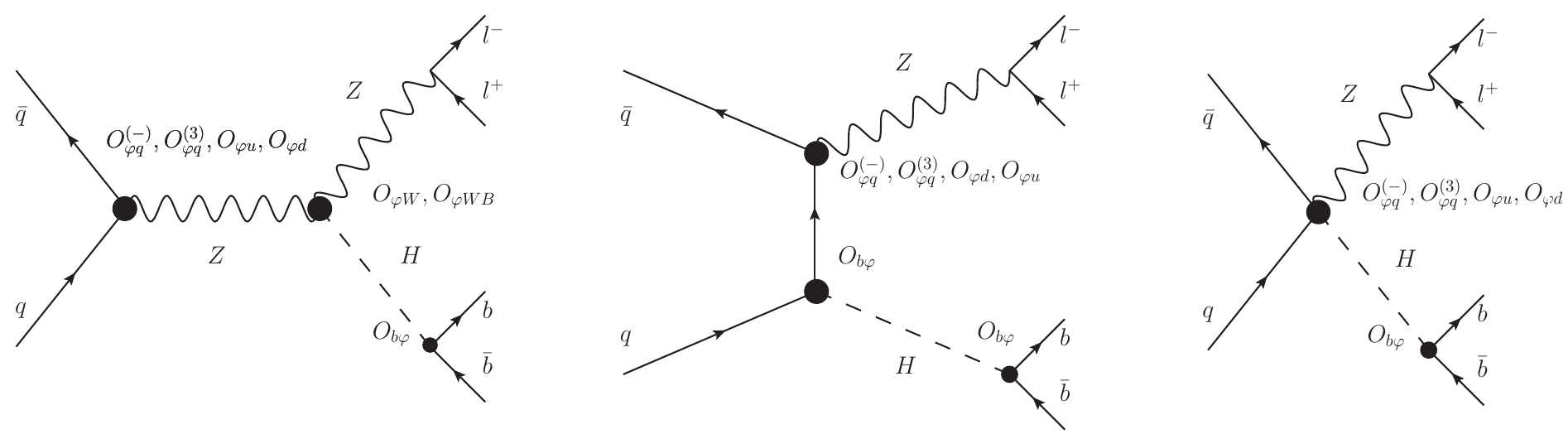}
    \caption{Same as Fig.~\ref{fig:tt_production} for
      Higgs associated production with a $Z$-boson,
      $hZ\to b\bar{b}\ell^+\ell^-$, indicating representative
    corrections arising from the SMEFT operators considered.
    For this process SMEFT effects modify also
    particle decays, in particular via the Yukawa interaction relevant
    for $h\to b\bar{b}$.
}
    \label{fig:zh_production}
\end{figure}

The second process that we consider is Higgs
production in association with a $Z$-boson in the $b\bar{b}\ell^+\ell^-$
final state, for which representative Feynman diagrams indicating the impact
of the EFT operators considered are displayed in Fig.~\ref{fig:zh_production}.
Following the same strategy as for top quark pair production,
the bottom panel of Fig.~\ref{fig:fisher_heatmap}
indicates in red the selected $n_{\rm eft}=7$ operators with the largest
value of the Fisher information matrix when evaluated on the 13 TeV LHC $hZ$ production data.
The definition of these operators, again consistent with those in~\cite{Ethier:2021bye},
is listed in Table~\ref{tab:op_defn_zh}
and includes the bottom Yukawa coupling  $\mathcal{O}_{b \varphi}$,
purely bosonic operators such as $\mathcal{O}_{\varphi W}$
and  $\mathcal{O}_{\varphi WB}$, and operators modifying the couplings
between quarks and vector bosons such as  $\mathcal{O}_{\varphi u}$
and  $\mathcal{O}_{\varphi d}$.

\begin{table}[t]
  \centering
  \renewcommand{\arraystretch}{1.5}
    \begin{tabular}{c|c|c|c}
      \toprule
      \small
        $\qquad $ operator  $\qquad $ &\verb|SMEFTsim| &  $\qquad $ \verb|SMEFiT|  $\qquad $ &Definition  \\
         \midrule
         $\mathcal{O}_{\varphi u}$& \verb|cHu| &\verb|cpui| &$\sum_{i=1,2}(\varphi^{\dagger}iD_\mu \varphi)(\bar{u}_i\gamma^\mu u_i)$ \\
         $\mathcal{O}_{\varphi d}$& \verb|cHd|&\verb|cpdi| &$\sum_{i=1,2}(\varphi^{\dagger}iD_\mu \varphi)(\bar{d}_i\gamma^\mu d_i)$ \\
         $\mathcal{O}_{\varphi q}^{(1)}$& \verb|cHj1|&$-$ &$\sum_{i=1,2}i(\varphi^{\dagger}\lra{D}_\mu \varphi)(\bar{q}_i\gamma^\mu q_i)$ \\
         $\mathcal{O}_{\varphi q}^{(3)}$& \verb|cHj3|& \verb|c3pq| &$\sum_{i=1,2}i(\varphi^{\dagger}\lra{D}_\mu \tau_I \varphi)(\bar{q}_i\gamma^\mu \tau^I q_i)$ \\
         $\mathcal{O}_{\varphi q}^{(-)}$&
         \verb|cHj1|\;$-$\;\verb|cHj3| &\verb|cpqMi|&$-$\\
         $\mathcal{O}_{b \varphi}$&
         \verb|cbHRe| &\verb|cbp|&$(\varphi^\dagger\varphi)\bar{Q}b\varphi + \mathrm{h.c.}$ \\
         $\mathcal{O}_{\varphi W}$&
         \verb|cHW| &\verb|cpW|&$(\varphi^\dagger \varphi)W_I^{\mu\nu}W_{\mu\nu}^I$ \\
         $\mathcal{O}_{\varphi WB}$&
         \verb|cHWB| &\verb|cpWB|&$(\varphi^\dagger\tau_I \varphi)B^{\mu\nu}W^I_{\mu\nu}$ \\
         \bottomrule
    \end{tabular}
    \vspace{0.3cm}
    \caption{Same as Table~\ref{tab:op_defn_tt} for Higgs associated production
    in the $hZ \to b\bar{b}\ell^+\ell^-$ final state.
    Only the operators selected by the Fisher information analysis of Fig.~\ref{fig:fisher_heatmap}
    are displayed.
}
    \label{tab:op_defn_zh}
\end{table}

The selection and acceptance cuts
imposed on the final-state particles of the $hZ\to b\bar{b}\ell^+\ell^-$ process
are collected in Table~\ref{tab:zh_cuts} and have been
adapted from the ATLAS Run II analysis~\cite{ATLAS:2017cen}.
In addition to these cuts, another cut on the jet cone radius $\Delta R$ in the $b\bar{b}$ system
is applied depending on  the value of $p_T^Z$, with
$\Delta R(b_1, b_2) < 3.0, 1.8, 1.2$ for $p_T^Z\in (75, 150]$~GeV,
  $(150, 200]$~GeV, and $(200, \infty)$ GeV respectively.
The  array of kinematic features ${\boldsymbol{x}}$
 for this process is composed of 
the following $n_k=7$ features: the transverse momentum
of the $Z$ boson $p_T^Z$, that of the $b$-quark $p_T^b$,
that of the $b\bar{b}$ pair $p_T^{b\bar{b}}$, the angular separation $\Delta R_{b\bar{b}}$ of the $b$-quarks,
their azimuthal angle separation $\Delta\phi_{b, b\bar{b}}$,
the rapidity difference between the dilepton
and the $b\bar{b}$ system $\Delta\eta_{Z, b\bar{b}}$,
and the azimuthal angle separation $\Delta\phi_{\ell b}$.
Again, most of these features are correlated among them and hence there
will be a degree of redundancy in the analysis.

\begin{table}[t]
    \centering
    \begin{tabular}{c|c}
        \toprule
       $\qquad$  kinematic feature $\qquad$  & $\qquad$ $\qquad$ cut$\qquad$$\qquad$  \\
         \midrule
         leading $b$-tagged jet $p^b_T$& $p^b_T >45$~GeV\\ 
         $m_{b\bar{b}}$&  $115<m_{b\bar{b}}<135$ GeV\\
         $m_{\ell\bar{\ell}}$& $81<m_{\ell\bar{\ell}}<101$ GeV\\
         $p_T^{\ell\bar{\ell}}$& $p_T^{\ell\bar{\ell}}> 75$~GeV\\
         $p_T^{\ell_1}$&$p_T^{\ell_1}> 27$~GeV\\
         $p_T^{\ell_2}$&$p_T^{\ell_1}> 7$~GeV\\
              \bottomrule
                  \end{tabular}
    \vspace{0.3cm}
    \caption{Selection and acceptance cuts
      imposed on the final-state particles of the $hZ\to b\bar{b}\ell^+\ell^-$ process.
      \label{tab:zh_cuts}
   }
\end{table}

\subsection{Inputs to the neural network training}
\label{sec:inputs_traiing}

\begin{figure}[htbp]
    \centering
    \includegraphics[width=0.99\textwidth]{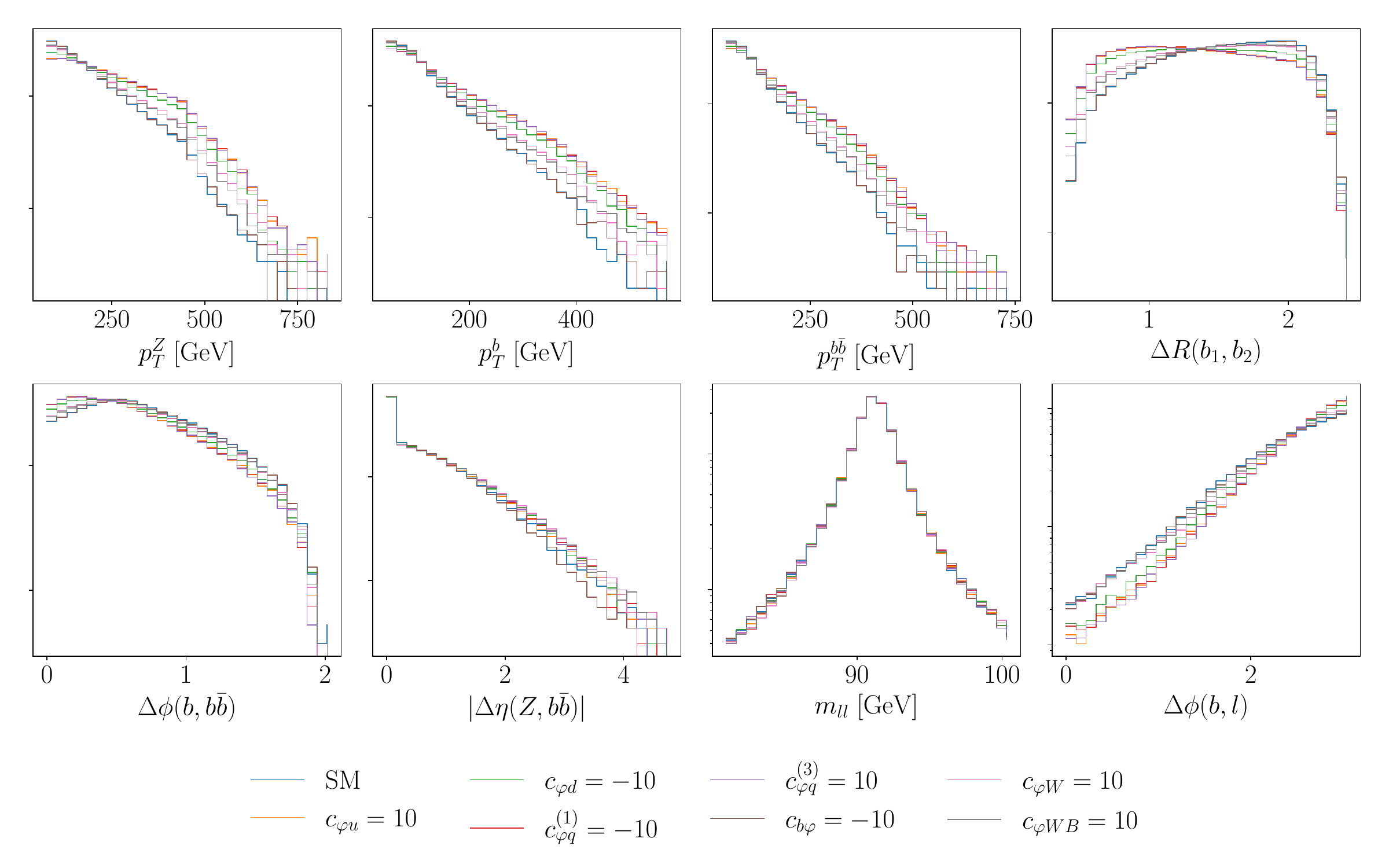}
    \includegraphics[width=0.99\textwidth]{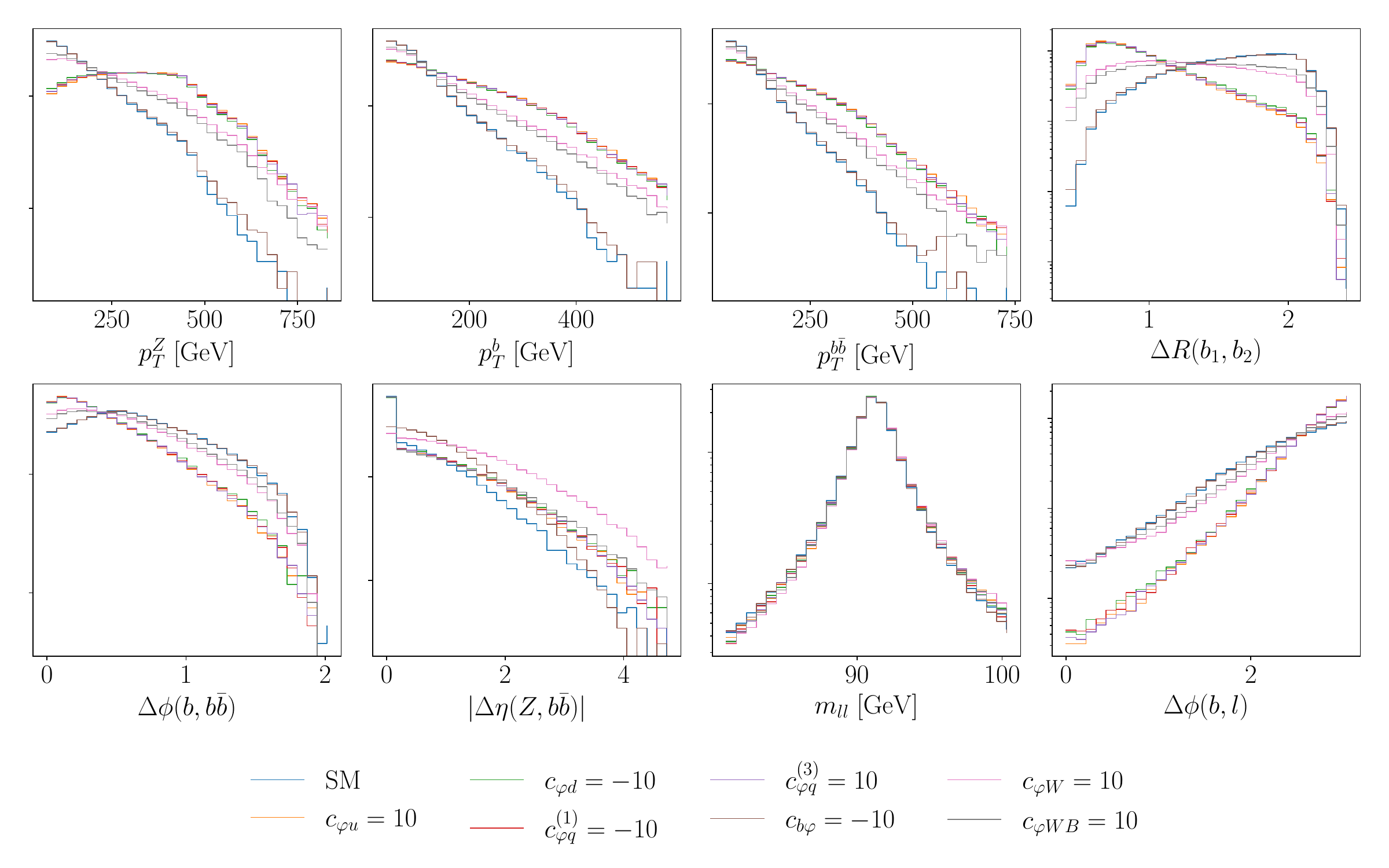}
    \caption{Differential distributions
      in the kinematic features used
      to parametrize the likelihood ratio in the 
      $hZ\to b\bar{b}\ell^+\ell^-$ process.
      We compare the SM predictions with those obtained in the SMEFT
      when individual operators are activated for coefficient values
      used for the neural network training.
      Results are shown separately
      at the linear-only level (upper) and quadratic-only (bottom panels) level
      and normalized to their fiducial
      cross-sections.
    }
    \label{fig:zh_dist}
\end{figure}

\begin{figure}[htbp]
    \centering
    \includegraphics[width=\textwidth]{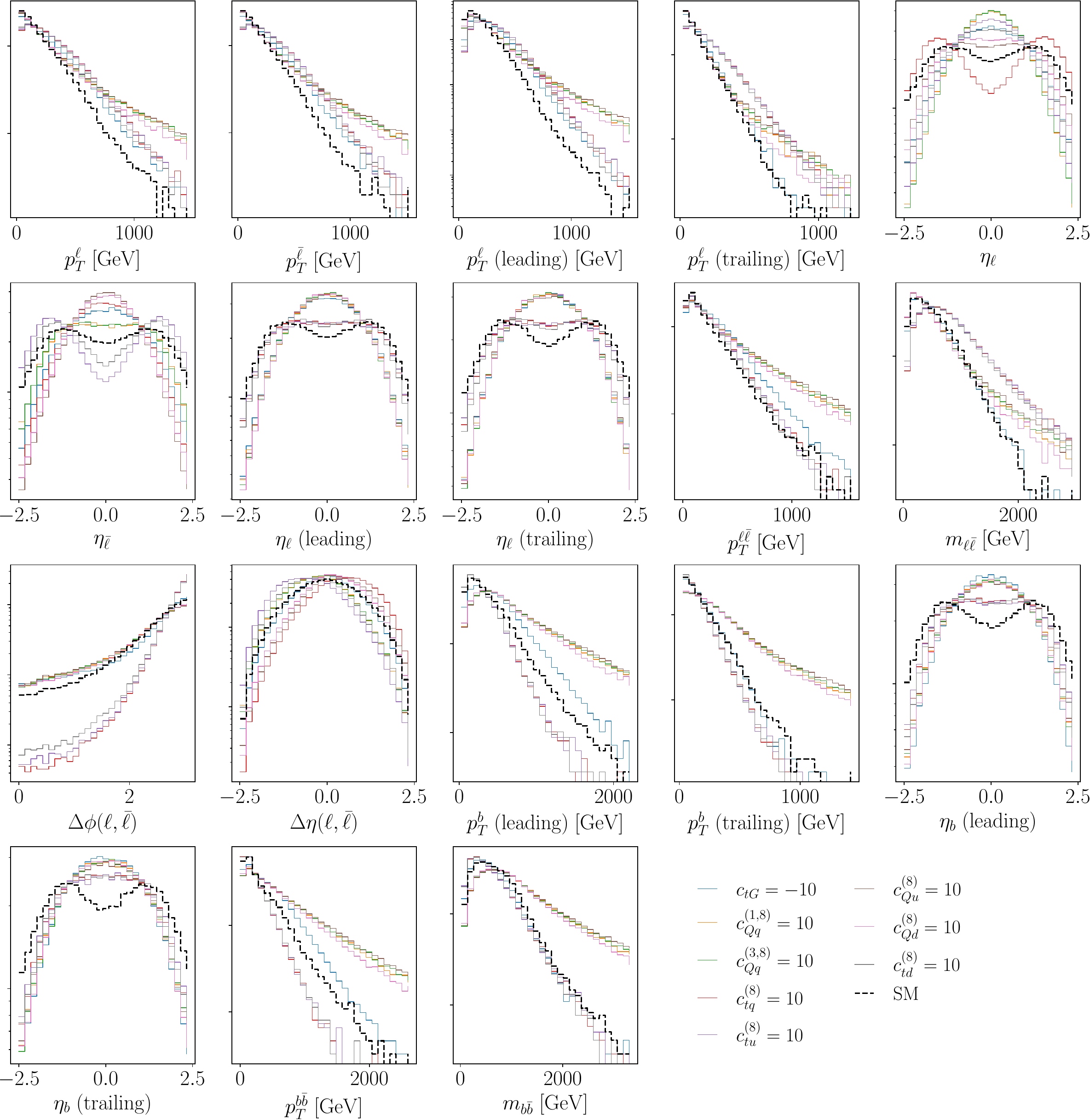}
    \caption{Same as the bottom panels in Fig.~\ref{fig:zh_dist} (quadratic-only EFT effects included) for
      the $n_k=18$ kinematic features entering the parametrization
      of the likelihood ratio in the $t\bar{t}\to b\bar{b}\ell^+\ell^- \nu_\ell \bar{\nu}_{\ell}$ process.
    }
    \label{fig:tt_quad_dist}
\end{figure}

Fig.~\ref{fig:zh_dist} displays the differential distributions
in the kinematic features used
to parametrize the likelihood ratio Eq.~(\ref{eq:EFT_structure_v4}) in the 
$hZ\to b\bar{b}\ell^+\ell^-$ process.
We compare the SM predictions with those obtained in the SMEFT
when individual operators are activated for the values of
the Wilson coefficients used for the neural network training.
Results are shown separately
at the linear-only and quadratic-only level,
to highlight how in our approach the learning strategy separates the training of the linear
from the quadratic cross-section ratios, see also Sect.~\ref{sec:methodology_xsec_param}.
In order to illustrate shape (rather than normalization)
differences of the NN inputs,
all distributions are normalized by their fiducial
cross-sections.
The corresponding comparisons at the level
of the $t\bar{t}\to b\bar{b}\ell^+\ell^- \nu_\ell \bar{\nu}_{\ell}$ process,
displaying the complete set of
$n_k=18$ kinematic features used to train the cross-section ratio at the quadratic-only level,
are shown in Fig.~\ref{fig:tt_quad_dist}.

From Figs.~\ref{fig:zh_dist} and~\ref{fig:tt_quad_dist}
one can observe how each operator modifies the qualitative shape of the various kinematic
features in different ways.
Furthermore, in general the EFT quadratic-only corrections enhance the shift
with respect to the SM distributions as compared to the linear ones.
The complementarity of the information provided by each kinematic feature motivates
the inclusion of as many final-state variables as possible when constructing unbinned
observables, though as mentioned above the limiting sensitivity will typically be saturated
before reaching the total number of kinematic features used for the training.

\begin{figure}[t]
    \centering
    \includegraphics[width=\textwidth]{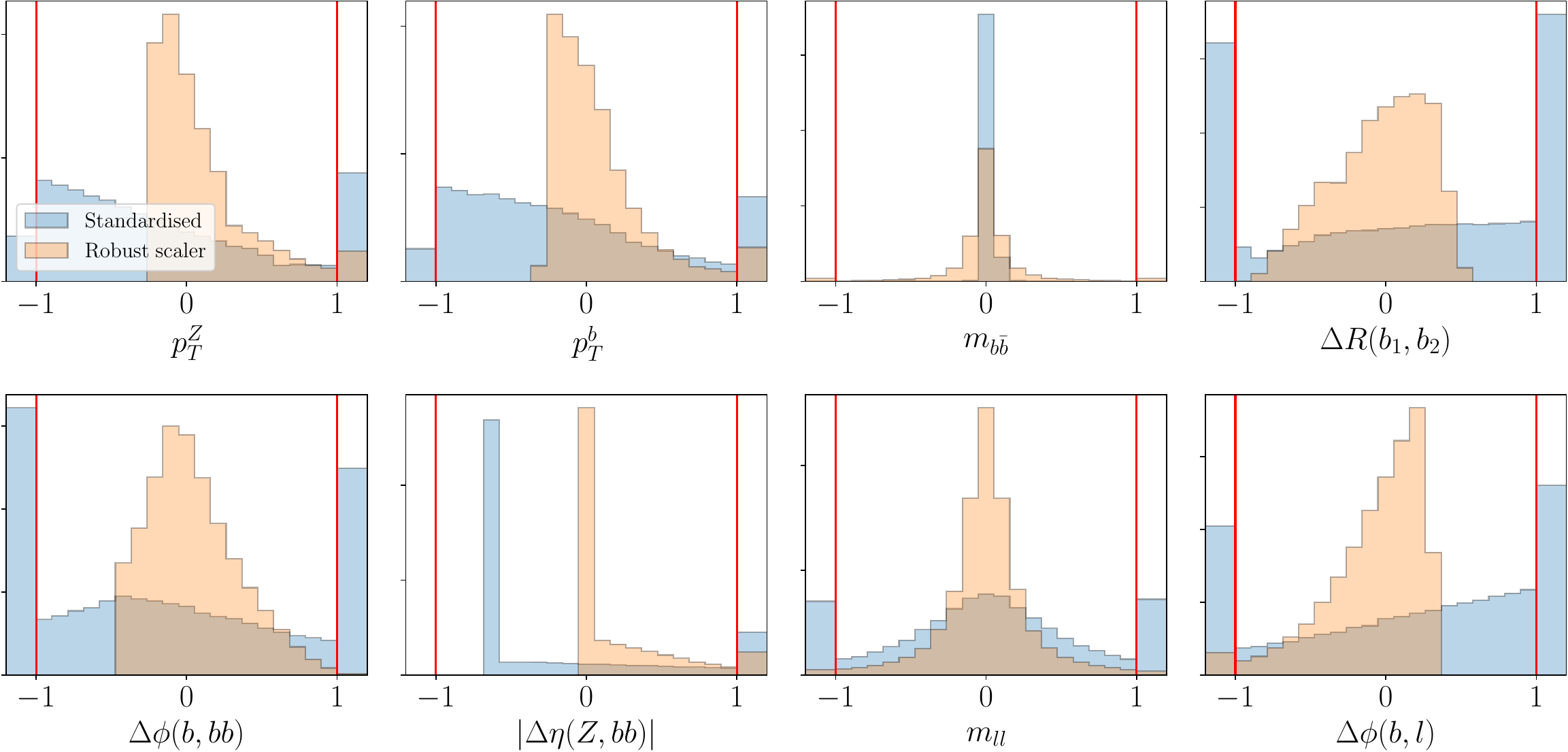}
    \caption{Comparison between two different data  preprocessing schemes
      applied to the input features in the neural network training
      of the $hZ\to b\bar{b}\ell^+\ell^-$ likelihood:
     a Standardized Gaussian scaler and a robust scaler adopting a $95\%$ inter-quantile range.
}
    \label{fig:feature_scaling}
\end{figure}

As mentioned in Sect.~\ref{sec:nntraining}, an efficient neural network
training strategy demands that the kinematic features $\boldsymbol{x}$ entering
the evaluation of
the $r_\sigma(\boldsymbol{x}, \boldsymbol{c})$
cross-section ratios are preprocessed
to ensure that the input information is provided to the neural networks
in the region of maximal sensitivity.
That is, all features should be transformed 
to a common range and their distribution within this range should be reasonably similar.
Here we use a robust scaler to ensure that this condition is satisfied.
Fig.~\ref{fig:feature_scaling} displays the comparison between two different  preprocessing schemes
applied to the input features before the neural network training
of the $hZ\to b\bar{b}\ell^+\ell^-$ likelihood is carried out.
We display results for a 
Standardized Gaussian scaler and for a robust scaler:
the latter subtracts the median and scales to the $95\%$ inter-quantile range, while the former
rescales all features to have zero mean and unit variance.
The robust scaler leads to input feature distributions peaked around zero
with their bulk contained within the $[-1,1]$ region,
which is not  the case in general for the Standardized Gaussian scaler.
Therefore our default robust scaler facilitates the incorporation of new kinematic features,
since the shapes of the input distributions are such that their bulk belongs
to the high-sensitivity region of the neural networks.

Table~\ref{tab:settings_nntrain} summarizes
the settings adopted for the neural network training of the
likelihood ratio function Eq.~(\ref{eq:EFT_structure_v4}) for the processes
considered.
For each process, we indicate the number of replicas $N_{\rm rep}$ generated,
the values of the EFT coefficients that enter the training as specified in Eqns.~(\ref{eq:nn_param_lin})
and~(\ref{eq:nn_param_quad}),
the number of Monte Carlo events generated $\widetilde{N}_{\rm ev}$ for each replica, and
the number of neural networks to be trained per replica $N_{\rm nn}$.
The values of the Wilson coefficients are chosen to be sufficiently large so as to mitigate 
the effect of MC errors that might otherwise dominate the SM-EFT discrepancy.  Furthermore, 
the sign of each Wilson coefficient is chosen such that the effect of the EFT is an enhancement relative to the SM,
and therefore the differential cross sections are consistently positive.  For example, in the case of negative EFT-SM 
interference, we select negative values of Wilson coefficients.
Cross-section positivity must also be maintained during training of the neural networks, and this 
is further discussed in Sect.~\ref{sec:methodology_xsec_param}.
The last column indicates the total number of trainings required to assemble the full parametrization including
the $N_{\rm rep}$ replicas, namely $\#{\rm trainings}=N_{\rm rep}\times N_{\rm nn}$.
For parton- and particle-level top-quark pair production and for $hZ$ production
our procedure requires the training of 200, 1000, and 1500 networks respectively
in the case of the quadratic EFT analysis.\footnote{We note that the actual value of $\#{\rm trainings}$
  can differ from the maximum value $N_{\rm rep}\times N_{\rm nn}$ since some quadratic cross-terms vanish. }
As discussed in Sect.~\ref{sec:trainingMethodology} these trainings are parallelizable
and the overall computational overhead remains moderate.

The values listed in
the last two columns of Table~\ref{tab:settings_nntrain} correspond to the case of quadratic EFT fits,
since as will be explained in Sect.~\ref{sec:results} at the linear level the presence of degenerate
directions requires restricting the subset of operators for which inference can be performed.
For each process, the total number of Monte Carlo events in the SMEFT that need to be generated is
therefore $\widetilde{N}_{\rm ev}\times N_{\rm rep}\times N_{\rm nn}$, and in addition the training needs
a balanced SM sample composed by $\widetilde{N}_{\rm ev}\times N_{\rm rep}$ events.
For example, in the case of $hZ$ production the total number of SMEFT events to be generated
is $10^5\times 50 \times 30 = 1.5\times 10^8$ events.

\begin{table}[htbp]
  \centering
  \renewcommand{\arraystretch}{1.5}
    \begin{tabular}{c|c|c|c|c|c}
      \toprule
      \small
        $\qquad $ Process  $\qquad $ & $\quad N_{\rm rep}\quad $ &  $c_j^{(\rm tr)}$ & $\widetilde{N}_{\rm ev}$~(per replica) & $\quad N_{\rm nn}\quad $  & \#trainings\\
         \midrule
         $pp \to t\bar{t}$&50&\begin{tabular}{l r}$c_{tG}=-10$ & \\
           $c_{tu}^{(8)}=10$
         \end{tabular}
         &$10^5$& 4  & 200 \\
         \midrule
         $pp\to t\bar{t}\to b\bar{b}\ell^+\ell^- \nu_\ell \bar{\nu}_{\ell}$ &25&
         \begin{tabular}{l r}
           $c_{td}^{(8)}=10$ &\\
           $c_{Qd}^{(8)}=10$&\\
           $c_{Qq}^{(1,8)}=10$&\\
           $c_{Qq}^{(3,8)}=10$&\\
           $c_{Qu}^{(8)}=10$&\\
           $c_{tG}=-10$&\\
           $c_{qt}^{(8)}=10$&\\
           $c_{tu}^{(8)}=10$& \\
         \end{tabular}& $10^5$ &$40$ & 1000\\
         \midrule
         $pp\to hZ\to b\bar{b}\ell^+\ell^-$ &50&
         \begin{tabular}{lr}
           $c_{\varphi u}=10$&\\
           $c_{\varphi d}=-10$&\\
           $c_{\varphi q}^{(1)}=-10$&\\
           $c_{\varphi q}^{(3)}=10$&\\
           $c_{b\varphi}=-10$&\\
           $c_{\varphi W}=10$&\\
           $c_{\varphi WB}=10$ & \\
         \end{tabular}& $10^5$ &$30$ &  1500\\
         \bottomrule
    \end{tabular}
    \vspace{0.3cm}
    \caption{Overview of the  settings for the neural network training of the
      likelihood ratio Eq.~(\ref{eq:EFT_structure_v4}).
      For each process, we indicate the number of replicas $N_{\rm rep}$,
      the values of the EFT coefficients that enter the training as specified in Eqns.~(\ref{eq:nn_param_lin})
      and~(\ref{eq:nn_param_quad}),
      the number of Monte Carlo events generated $\widetilde{N}_{\rm ev}$ for each replica, and
      the number of neural networks to be trained per replica $N_{\rm nn}$
      taking into account that  some EFT cross-sections vanish at the linear level.
      The values of the Wilson coefficients are chosen such that the EFT has a large effect relative to the SM, mitigating the 
	effect of MC errors that could otherwise dominate the SM-EFT discrepancy.  The signs of the Wilson coefficients are chosen such that 
	the EFT always leads to an enhancement relative to the SM; for example, negative Wilson coefficients are chosen
	to compensate for negative SM-EFT interference.  This ensures positive differential cross-sections throughout the training samples.
      The last column indicates the total number of trainings
      required, $\#{\rm trainings}=N_{\rm rep}\times N_{\rm nn}$.
      The last two columns correspond to the case of quadratic EFT fits;
      at the linear EFT level the presence of quasi-flat directions
      restricts the subset of operators for which inference can be performed.
      For each process, the total number of Monte Carlo EFT events generated is
      $\widetilde{N}_{\rm ev}\times N_{\rm rep}\times N_{\rm nn}$, and in addition we need
      balanced SM samples requiring $\widetilde{N}_{\rm ev}\times N_{\rm rep}$ events.
}
    \label{tab:settings_nntrain}
\end{table}

\section{EFT constraints from unbinned multivariate observables}
\label{sec:results}

We now present the
constraints on the EFT parameter space provided by the unbinned observables
constructed in Sect.~\ref{sec:trainingMethodology}
in comparison with those provided by their binned counterparts.
We study the dependence of these results on the choice of binning and
on the kinematic features.
We also quantify how much the EFT constraints are modified
when restricting the analysis to linear 
$\mathcal{O}(\Lambda^{-2})$ effects as compared to when the quadratic
$\mathcal{O}(\Lambda^{-4})$ contributions are also included.

First, we describe the method adopted to infer the posterior distributions
of the EFT parameters for a given observable, either binned or unbinned.
Second, we present results for parton-level inclusive top quark pair production, described in
Sect.~\ref{subsec:ttbar_theorysim_parton}. The motivation for this is to validate our
machine learning methodology by comparing it with the results
of parameter inference based on the analytical calculation of the likelihood.
Then we consider the analogous process, now at the particle level in the dilepton final
state (Sect.~\ref{subsec:ttbar_theorysim}), and quantify the information gain resulting from
unbinned observables and its dependence on the choice of kinematic features
used in the training.
This is
followed by the  corresponding analysis for $hZ$ production (Sect.~\ref{subsec:hz_theorysim}).
Finally, we will discuss the impact of methodological uncertainties, discussed in 
Sect.~\ref{sec:nntraining}, on the constraints we obtain on the EFT parameter space.
The results presented here can be reproduced and extended to other processes by means of the {\sc\small ML4EFT}
framework, summarized in App.~\ref{app:code}.

\subsection{EFT parameter inference}
\label{subsec:EFT_inference}
For each of the LHC processes considered in Sect.~\ref{sec:pseudodata}, Monte Carlo samples
in the SM and the SMEFT are generated in order to train the decision boundary
$g({\boldsymbol x},\boldsymbol{c})$ from the minimization of the cross-entropy loss
function Eq.~(\ref{eq:loss_CE}). See Tables~\ref{tab:tt_cuts} and \ref{tab:zh_cuts} for the pseudodata generation settings.
The outcome of the neural network training is a parametrization of the cross-section ratio
$\hat{r}_{\sigma}(\boldsymbol{x}, \boldsymbol{c}) $,
Eq.~(\ref{eq:EFT_structure_v4}), which in the limit of large statistics and perfect training
reproduces the true result $r_{\sigma}(\boldsymbol{x}, \boldsymbol{c}) $,
Eq.~(\ref{eq:EFT_structure_v2}).
To account for finite-sample and finite network flexibility effects,
we use the Monte Carlo replica method as described in Sect.~\ref{sec:nntraining} to
estimate the associated methodological uncertainties. 
Therefore, the actual requirement that defines a satisfactory neural
network parametrization $\hat{r}_{\sigma}$ is that it reproduces the exact result $r_{\sigma}$,
within the 68\% CL replica uncertainties evaluated from the ensemble Eq.~(\ref{eq:EFT_structure_v5}).

From the exact result for the cross-section ratio $r_{\sigma}(\boldsymbol{x}, \boldsymbol{c}) $,
or alternatively its ML representation $\hat{r}_{\sigma}(\boldsymbol{x}, \boldsymbol{c})$,
one can carry out inference on the EFT Wilson coefficients by means of the profile
likelihood ratio Eq.~(\ref{eq:plr_5}) applied to the $N_{\rm ev}$ generated pseudo-data events.
We emphasize that the $N_{\rm ev}$ events entering
in Eq.~(\ref{eq:plr_5}), or alternatively in Eq.~(\ref{eq:plr_6}), are the physical events
expected after acceptance and selection cuts for a given integrated luminosity $\mathcal{L}_{\rm int}$,
see also the discussion in Sect.~\ref{sec:pipeline}.
The Monte Carlo events used
to train the ML classifier are instead  different: in general, the classifier is trained
on a larger event sample than the one expected for the actual measurement.
Once the machine learning parametrization of the cross-section ratio, Eq.~(\ref{eq:EFT_structure_v4}),
has been determined alongside its replica uncertainties,
the profile likelihood ratio Eq.~(\ref{eq:plr_5}) can be used to infer the posterior
probability distribution in the EFT parameter space and thus determine confidence level
intervals associated to the unbinned observables.
The same method is applied to binned likelihoods and eventually
to extended likelihoods which combine the information provided by
binned and unbinned observables, Eq.~(\ref{eq:log_likelihood_global}).

In this work, the EFT parameter inference based on the machine learning parametrization of the
PLR is carried out by means of the Nested Sampling algorithm, in particular
via the MultiNest implementation~\cite{Feroz:2013hea}.
We recall that this
is the baseline sampling algorithm used in the {\sc\small SMEFiT} analysis~\cite{Ethier:2021bye}
to determine the posterior distributions in the  EFT parameter space composed of $n_{\rm eft}=36$
independent Wilson coefficients.
The choice of MultiNest is motivated from its previous applications
to EFT inference problems of comparable complexity to those relevant here, and also to
facilitate the integration of unbinned observables into global EFT fits.
The posterior distributions provided by MultiNest
are represented by $N_{s}$ samples in the EFT parameter space,
where $N_{s}$ is determined 
by requiring a given accuracy in the sampling procedure.
From this finite set of samples, contours of the full posterior can be reconstructed via the kernel density
estimator (KDE) method, also used e.g. in the context of Monte Carlo
PDF fits~\cite{Carrazza:2015hva,Butterworth:2015oua}.

Methodological uncertainties associated to the $N_{\rm rep}$ replicas
are propagated to the constraints on the EFT 
parameter space in two complementary ways.
Firstly,
after evaluating the neural networks on the $N_{\rm ev}$ events entering into the inference procedure,
we calculate the median of each $\mathrm{NN}^{(j)}_i(\boldsymbol{x})$ and $\mathrm{NN}^{(j, k)}_i(\boldsymbol{x})$ in
Eq.~(\ref{eq:EFT_structure_v5}).
The corresponding median profile likelihood ratio is then calculated and 
used in the Nested Sampling algorithm, from which contours of the 
posterior distribution are obtained.
The results shown in the following Sects.~\ref{subsec:top_parton},~\ref{subsec:top_particle} and~\ref{subsec:zh} have been produced using this method.
Alternatively, one can assess the impact of the 2-$\sigma$ methodological uncertainties on
these contours.  We do so by defining $N_{\rm rep}$ profile likelihood ratios, one from each of the $N_{\rm rep}$ replicas,
and performing Nested Sampling for each, resulting in $N_{\rm rep}$ sets of $N_{\rm s}$ 
samples from the EFT posterior distribution.  The samples are then combined, and the KDE method used to determine
contours of the posterior distribution.
In Sect.~\ref{subsec:eft_uncert} we will assess the differences observed between these two methodologies.

\subsection{Top-quark pair production: parton level results}
\label{subsec:top_parton}
\begin{figure}[t]
    \centering
    \includegraphics[width=0.78\textwidth]{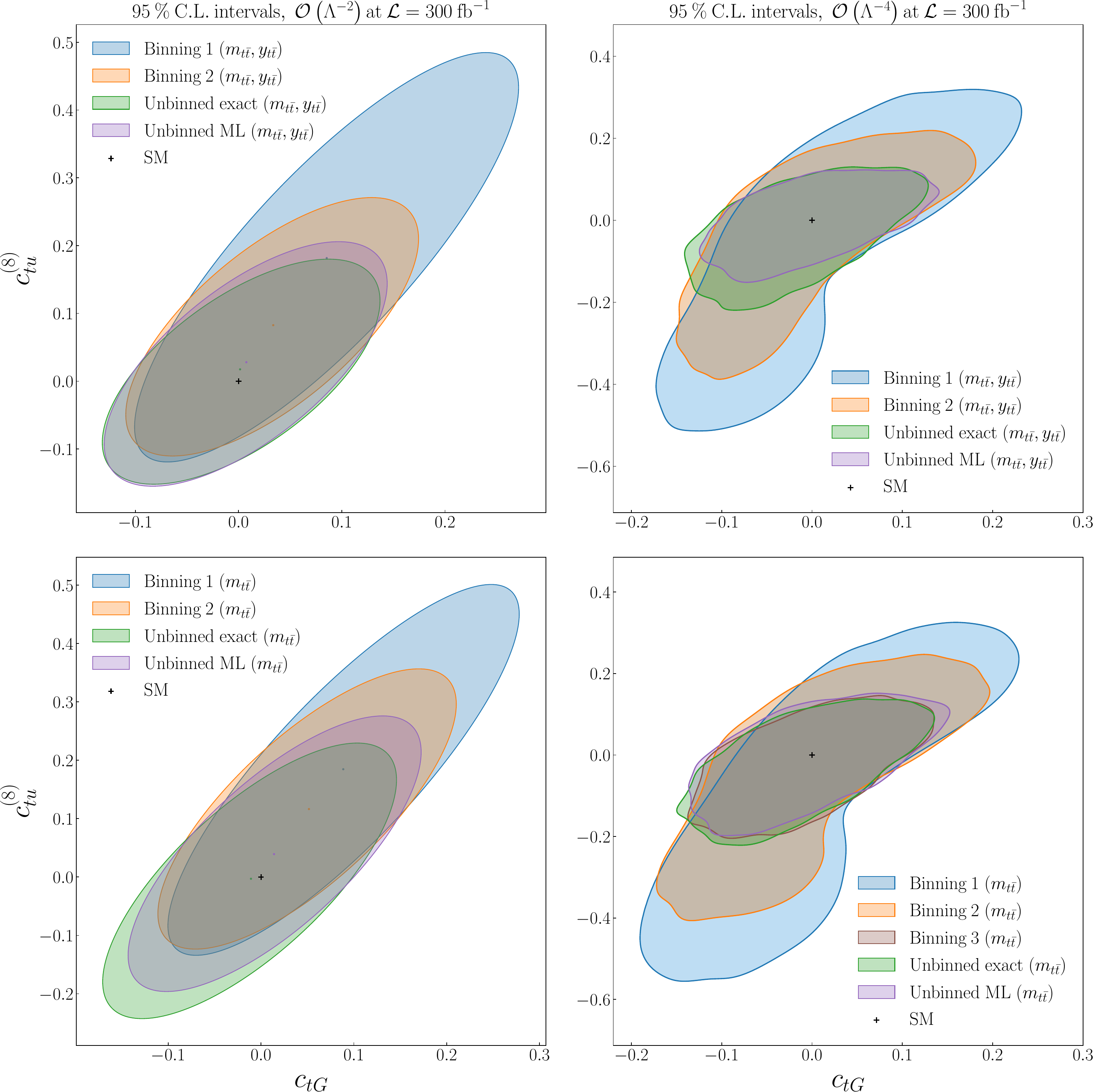}
    \caption{The 95\% CL regions in the $(c_{tG}, c_{tu}^{(8)})$
    plane obtained for the parton-level top quark
    pair production process described in Sect.~\ref{subsec:ttbar_theorysim_parton}.
    We present results based on  EFT calculations at
    the linear level (left) and also including quadratic contributions (right panels).
    The black cross indicates the SM result, which is the hypothesis assumed in the generation of
    the pseudo-data that enters the inference procedure.
    We compare the results
    from two binned analyses (Binning 1 and 2) with those of the corresponding unbinned observable,
    constructed either with the analytic likelihood calculation or with its
    machine learning parametrization. We demonstrate how the analytical calculation is reproduced 
    for a sufficiently fine binning (Binning 3).
    We display the contours corresponding
    to two different observables, the first built upon two
    kinematic features $(m_{t\bar{t}},y_{t\bar{t}})$ (upper) and the second
    based on the invariant mass $m_{t\bar{t}}$ as a single feature (lower panels).
    }
    \label{fig:contours_tt_parton}
\end{figure}

Fig.~\ref{fig:contours_tt_parton} displays the 95\% CL contours in the $(c_{tG}, c_{tu}^{(8)})$
plane obtained from the parton-level top quark
pair production process described in Sect.~\ref{subsec:ttbar_theorysim_parton}.
We compare the results based on EFT calculations at
the linear level with those also including quadratic contributions.
The black cross indicates the SM result, which is the hypothesis assumed in the generation of
the pseudo-data that enters the inference procedure.
The results shown here and in the remainder of this section correspond to a center-of-mass energy of $\sqrt{s}=14$ TeV
and an integrated luminosity of $\mathcal{L}=300$ fb$^{-1}$.
We show contours corresponding to
two binned analyses, the first based on a coarse binning (Binning 1) in two kinematic features
$m_{t\bar{t}}$ and $y_{t\bar{t}}$, defined by
\be
m_{t\bar{t}} \in \lc 1.45, 2.5, \infty\rp\;\mathrm{TeV}\, ,\quad
y_{t\bar{t}} \in \pm \lc 0, 1.5, 3.0  \rc \, , \nonumber 
\ee
and the second based on a finer binning (Binning 2), given by
\bea
m_{t\bar{t}} &\in& \lc 1.45, 1.5, 1.55, 1.6, 1.7, 1.8, 1.9, 2.0, 2.1, 2.2, 2.3, 2.4, 2.5, 2.6, 2.7, 2.8, 2.9, 3.0, \infty\rp\;\mathrm{TeV}\, , \nonumber \\
y_{t\bar{t}} &\in& \pm \lc 0, 0.3, 0.6, 0.9, 1.2, 3.0  \rc \, . \nonumber 
\eea
Furthermore, we show results for the  corresponding
observable
based on a single feature, the invariant mass $m_{t\bar{t}}$,
obtained by marginalizing the double differential binning over the rapidity $y_{t\bar{t}}$.
The contours derived from these two binned observables
in Fig.~\ref{fig:contours_tt_parton} are compared with the corresponding unbinned observables
constructed either with the analytic likelihood calculation (``Unbinned exact'') or with its
machine learning parametrization (``Unbinned ML'').

For the two-feature observable in the quadratic case, upper right panel in Fig.~\ref{fig:contours_tt_parton},
we observe how the 95\% CL contours shrink first when moving from a coarser binning
to a finer binning and then further when using the unbinned observable.
Good agreement is found between the inference carried out with the exact likelihood calculation
and with its ML interpolation, with the small remaining differences covered by the spread
of the replicas as discussed below.
We observe that the constraints
on the chromomagnetic operator
$c_{tG}$ are similar for the various observables considered, as expected, since
it is mostly determined from the absolute normalization
of the $t\bar{t}$ fiducial cross-section and hence adding kinematic information on the shape
of the distributions does not provide significant extra information.
The main improvement is seen in the constraints on the two-light-two-heavy operator $c_{tu}^{(8)}$,
which benefits from exploiting the kinematic information.

By comparing with the corresponding analysis based on the single-feature $m_{t\bar{t}}$ observable
(bottom right panel) one finds a moderate worsening of the obtained bounds.
No major
differences are observed, however, indicating how in this case the addition of $y_{t\bar{t}}$ as kinematic feature does
not significantly modify the inference results.
We have verified that good agreement is found between the exact likelihood calculation and 
the binned approach when a sufficiently fine binning is taken, as expected,
because in the limit of infinitely fine bins the binned approach will exactly reproduce the
analytical calculation. This is illustrated by Binning 3, given by
\bea 
\nonumber  m_{t\bar{t}} &\in& \lc 1.45, 1.5, 1.55, 1.6, 1.7, 1.8, \dots, 4.8, 4.9, 5.0, \infty\rp\;\mathrm{TeV}.
\eea
We have also verified that in this relatively simple scenario (a $2\to 2$ process)
the addition of a third independent kinematic feature such as $p_T^t$ to
the unbinned observable does not affect the outcome.
As we show next, for a more realistic particle-level configuration it is instead
clearly beneficial to increase the number of  kinematic features 
considered in the EFT parameter inference.

In the case of the linear EFT level results, displayed in left panels in Fig.~\ref{fig:contours_tt_parton},
we note that there is a strong correlation between the two coefficients for the case
of the coarse binning, indicating a quasi-flat direction.
This is removed by using first a finer binning and then the unbinned
observable, which in this case provides only a moderate improvement as compared
to the finer binning.
As in the case of the quadratic analysis, adding a second feature, $y_{t\bar{t}}$, 
does not significantly modify the results relative to the analysis of a single feature $m_{t\bar{t}}$.

\subsection{Top-quark pair production: particle level results}
\label{subsec:top_particle}
We assess next the bounds on the EFT parameter space obtained from unbinned observables in the case
of particle-level top quark pair production, specifically in the dilepton final state
described in Sect.~\ref{subsec:ttbar_theorysim},
$pp \to t\bar{t} \to b \ell^+ \nu_{\ell} \bar{b} \ell^- \bar{\nu}_{\ell}$.
As discussed there, this process is most sensitive to the $n_{\rm eft}=8$ operators
with the highest Fisher information in
inclusive $t\bar{t}$ production, as 
listed in Table~\ref{tab:op_defn_tt}.
While for the quadratic EFT analysis it is possible to derive the posterior
distribution associated to the full set of $n_{\rm eft}=8$ operators,
at the linear level there are quasi-flat directions that destabilize the ML training
of the likelihood ratio and the subsequent parameter inference.
For this reason, in the linear EFT analysis of this process we consider a subset of $n_{\rm eft}=5$ operators
that can be simultaneously constrained from inclusive
top-quark pair production~\cite{Brivio:2019ius}, given by
\be
\label{eq:subset_operator_topparticle_linear}
c_{tG}, c_{qt}^{(8)}, c_{Qu}^{(8)}, c_{Qq}^{(1,8)}, c_{td}^{(8)} \, .
\ee

\begin{figure}[htbp]
    \centering
    \includegraphics[width=0.85\textwidth]{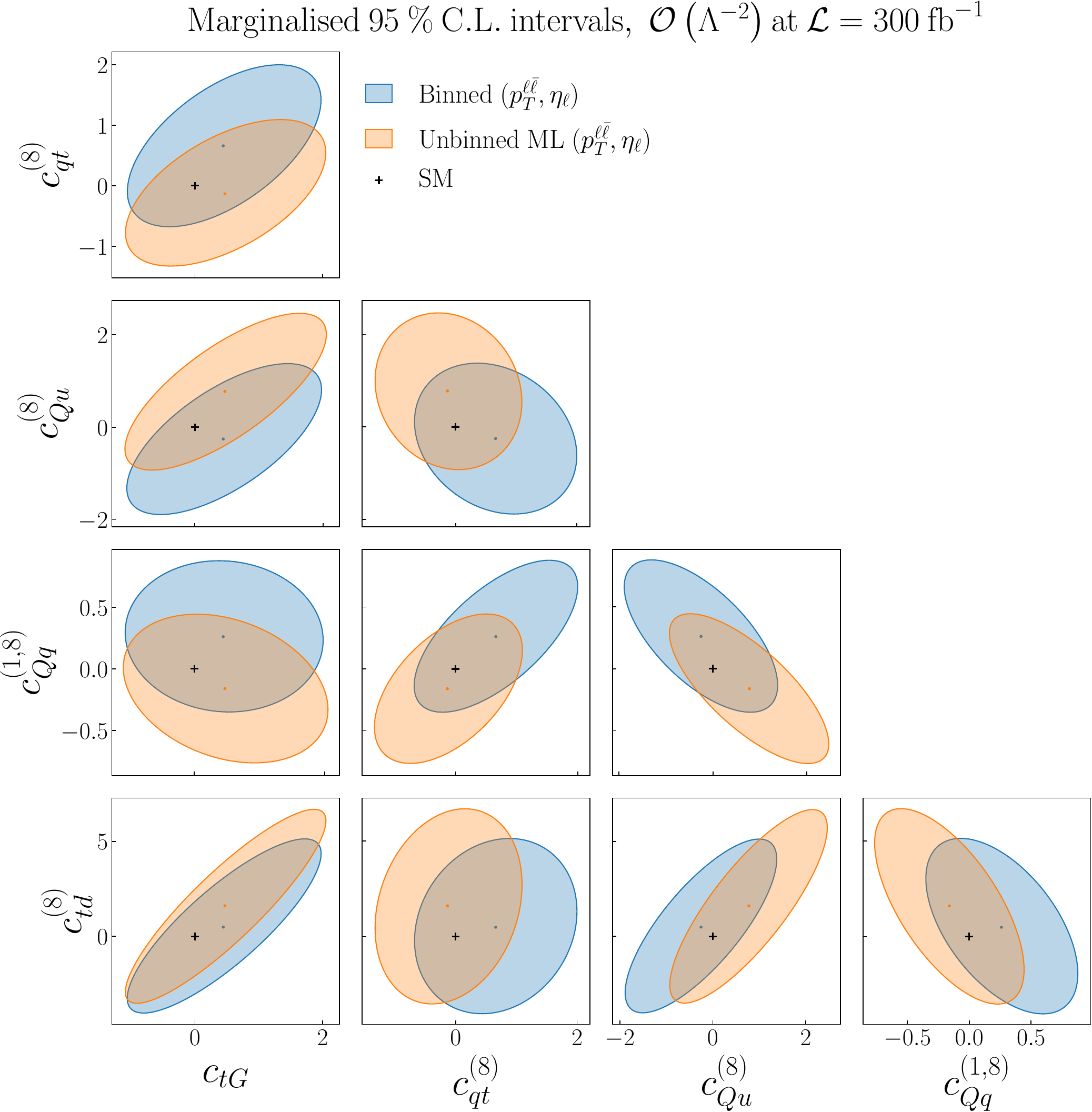}
    \caption{Pair-wise 95\% CL  contours for the 
      Wilson coefficients entering  top quark pair production
      in the dilepton final state, see Sect.~\ref{subsec:ttbar_theorysim} for
      more details.
      These contours are obtained by marginalizing over the full posterior
      distribution provided by Nested Sampling.
      We consider here $n_{\rm eft}=5$ Wilson coefficients that
      can be simultaneously constrained from inclusive
      top-quark pair production at the linear level in the EFT expansion.
      We compare the results obtained from both binned
      and unbinned ML  observables constructed
      on the $(p_T^{\ell\bar{\ell}},\eta_\ell)$ kinematic features.
      As in Fig.~\ref{fig:contours_tt_parton}, the black cross indicates the SM values used to generate
      the pseudo-data that enters the inference.
      The comparison of the  unbinned ML
      observable trained on $(p_T^{\ell\bar{\ell}},\eta_\ell)$
      with its counterpart trained on the full set of $n_{k}=18$ kinematic features
      is displayed in Fig.~\ref{fig:tt_glob_lin_nn}.
    }
    \label{fig:tt_glob_lin_nn_binned}
\end{figure}

For this process we construct $n_k=18$  kinematic features
built from the four-vectors
of the final-state leptons and $b-$quarks.
Considering further kinematic variables such as the missing $E_T$ would
be redundant and not provide additional information.
The distribution of these kinematic features in the SM and when
quadratic EFT corrections are accounted for is displayed in Fig.~\ref{fig:tt_quad_dist}, showing
how different features provide complementary information to constrain the EFT and
thus making a fully-fledged multivariate analysis both interesting and necessary in order to
fully capture the EFT effects.
For example, the pseudo-rapidity distributions bend around $\eta=0$ corresponding to the direction transverse to the beam pipe,
while the transverse momenta are sensitive to energy growing effects in the
high-$p_T$ tails of the distributions.

Fig.~\ref{fig:tt_glob_lin_nn_binned} displays
the pair-wise 95\% CL intervals for the $n_{\rm eft}=5$ Wilson coefficients
in Eq.~(\ref{eq:subset_operator_topparticle_linear}) relevant
for the description of top quark pair production at the linear $\mathcal{O}(\Lambda^{-2})$
level.
The black cross indicates the SM scenario used to generate
the pseudo-data that enters the inference.
The  95\% CL contours shown are obtained from the full posterior
distribution provided by Nested Sampling, marginalizing over the 
remaining Wilson coefficients for each of the operator pairs, with the ellipses drawn from the $N_s$ posterior samples provided by MultiNest.
These are compared with 
the bounds provided by a binned observable
based on the dilepton transverse momentum
$p_T^{\ell\bar{\ell}}$ and the lepton pseudorapidity  $\eta_\ell$
as kinematic features, where the binning is defined as
\bea
p_T^{\ell\bar{\ell}} &\in& \lc 0,10,20,40,60, 100, 150, 400, \infty\rp \, \textrm{GeV}, \nonumber \\
\eta^{\ell} &\in& \lc 0, 0.3, 0.6, 0.9, 1.2, 1.5, 1.8, 2.1, 2.5\rc \, . \nonumber 
\eea
As for the parton-level case, a cut in the invariant mass $m_{t\bar{t}}$
is applied to ensure that Eq.~(\ref{eq:condition_luminosity})
is satisfied and that statistical uncertainties dominate.
Note that for all observables the pseudo-dataset used for the EFT parameter inference
is the same and was not used during training.

\begin{figure}[htbp]
    \centering
    \includegraphics[width=0.85\textwidth]{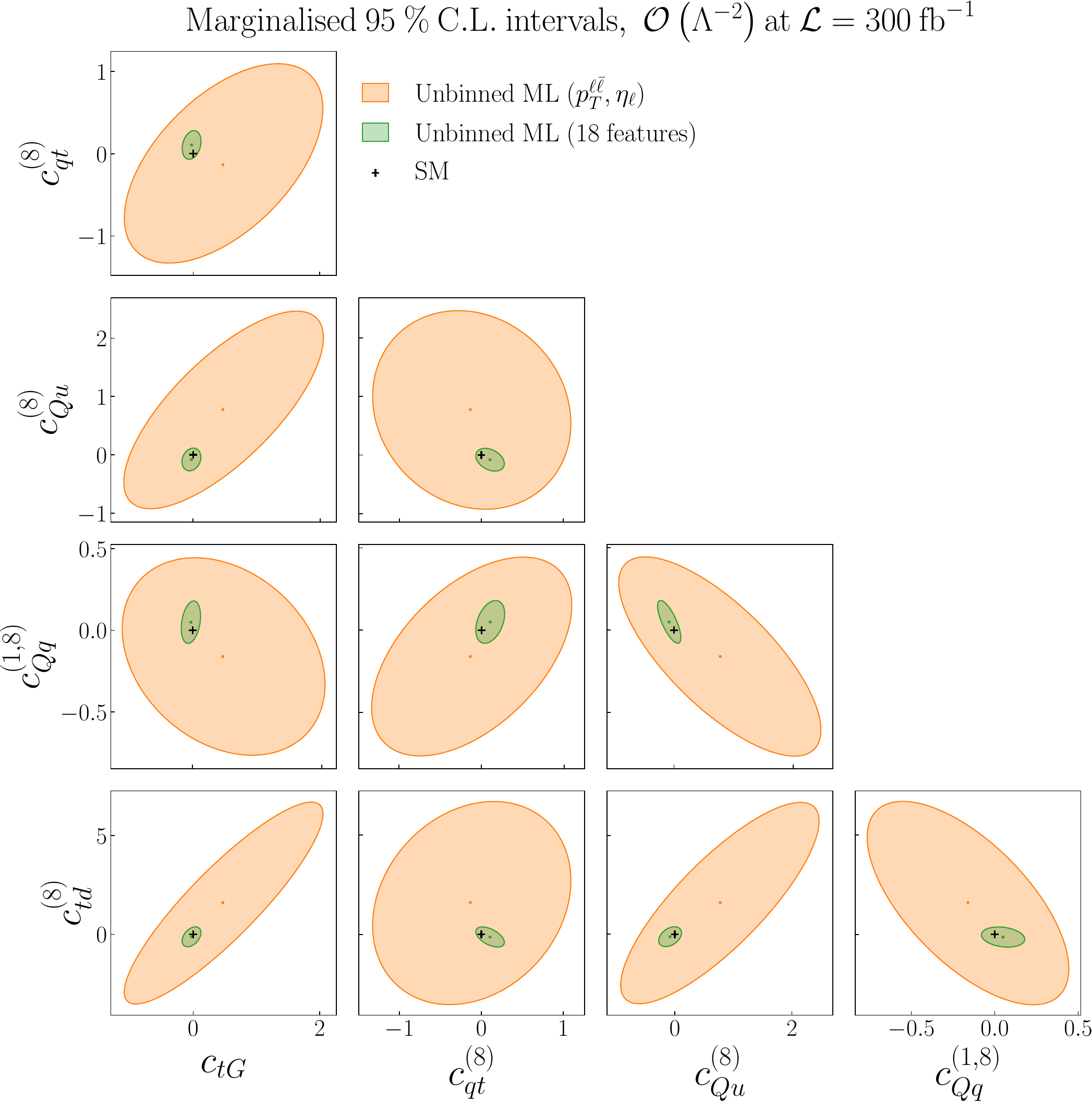}
    \caption{Same as Fig.~\ref{fig:tt_glob_lin_nn_binned} comparing
      the bounds on the EFT coefficients from the ML unbinned observables
trained on $(p_T^{\ell\bar{\ell}},\eta_\ell)$ and on
the full set of $n_k=18$ kinematic features listed in Sect.~\ref{subsec:ttbar_theorysim},
see also Fig.~\ref{fig:tt_quad_dist}.}
    \label{fig:tt_glob_lin_nn}
\end{figure}

From the comparisons in Fig.~\ref{fig:tt_glob_lin_nn_binned}
one can observe how, for all observables and all operator pairings, the  95\% CL intervals
include the SM values used to generate the pseudo-data.
The bounds obtained from the binned and unbinned observables are in general
similar and compatible, and no major improvement is obtained with the adoption of the latter
in this case.
However, here we consider only $n_k=2$ kinematic features and hence ignore the potentially useful
information contained in other variables that can be constructed from the final state kinematics.
In order to assess their impact, Fig.~\ref{fig:tt_glob_lin_nn} displays
the same pair-wise marginalized comparison,
now between  ML unbinned observables with only $p_T^{\ell\bar{\ell}}$ and  $\eta_\ell$
as input features and with the full set of $n_k=18$ kinematic variables displayed
in Fig.~\ref{fig:tt_quad_dist}.
We now find a very significant change in the bounds on the
EFT parameter space, improving by up
to an order of magnitude or better in all cases.
Again, the 95\% CL contours include the SM hypothesis used to generate
the pseudo-data.
Furthermore, the inclusion of the full set of kinematic features
 reduces the correlations between operator pairs that arise
when only $p_T^{\ell\bar{\ell}}$ and $\eta_\ell$ are considered, indicating a breaking
of degeneracies in parameter space.
This result indicates that a multivariate analysis improves the information
on the EFT parameter space that can be extracted from this process
as compared to that obtained from a subset of kinematic features.

\begin{figure}[htbp]
    \centering
    \includegraphics[width=\textwidth]{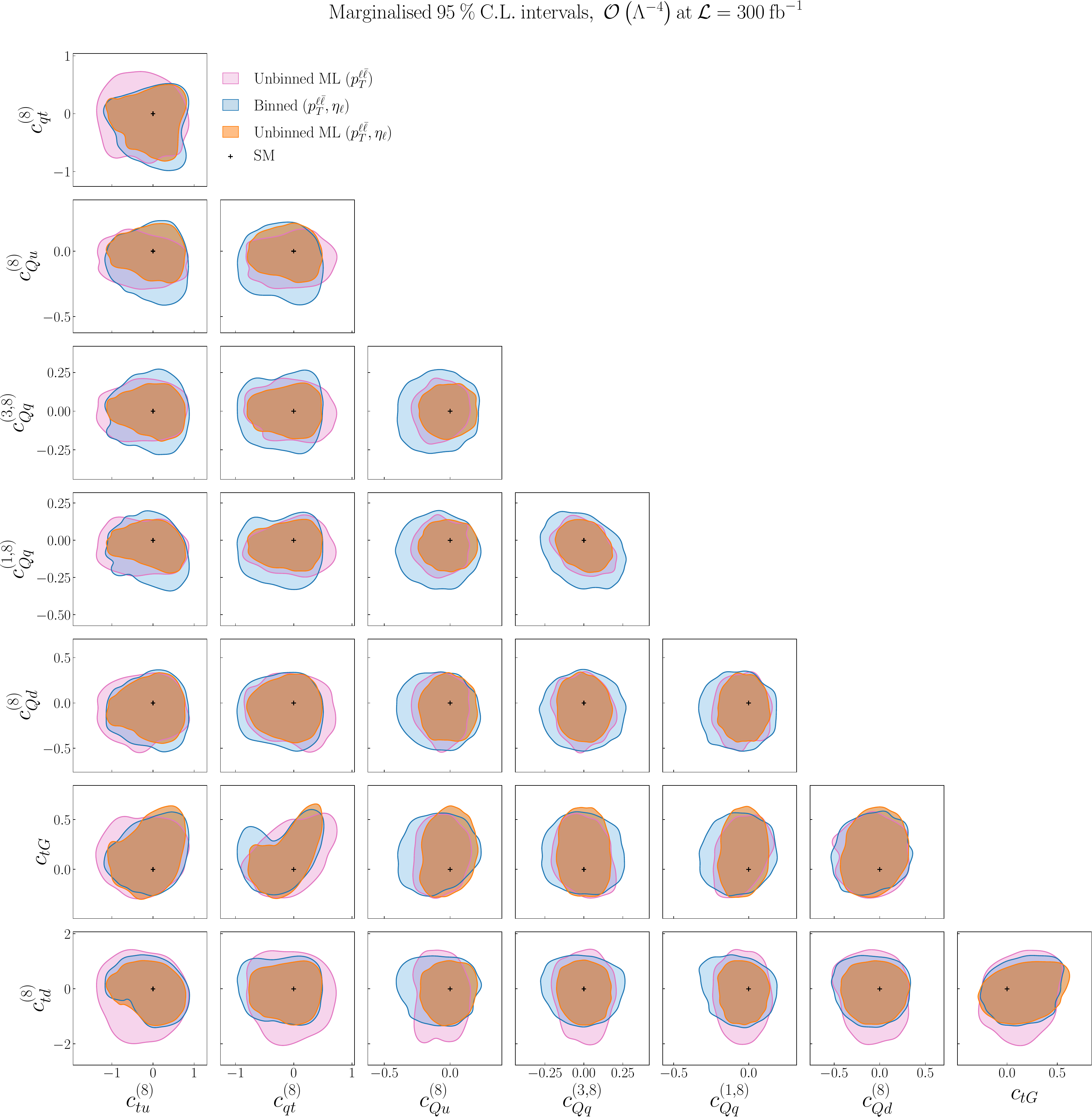}
    \caption{Same as Fig.~\ref{fig:tt_glob_lin_nn_binned} in the case of the EFT analysis
      carried out at the quadratic $\mathcal{O}(\Lambda^{-4})$ level.
      We display the results for pair-wise contours obtained
      from the marginalization of the posterior distribution
      in the space of $n_{\rm eft}=8$ Wilson coefficients.
      In comparison with the linear EFT analysis, it becomes possible
      to constrain three more coefficients from the same process once quadratic
      corrections are accounted for.
      We also include in this comparison the results obtained
      from the unbinned ML analysis based on $p_T^{\ell\bar{\ell}}$
      as the single input feature.
    }
    \label{fig:tt_glob_quad_nn_binned}
\end{figure}

We note that that the results displayed in Figs.~\ref{fig:tt_glob_lin_nn_binned}
and~\ref{fig:tt_glob_lin_nn} are expected to differ should the starting point be a global
EFT analysis rather than a flat prior as is the case in the present proof-of-concept analysis.
For instance, the two-light-two-heavy operators entering $t\bar{t}$ production
at the linear level are highly correlated,
which means that adopting a multivariate analysis leads to a sizable effect
partly because it breaks degeneracies in the parameter space.
Hence, the actual impact  of unbinned multivariate observables
for EFT analyses depends on which other datasets and processes are considered
and can only be assessed on a case-by-case basis.
In this respect, beyond the implications for the specific processes considered in this work,
what our framework provides is a robust method to quantify the information on
the EFT parameters provided by different types
of observables constructed on exactly the same dataset: binned versus unbinned,
different choices of binnings, and different numbers and types of kinematic
features.

\begin{figure}[htbp]
    \centering
    \includegraphics[width=\textwidth]{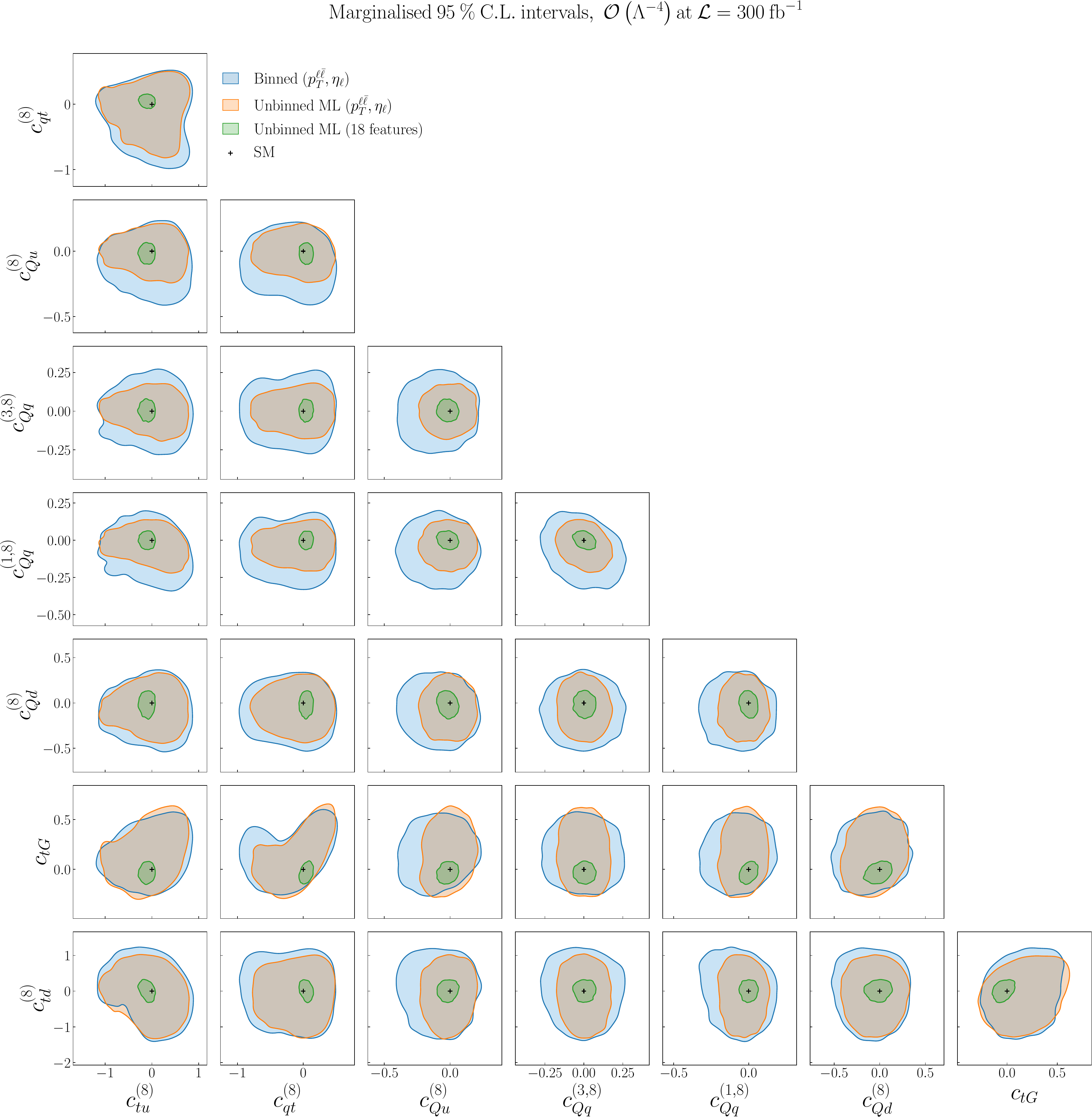}
    \caption{Same as Fig.~\ref{fig:tt_glob_quad_nn_binned}
      comparing the bounds obtained
      from the binned and unbinned observables built upon two
      kinematic features,  $p_T^{\ell\bar{\ell}}$ and $\eta_\ell$, with the corresponding
      results for the unbinned ML observable trained on the full set of
      $n_k=18$ kinematic features.
      \label{fig:tt_glob_quad_nn}}
\end{figure}

As is well known, in top quark pair production quadratic
EFT corrections are important for most operators, in particular due to energy-growing effects
in the tails of distributions.
These sizable effects are highlighted by the distortions with respect to the SM
distributions induced by quadratic EFT effects, shown in
Fig.~\ref{fig:tt_quad_dist}, in the kinematic features
used to train the ML classifier for this process.
In order to investigate how the results based on linear EFT calculations vary once
quadratic corrections are considered,
in Figs.~\ref{fig:tt_glob_quad_nn_binned} and~\ref{fig:tt_glob_quad_nn}
we present the analogous comparisons to Figs.~\ref{fig:tt_glob_lin_nn_binned}
and~\ref{fig:tt_glob_lin_nn} respectively in the case of EFT calculations that also
include the quadratic corrections.
We display the results for pair-wise contours obtained
from the marginalization of the posterior distribution
in the space of the full set of $n_{\rm eft}=8$ Wilson coefficients,
given that once quadratic corrections are accounted for it becomes
possible to constrain the full set of relevant operators simultaneously.
We also include in Fig.~\ref{fig:tt_glob_quad_nn_binned}  the results obtained
from the unbinned ML analysis based on $p_T^{\ell\bar{\ell}}$
as the single input feature, while Fig.~\ref{fig:tt_glob_quad_nn}
also displays the two-feature binned contours as reference.
We note that once quadratic effects are accounted for the 95\% CL contours
will in general not be elliptic, and may even be composed of disjoint regions in the case
of degenerate maxima.

Comparing the constraints provided by the binned $(p_T^{\ell\bar{\ell}},\eta_\ell)$ observable
in Fig.~\ref{fig:tt_glob_quad_nn_binned} with those in Figs.~\ref{fig:tt_glob_lin_nn_binned},
one observes how the bounds on the EFT coefficients are improved
when accounting for the quadratic EFT corrections.
This improvement is consistent with the large quadratic corrections
to the kinematic distributions entering the likelihood ratio of this process which lead
to an enhanced sensitivity,
 and with previous studies~\cite{Ethier:2021bye,Hartland:2019bjb} within global SMEFT analyses.
From the results Fig.~\ref{fig:tt_glob_quad_nn_binned} we find that
for all operator pairs considered, the most stringent bounds are provided
by the unbinned observable built upon the $(p_T^{\ell\bar{\ell}},\eta_\ell)$
pair of kinematic features.
Furthermore, one can identify the factor which brings in more information: either
using the full event-by-event kinematics for a given set of features, or increasing
the number of kinematic features being considered.
For some operator combinations, the dominant effect turns out to be
that of adding
a second kinematic feature, for instance in the case of $(c_{tu}^{(8)},c_{td}^{(8)})$
binned and unbinned observables coincide when using the $(p_T^{\ell\bar{\ell}},\eta_\ell)$
features while the unbinned $p_T^{\ell\bar{\ell}}$ observable results in larger
uncertainties.
For other cases, it is instead the information provided by the event-by-event kinematics
which dominates, for example in the $(c_{Qq}^{(3,8)},c_{Qu}^{(8)})$ plane
the constraints from binned $(p_T^{\ell\bar{\ell}},\eta_\ell)$ are clearly looser
than those from their unbinned counterparts.

The comparison between the constraints on the EFT parameter space provided by the unbinned observable
based on two kinematic features, $(p_T^{\ell\bar{\ell}},\eta_\ell)$,
and those provided by its counterpart based on the full set $n_k=18$ kinematic
features is displayed in Fig.~\ref{fig:tt_glob_quad_nn}.
It confirms that at the quadratic level there is also a marked gain in constraining power
obtained from increasing the number of kinematic variables considered in the analysis.
For reference, the plot also displays the bounds obtained with the two-feature binned observable
in the same process.
Interestingly, when considering the full set of kinematic features, one obtains
posterior contours which display a reduced operator correlation,
highlighting how the unbinned multivariate operator is especially effective
at removing (quasi)-degeneracies in the EFT coefficients.
As for the linear case, the improvement in the bounds obtained
with the multivariate can reach up to an order
of magnitude as compared to the two-feature unbinned analysis.

One can compare the bounds obtained in the present study with those from the corresponding
{\sc\small SMEFiT} global analysis of LHC data.
For the same EFT settings, the
95\% CL marginalized bounds on the chromomagnetic operator
$c_{tg}$ are $\lc 0.062,0.24\rc$ in {\sc\small SMEFiT},
to be compared with a bound of
$\Delta c_{tg} \simeq 0.4$ obtained from the binned and unbinned $(p_T^{\ell\bar{\ell}},\eta_\ell)$ observable
and that of $\Delta c_{tg} \simeq 0.1$ from the multivariate unbinned observable trained using
all features.
While there are too many differences in the input dataset and other settings
to make possible a consistent comparison, this initial estimate suggests that unbinned
multivariate measurements
based on Run III data could provide competitive constraints on the EFT parameter space
as compared to the available binned observables.

All in all, we find that for inclusive top quark pair production,
deploying unbinned multivariate observables makes it possible to tap into a source of information
on the EFT parameter space which is not fully exploited in binned observables,
both at the level of linear and quadratic EFT analyses, and that increasing
the number of  kinematic features considered in the likelihood parametrization
enhances the constraining power of the measurement.

\subsection{Higgs associated production with vector bosons}
\label{subsec:zh}
We consider next the impact of unbinned multivariate
observables in Higgs associated production with $Z$ bosons
in the $pp \rightarrow Z h \rightarrow b\bar{b}\ell^+\ell^-$ channel.
The pseudo-data for this process has been generated with the settings and fiducial
cuts described in Sect.~\ref{subsec:hz_theorysim}, and the list of operators
constrained is defined in Table \ref{tab:op_defn_zh}.
As for the case of $t\bar{t}$ production, we compare the bounds provided by binned and by unbinned
observables at the level of linear and quadratic EFT calculations, and study the dependence of the results
on the number of kinematic features used to parametrize the likelihood ratio.

\begin{figure}[htbp]
    \centering
    \includegraphics[width=\textwidth]{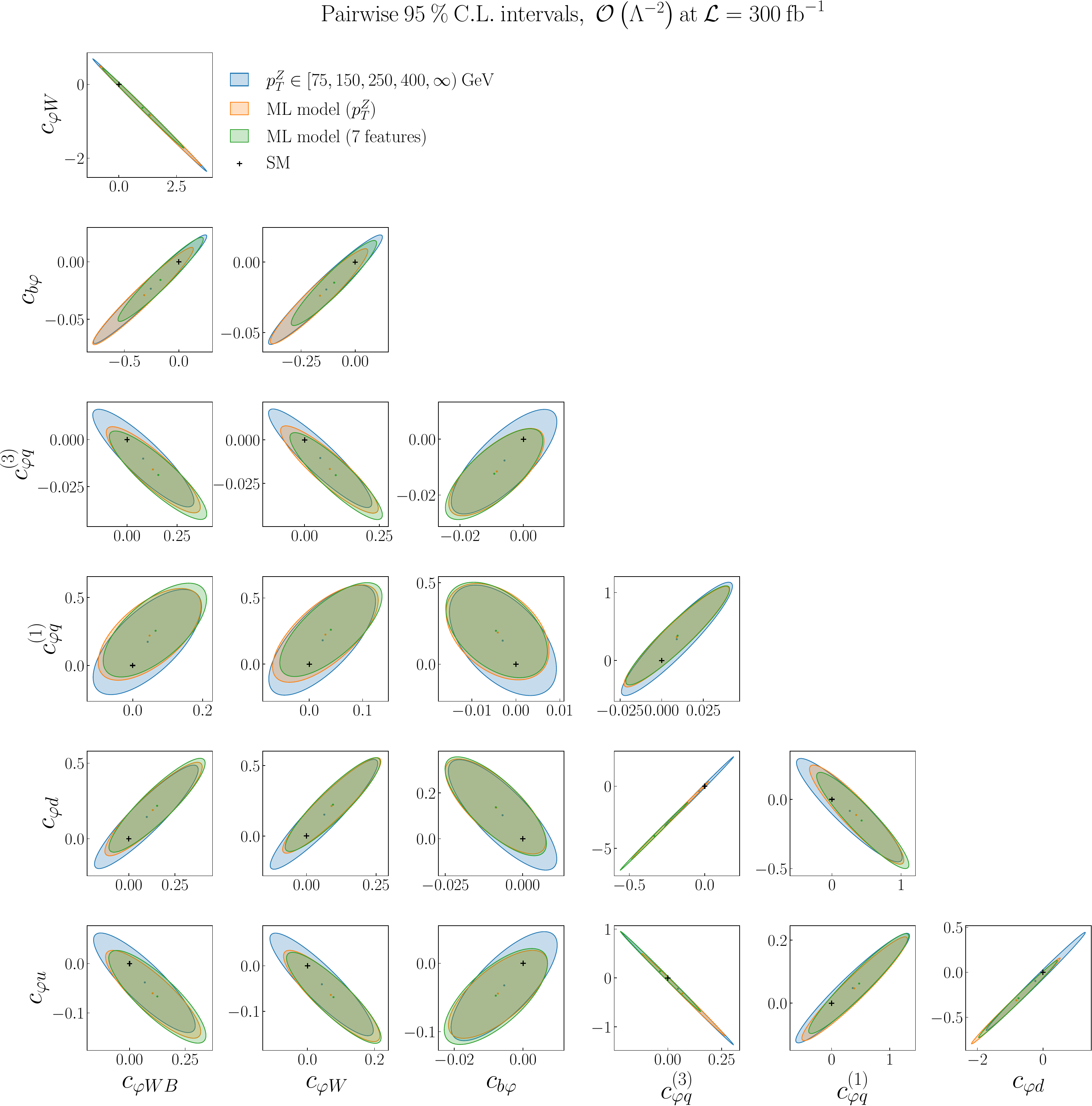}
    \caption{Pair-wise 95\% CL contours obtained from two-parameter fits
      in case of $pp\rightarrow hZ \rightarrow b\bar{b}\ell^+\ell^-$ at the linear level in the EFT expansion
      for the $n_{\rm eft}=7$ Wilson coefficients relevant for the description of this process.
      For each panel, the contribution from the operators not displayed is set to zero.
      We compare the bounds obtained from a binned $p_T^Z$ distribution with those from two ML unbinned
      observables, first when only $p_T^Z$ is used for the training and then when $n_k=7$ kinematic features
      are used to parametrize the likelihood ratio. }
    \label{fig:zh_pairwise_lin_nn_binned}
\end{figure}

Fig.~\ref{fig:zh_pairwise_lin_nn_binned} displays 
the pair-wise 95\% CL contours obtained from two-parameter fits
of $pp\rightarrow hZ \rightarrow b\bar{b}\ell^+\ell^-$ pseudo-data at the linear level in the EFT expansion.
In these two-parameter fits, for each  panel the contribution from the operators not being shown is set to zero.
The choice of displaying the outcome of two-parameter fits, rather than a full marginalized analysis,
is motivated by the fact that the linear EFT analysis
is not stable when all coefficients are simultaneously
considered due to quasi-flat directions and strong operator correlations.
This restriction can be lifted once we account for quadratic effects.

We show the bounds obtained from a binned $p_T^Z$ distribution, with $p_T^Z$ being the dilepton
transverse momentum,
where the choice of binning is given by
\be
p_T^Z \in [75, 150, 250, 400, \infty ) \, ,
\ee
consistent with the definitions entering the Simplified Template Cross Section (STXS)~\cite{Berger:2019wnu}
adopted by the LHC Higgs Working Group.
These binned bounds are compared with those from two ML unbinned
observables, first when only the dilepton transverse momentum
$p_T^Z$ is used for the training and second when the full set of $n_k=7$ kinematic features
is used to parametrize the likelihood ratio. 
As in the case of top quark production,
comparing the constraints from the binned observable with those from the unbinned
one trained on the same feature  quantifies the information loss incurred by the binning procedure,
while comparing one feature with multivariate unbinned observables determines
the information loss associated to the use of a restricted set of kinematic features

\begin{figure}[htbp]
    \centering
    \includegraphics[width=\textwidth]{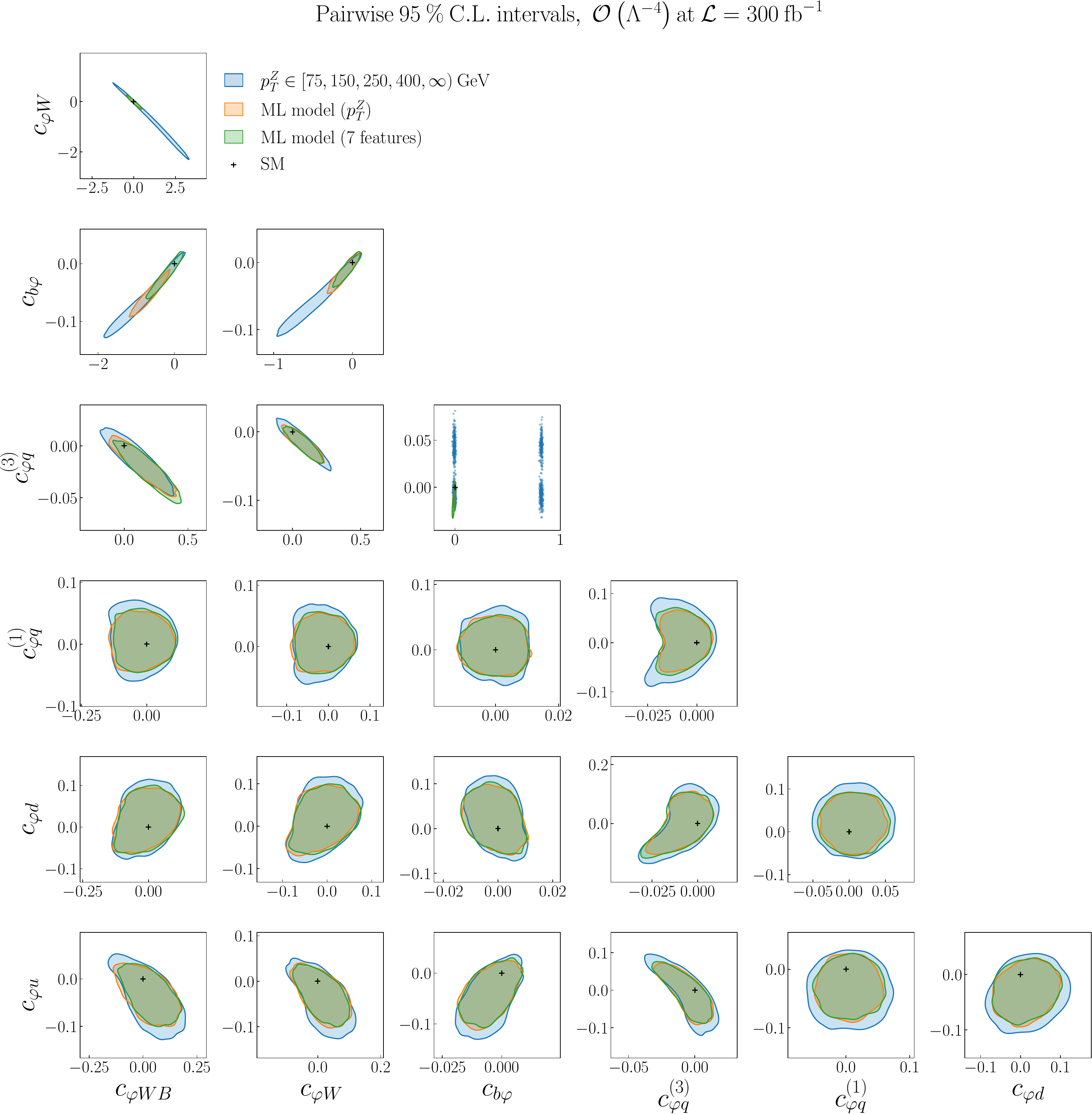}
    \caption{Same as Fig.~\ref{fig:zh_pairwise_lin_nn_binned}
      for the case in which
    the EFT calculations also include the quadratic $\mathcal{O}(\Lambda^{-4})$ corrections.}
    \label{fig:contours_zh_quad_pt_all_pairwise}
\end{figure}

At this linear EFT level, the improvement in the constraints provided
by  unbinned observables as compared to their binned $p_T^Z$ counterparts is moderate,
yet appreciable, for most operator pairs.
The most stringent bounds
are those associated with unbinned observables,
and the constraints provided by the multivariate unbinned observable based
on 7 kinematic features are similar to those associated
to the single-feature version  trained on $p_T^Z$.
One observes strong correlations between all operators pairs
being considered, indicating quasi-flat directions:
these will be broken either by quadratic corrections or by adding other processes
sensitive to the same SMEFT operators, such as vector-boson scattering and diboson data~\cite{Ethier:2021ydt}.
The analysis of Fig.~\ref{fig:zh_pairwise_lin_nn_binned}  indicates
that for the $hZ \rightarrow b\bar{b}\ell^+\ell^-$ process the chosen binning in $p_T^Z$
provides the bulk of the information on the EFT operators, with minor additional
improvements provided by the unbinned multivariate analysis.

Therefore, for this specific process the formalism  developed
in this work provides a diagnosis tool that
determines when and under which circumstances a binned analysis provides
information on the EFT parameter space close to the optimal one, represented here
by the bounds associated to the unbinned multivariate observable.
This finding illustrates the two-fold applicability of our formalism: whenever unbinned multivariate observables
markedly improve over their binned counterparts, they can be included in the global SMEFT
fit, otherwise one obtains  quantitative
confirmation that the information provided by the specific settings of the binned observables is sufficiently close
to the optimal amount.

\begin{figure}[htbp]
    \centering
    \includegraphics[width=\textwidth]{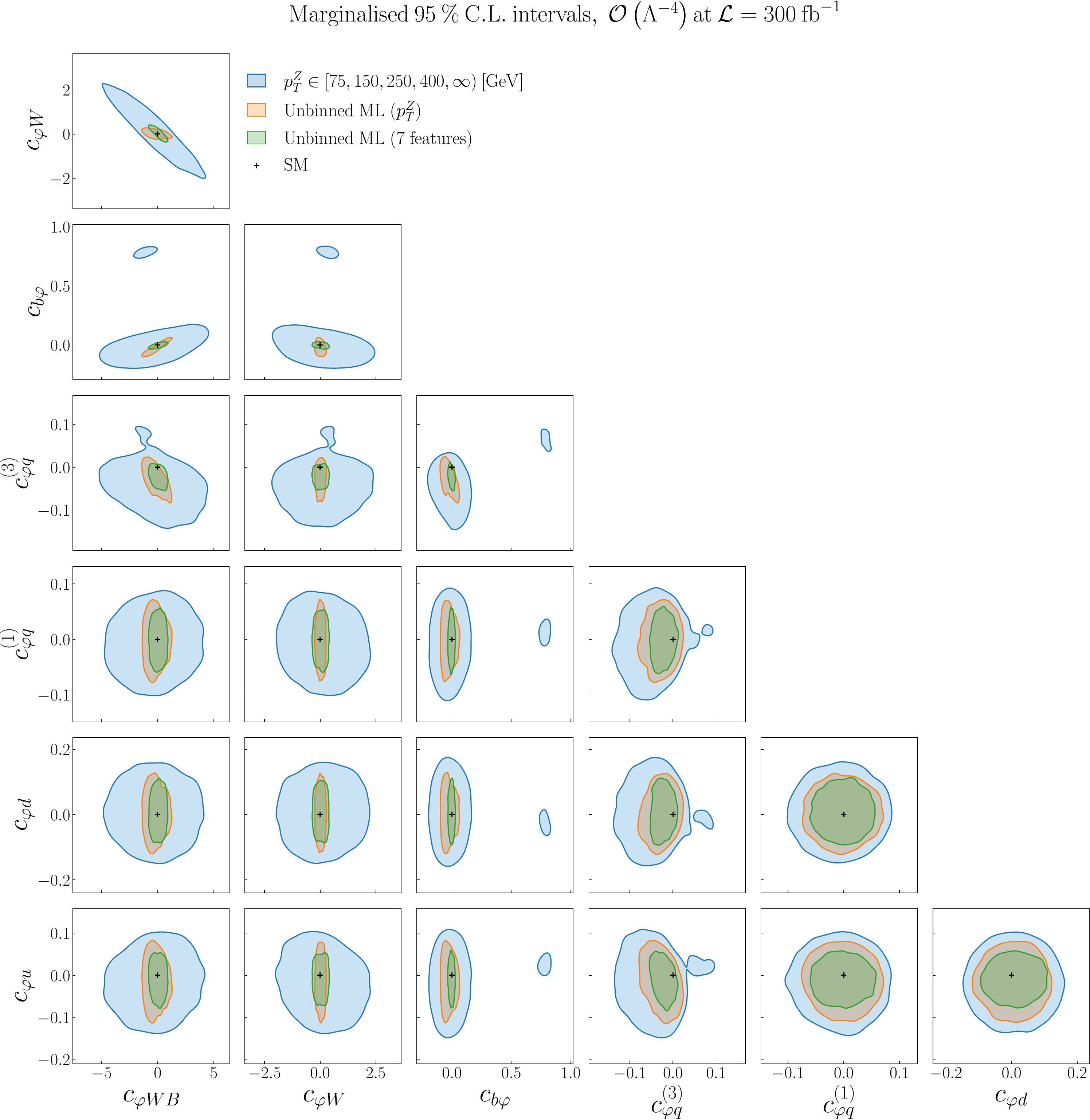}
    \caption{Same as Fig.~\ref{fig:contours_zh_quad_pt_all_pairwise} now
      where the $95\%$ CL contours are obtained from the marginalisation over the full
      posterior probability distribution, rather than from two-parameter fits. }
    \label{fig:contours_zh_quad_pt_all_marg}
\end{figure}

The results of the corresponding analyses of the impact
of unbinned observables in the $hZ \rightarrow b\bar{b}\ell^+\ell^-$ process
in the presence of quadratic EFT corrections 
are shown in Fig.~\ref{fig:contours_zh_quad_pt_all_pairwise}
for the case of two-parameter fits, and in Fig.~\ref{fig:contours_zh_quad_pt_all_marg}
for the case  where all relevant operators are considered simultaneously,
producing marginalized constraints on each operator pair.
As mentioned above, the latter is possible since quadratic corrections break up
the EFT parameter degeneracies that
destabilize the inference procedure
in the linear case.

Considering first the results of Fig.~\ref{fig:contours_zh_quad_pt_all_pairwise},
one finds that on the one hand the improvement associated to the use of unbinned observables
is more marked as compared to the linear EFT case.
On the other hand, the addition of other kinematic features on top
of $p_T^Z$ does not modify the discrimination power of
the unbinned observables, as also observed for the linear analysis.
For specific operator pairs, there is a clear reduction
in the parameter bounds based on the unbinned observables,
for instance in the case of the  $\lp c_{\varphi W}, c_{\varphi W B}\rp$ pair.
We note that enhanced discrimination in the presence of quadratic EFT corrections
can be partially attributed to energy-growing effects in the $p_T^Z$ distribution which are more pronounced
as compared to the linear level.

Another benefit associated to unbinned observables in this specific process
is that of breaking degeneracies leading to a double maximum
structure in the posterior distribution.
Indeed, the unbinned observables
removes the second minimum present in the binned analysis arising from a bimodal distribution
in the ($c_{\varphi q}^{(3)}$, $c_{b\varphi}$)-plane
and 
corresponding to the scenario where the quadratic EFT contribution becomes of the same
magnitude and opposite sign as the linear term and hence canceling it.
We also note that, as opposed to the $t\bar{t}$ production case,
for the $hZ$ process, even with unbinned observables, one ends up with some operator pairs
that exhibit relatively large correlations which would thus only be broken by
adding other processes to the fit.
The results of Fig.~\ref{fig:contours_zh_quad_pt_all_pairwise} highlight how
in general the relevance of considering unbinned observables
depends not only on the processes and on the order in the EFT expansion,
but also on the specific directions in the EFT parameter space in which one is interested.

The results of  Fig.~\ref{fig:contours_zh_quad_pt_all_pairwise}
display good agreement between unbinned observables trained
only on $p_T^Z$ and on the full set of $n_k=7$ kinematic features,  suggesting that a multivariate analysis
does not appear to be beneficial for this process.
However, this interpretation turns out to be an artifact of restricting the analysis
to the case of two-parameter fits, and a different picture emerges
in the case of the marginalized analysis results displayed in Fig.~\ref{fig:contours_zh_quad_pt_all_marg}.
The impact on the EFT
parameter space of adopting unbinned observables  is more important now,
leading in some cases to stringent bounds improved by up to a factor 5 as compared
to the binned results, for instance for the $c_{\varphi W}$ operator.
Furthermore, one can also appreciate  the improved constraints associated to the multivariate
observable based on the full set of kinematic features as compared to the single-feature unbinned observable.
We also note how the degenerate minimum associated to the bottom Yukawa operator
 present in the binned analysis, now for all operator pairs, goes away when
unbinned observables are considered.
The comparison of Figs.~\ref{fig:contours_zh_quad_pt_all_pairwise}
and~\ref{fig:contours_zh_quad_pt_all_marg} emphasizes that bounds obtained
in two-parameter fits,  neglecting the effects of other operators, will
in general overestimate the EFT constraints associated to a given dataset.

To summarize, as in the case of top quark pair
production, for $hZ$ production the use
of unbinned multivariate observables is advantageous in
constraining the EFT parameter space.
Our analysis indicates that the choice of $p_T^Z$ binning adopted
by the STXS analysis is close to being optimal, especially in the linear case, at least
for the assumed integrated luminosity and for two-parameter fits.
When considering a full marginalized analysis in the presence of quadratic
EFT corrections, marked improvements associated first to the use of unbinned observables
and second to that of multiple kinematic features are obtained.
The use of unbinned observables also results in the removal of degenerate minima corresponding
to solutions where the quadratic EFT correction approximately cancels out the linear one.
The comparison between the output of the two-parameter and the marginalized fits
emphasizes the crucial importance of carrying out global SMEFT interpretations,
since setting to zero a subset of operators that contribute
to a given process is likely to overestimate its impact in the EFT parameter space.

\subsection{Methodological uncertainties}
\label{subsec:eft_uncert}

The results for the unbinned observables derived so far
are obtained from  EFT parameter inference carried out with Eq.~(\ref{eq:EFT_structure_v4}), the parametrization of the
likelihood ratio, as described in Sect.~\ref{subsec:EFT_inference}.
As explained in Sect.~\ref{sec:nntraining}, within our approach, rather than a single best model,
we produce a distribution of models, denoted as replicas, each of them trained on a different set
of Monte Carlo events.
Hence the end result of our procedure is Eq.~(\ref{eq:EFT_structure_v5}),
the representation of the probability distribution of the likelihood ratio
$\{\hat{r}^{(i)}_{\sigma}(\boldsymbol{x}, \boldsymbol{c})\}$ composed of $N_{\rm rep}$ equiprobable replicas.
The spread of this distribution provides a measure of methodological and
procedural uncertainties
associated e.g. to finite training datasets and inefficiencies of the optimization and
stopping algorithms.
Results presented in Sects.~\ref{subsec:top_parton}--\ref{subsec:zh}  are based on using the median of the
replica distribution
$\{\hat{r}^{(i)}_{\sigma}(\boldsymbol{x}, \boldsymbol{c})\}$ to determine the bounds in the EFT
parameter space, and here we assess the impact of methodological
uncertainties by displaying results for parameter inference based on the full
distribution of replicas of the likelihood ratio parametrization.

\begin{figure}[htbp]
    \centering
    \includegraphics[width=\textwidth]{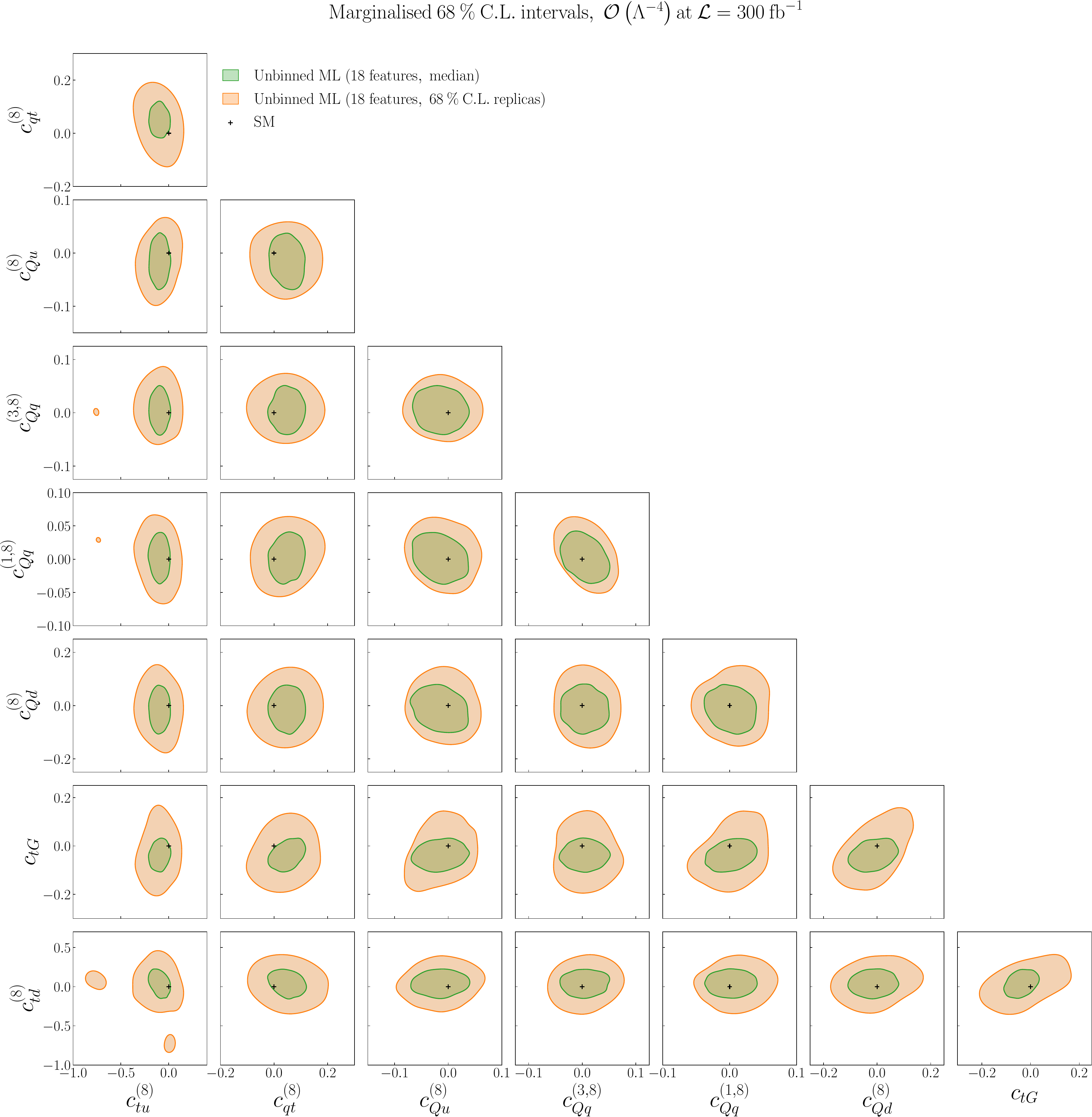}
    \caption{Same as Fig.~\ref{fig:tt_glob_quad_nn} for the 68\% CL contours
       in the EFT parameters obtained with the unbinned multivariate
       observable for particle-level top-quark pair production.
       We compare the bounds obtained from the median of the replica
       distribution of the likelihood ratio
       parametrization, as  done in Fig.~\ref{fig:tt_glob_quad_nn},
       with the corresponding bounds obtained taking into account the full replica
       distribution.
    }
    \label{fig:tt_quad_envelope}
\end{figure}

\begin{figure}[htbp]
    \centering
    \includegraphics[width=\textwidth]{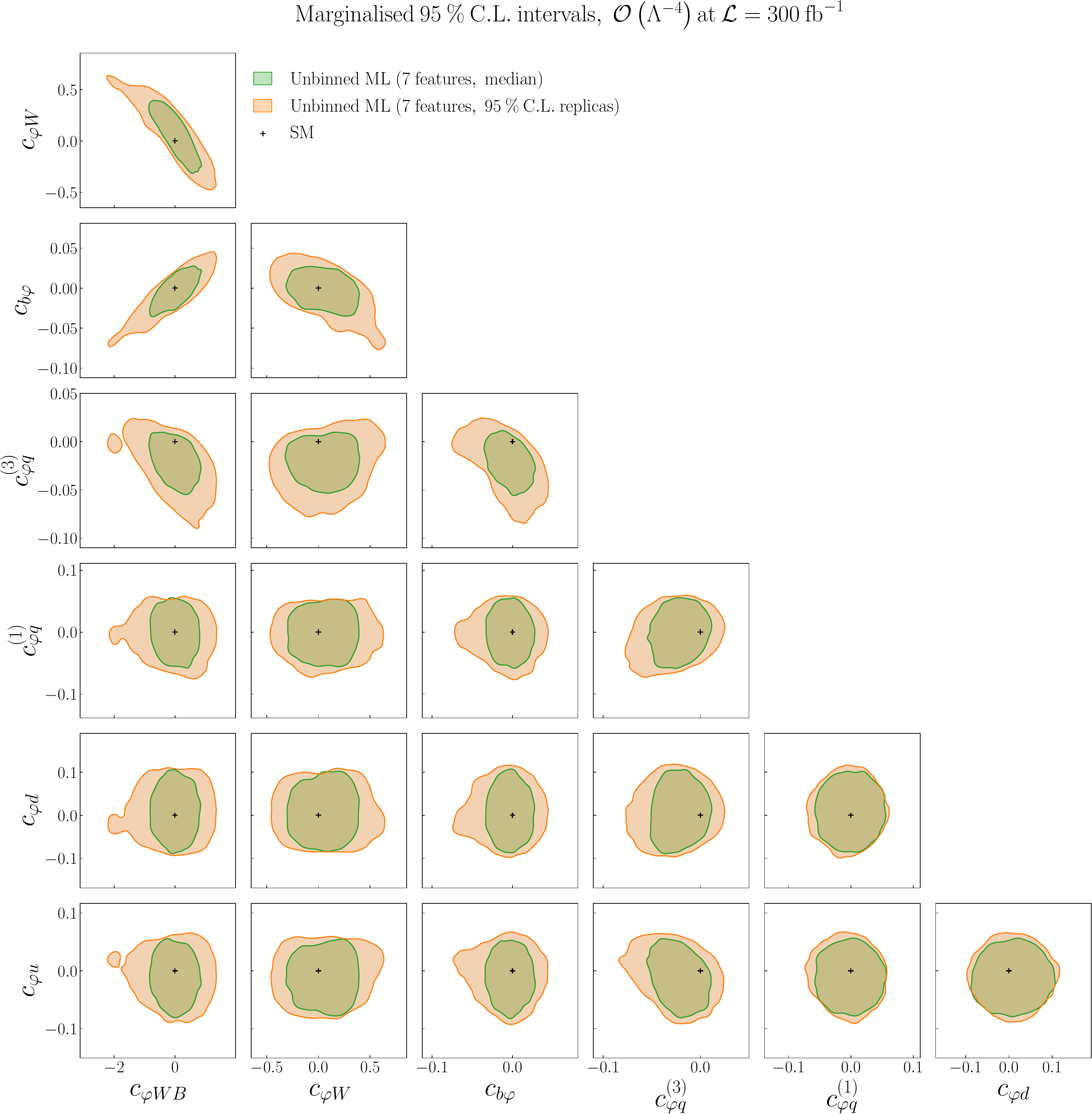}
	\caption{
	  Same as Fig.~\ref{fig:tt_quad_envelope} for Higgs associated production
          with a $Z$ boson, now for the 95\% CL contours.
    }
    \label{fig:zh_quad_envelope}
\end{figure}

Figs.~\ref{fig:tt_quad_envelope} and~\ref{fig:zh_quad_envelope} display
the bounds in the parameter space obtained for
particle-level top-quark pair and $hZ$ associated production,
respectively, from the unbinned multivariate
observables based on the complete set of kinematic features and in the case of theory
simulations that include quadratic EFT corrections.
We compare results for parameter inference based on the median of
the replica distribution
$\{\hat{r}^{(i)}_{\sigma}(\boldsymbol{x}, \boldsymbol{c})\}$, which coincide with
those of Figs.~\ref{fig:tt_glob_quad_nn}  and~\ref{fig:contours_zh_quad_pt_all_marg},
with results based on the full distribution of replicas.\footnote{We note that for top quark
  pair production we display the 68\% CL contours in both cases.
  The reason is that for $N_{\rm rep}=25$  the evaluation of
 95\% CL contours over the replica distribution becomes overly sensitive to outliers.}
Namely, in the latter case  one starts from the $N_{\rm rep}$ 
individual replicas of the profile likelihood ratio functions and performs Nested Sampling inference
on each of them, to subsequently combine the resulting samples and estimate the
posterior distribution by means of the KDE method.
In this manner, the differences between the contours  shown
in Figs.~\ref{fig:tt_quad_envelope} and~\ref{fig:zh_quad_envelope} provide an estimate of how
methodological uncertainties associated to the machine learning training procedure
impact the derived bounds in the EFT parameter space.

Inspection of Figs.~\ref{fig:tt_quad_envelope} and~\ref{fig:zh_quad_envelope} and
comparison with the corresponding bounds displayed in
Figs.~\ref{fig:tt_glob_quad_nn}  and~\ref{fig:contours_zh_quad_pt_all_marg} confirm that
these procedural uncertainties, as estimated with the replica method, do not modify
the qualitative results obtained from the median of the likelihood ratio.
In particular, the observation that bounds obtained using multivariate unbinned
observables are much stronger than those based on unbinned  models based on one or two
kinematic features,
which is common to both the $t\bar{t}$ and $hZ$ processes, remains valid once
replica uncertainties are accounted for.
Furthermore, we note that in principle it is possible to reduce the spread of
the replica distribution by training on higher-statistics samples and adopting more
stringent stopping criteria.
In this respect, our approach provides a strategy to quantify under which
conditions the computational overhead required to achieve more accurate
machine learning trainings  is justified from the point of view of the
impact on the EFT coefficients.
This said, these results also indicate that in realistic scenarios
based on finite training samples methodological uncertainties cannot
be neglected, and should be accounted for in studies of the
impact of ML-based observables in EFT analyses.

\section{Summary and outlook}
\label{sec:summary}

The general problem of identifying novel types of measurements which, given
a theoretical framework,
provide enhanced or even maximal sensitivity to the parameters of interest
 is ubiquitous in modern particle physics.
 Constructing 
 such optimized observables has a two-fold motivation:
 on the one hand, to achieve
 the most stringent constraints
 on the model parameters from a specific process,
 and on the other hand, to provide bounds
 on the maximum  amount of information that can be extracted
 from the same process.
Optimized observables can hence be used to  design
traditional observables in a way that approaches or saturates
the limiting sensitivity by highlighting the best choices
of binning and kinematic features.

In this work, we have presented a new framework
for the design of optimal observables for EFT applications
at the LHC, making use of machine learning techniques
to parametrize multivariate likelihoods for an arbitrary number
of higher-dimensional operators.
To illustrate the reach of our method,
we have constructed multivariate
unbinned observables for top-quark pair production
and Higgs production in association with a $Z$ boson.
We have demonstrated how these observables  either lead
to a significant improvement in the constraints
on the EFT parameter space, or indicate the conditions upon which
a traditional binned analysis already saturates the EFT sensitivity. 

The {\sc\small ML4EFT} framework  presented in this work and the accompanying simulated
event samples are made available as an open source code which
can be interfaced with existing global EFT fitting tools
such as {\sc\small SMEFiT}, {\sc\small FitMaker}~\cite{Ellis:2020unq},
{\sc\small HepFit}~\cite{DeBlas:2019ehy}, and {\sc\small Sfitter}~\cite{Brivio:2019ius}.
Its scaling behavior with the number of parameters,
together with its parallelization capabilities, 
make it suitable for its integration within global EFT fits which involve several tens
of independent Wilson coefficients.
While in its current implementation the parametrized likelihood ratios are provided
in terms of the output of the trained network replicas, work in progress aims
to tabulate this output in terms of  fast interpolation grids, as done
customarily in the case of PDF analyses~\cite{Buckley:2014ana,Bertone:2014zva,Carrazza:2020gss,Carli:2010rw}.
The resulting interpolated unbinned likelihood ratios can then be
combined with Gaussian or Poissonian binned measurements within a global fit.
While currently {\sc\small ML4EFT} can only be used
in combination with simulated Monte Carlo pseudo-data,
all of the ingredients required for the analysis of actual LHC measurements are already in place.

The results presented in this work could be extended along
several directions.
First, one could include experimental and theoretical
correlated systematic uncertainties, as required for the interpretation of measurements
which are not dominated by statistics.
Second, other machine learning algorithms could be adopted,
which may offer
performance advantages as compared to those used here:
one possibility could be graph neural networks~\cite{ArjonaMartinez:2018eah,Abdughani:2018wrw} which make possible
varying the kinematic features used for the training on an event-by-event basis.
Third, {\sc\small ML4EFT} could be applied to more realistic final states
with higher order corrections, such as by means
of NNLO+PS simulations
of $t\bar{t}$ and $hV$~\cite{Haisch:2022nwz,Mazzitelli:2021mmm,Zanoli:2021iyp}, and  accounting
for detector effects to bridge the gap between
the theory predictions and the experimentally accessible quantities.
It would also be interesting to compare
the performance of different approaches
to construct ML-assisted optimized observables
for EFT applications in specific benchmark scenarios.

Beyond hadron colliders, the framework developed here could also be
relevant to construct optimal observables for EFT analyses in the case of
high-energy lepton colliders, such as electron-positron
collisions at CLIC~\cite{CLICPhysicsWorkingGroup:2004qvu,Linssen:2012hp}
or at a multi-TeV muon collider~\cite{MuonCollider:2022xlm,Chen:2022msz,Buttazzo:2020uzc}.
As mentioned,
statistically optimal observables were actually first designed for electron-positron colliders, where the  simpler final state facilitates the calculation of the exact event
likelihood for parameter inference~\cite{DELPHI:2010ykq,Diehl:1993br,Durieux:2018tev}.
These methods may not be suitable to the high  multiplicity environment
of multi-TeV lepton colliders, in particular due to the complex pattern
of electroweak radiation.
Therefore, the {\sc\small ML4EFT}
method could provide a suitable alternative to construct unbinned multivariate observables
achieving maximal EFT sensitivity at high-energy lepton colliders.

In addition to applications to the SMEFT, our framework
could also be relevant to other types of theory interpretations of collider data such as
global PDF fits.
In particular, PDFs at large values of Bjorken-$x$ are
poorly constrained~\cite{Beenakker:2015rna}
due to the limited amount of experimental data available.
This lack of knowledge degrades the reach of searches for both resonant
and non-resonant new physics in the high-energy tail of
 differential distributions,
as  recently emphasized for the case of the forward-backward asymmetry
in neutral-current Drell-Yan production~\cite{afb}.
Given that measurements of these high-energy tails are often dominated by statistical
uncertainties, the {\sc\small ML4EFT} method could be used
to construct unbinned multivariate observables tailored
to constrain large-$x$ PDFs at the LHC.
Our method could also be applied in the context
of a joint extraction of PDFs and EFT coefficients~\cite{Greljo:2021kvv,Carrazza:2019sec,Iranipour:2022iak, Liu:2022plj, Gao:2022srd},
required to disentangle QCD effects
from BSM ones in kinematic regions where
they potentially overlap, such as large Bjorken-$x$.

\subsection*{Acknowledgments}
We are grateful to Gaia Grosso, Wouter Verkerke, and Andrea Wulzer for stimulating discussions.
The work of J.~t.~H is supported by the Dutch Science Council (NWO) via an ENW-KLEIN-2 project.
The  research of J.~R. is partially supported by NWO and by an ASDI grant
of The Netherlands eScience Center.
The research of V.~S. is supported by the Generalitat Valenciana PROMETEO/2021/083 and the Ministerio de Ciencia e
Innovacion PID2020-113644GB-I00.
The work of M.~M. is supported by
the European Research Council under the European
Union’s Horizon 2020 research and innovation Programme (grant agreement n.950246) and in part by STFC consolidated grant ST/T000694/1.
The work of R.~G.~A. is supported by the ERC Starting Grant REINVENT-714788 and by Grant DataSMEFT23 (PNRR - DM 247 08/22).

\appendix
\section{The {\sc\small ML4EFT} framework}
\label{app:code}

The {\sc\small ML4EFT} framework presented in this work
is an open source {\sc\small Python} code
designed to facilitate the integration of
unbinned multivariate observables into fits of Wilson coefficients
in the SMEFT.
It is based on  machine learning regression and classification techniques
to parameterize high-dimensional likelihood ratios as required to carry out
parameter inference in the context of 
global SMEFT analyses.
{\sc\small ML4EFT} is made available via the Python Package Index (pip) and can be installed directly 
by running 
\begin{center}
	{\tt pip install ml4eft}
\end{center}
or alternatively the code can be downloaded from its public GitHub repository,
\begin{center}
  \url{https://github.com/LHCfitNikhef/ML4EFT.git}
\end{center}
and then installed from source following the installation instructions.
The framework is documented in a dedicated website
\begin{center}
  \url{https://lhcfitnikhef.github.io/ML4EFT},
\end{center}
where, in addition, one can find a self-standing tutorial (which can also
be run in Google Colab) where the user is guided step by step in how
unbinned multivariate observables can be constructed given a choice of
EFT coefficients and of final-state kinematic features.

In the same website we also provide links to the main results of this paper
 presented in Sect.~\ref{sec:results},
including the likelihood ratio parametrizations that have been obtained
for the two processes ($t\bar{t}$ and $hZ$) considered here.
We also include there animations demonstrating the 
training of the neural networks, such as e.g.
\begin{center}
\url{https://lhcfitnikhef.github.io/ML4EFT/sphinx/build/html/results/ttbar_analysis_parton2.html#overview}
\end{center}
Additional unbinned multivariate observables to be constructed in the future using
our framework will be added to the same page.
Furthermore, we also plan to tabulate the neural network
outputs using fast grid techniques so that these observables can be  stand-alone integrated
in global fits without the need to link the actual {\sc\small ML4EFT} code.

\section{The unbinned Asimov data set}
\label{sec:continous_asimov}
\label{app:continous_asimov}

The Asimov data set was first introduced in~\cite{Cowan:2010js} as an efficient method of obtaining the distribution of the profile
likelihood ratio $q_{\boldsymbol{c}}$, Eq.~(\ref{eq:plr}), under the alternative hypothesis $\boldsymbol{c}' \neq \boldsymbol{c}$ without having to resort to computationally expensive MC simulations.
Since then, it has become part of the standard analysis toolkit at the LHC to
determine e.g. expected exclusion limits.
The analysis of~\cite{Cowan:2010js} considers
the  Asimov data set  applied to binned observables
and here, by adopting a simple EFT-like toy model, we present
its generalisation to a continuous Asimov data set.
This makes possible
determining expected exclusion limits in a sampling free way
by extending the results of~\cite{Cowan:2010js} to unbinned observables.
While we do not make use of this unbinned
Asimov data set in the results presented in this work,
this derivation can play an important role
in the general discussion of unbinned multivariate measurements for EFT
studies.

Let us consider a kinematic variable $x$ (e.g. the invariant mass) restricted to $x \in X = [1, 5]$ and distributed according to a probability distribution $f_\sigma(x, c)$.
The EFT toy model is chosen such that the effect of the EFT on the kinematic variable is an enhancement in the tail of the distribution $f_\sigma(x, c)$ with respect to the SM, as
is common in many EFT scenarios.
The distribution $f_\sigma(x, c)$ is defined as
\be 
f_\sigma(x, c) \equiv \frac{1}{\sigma(X, c)}\left[f_{\sigma}^{(\mathrm{sm})}(x) + 0.1c \cdot f_{\sigma}^{(\mathrm{eft})}(x)\right],
\label{eq:eft_toy_dist}
\ee 
with 
\be 
f_{\sigma}^{(\mathrm{sm})}(x) = \frac{1}{\sigma^{(\mathrm{sm})}}\frac{(1+x^2)\sqrt{x^2-1}}{x^4}
\ee 
and 
\be 
f_{\sigma}^{(\mathrm{eft})}(x) = \frac{1}{\sigma^{(\mathrm{eft})}}\frac{(x^2-1)^{3/2}}{x},
\ee 
where $f_{\sigma}^{(\mathrm{sm})}(x)$ and $f_{\sigma}^{(\mathrm{eft})}(x)$ are both separately normalized by $\sigma^{(\mathrm{sm})}$ and $\sigma^{(\mathrm{eft})}$ respectively such that the overall normalization constant $\sigma(X,c)$ is given by
\be 
\sigma(X, c) = 1 + 0.1 c.
\label{eq:fid_xsec_toy}
\ee 
The number of expected events $\nu(c)$ is determined from $\sigma(X, c)$ and the luminosity $L$, which we set to $L=1000$ unless otherwise specified.  We work with dimensionless parameters throughout this toy example.

\begin{figure}
\centering
\begin{subfigure}{.5\textwidth}
  \centering
  \includegraphics[width=\linewidth]{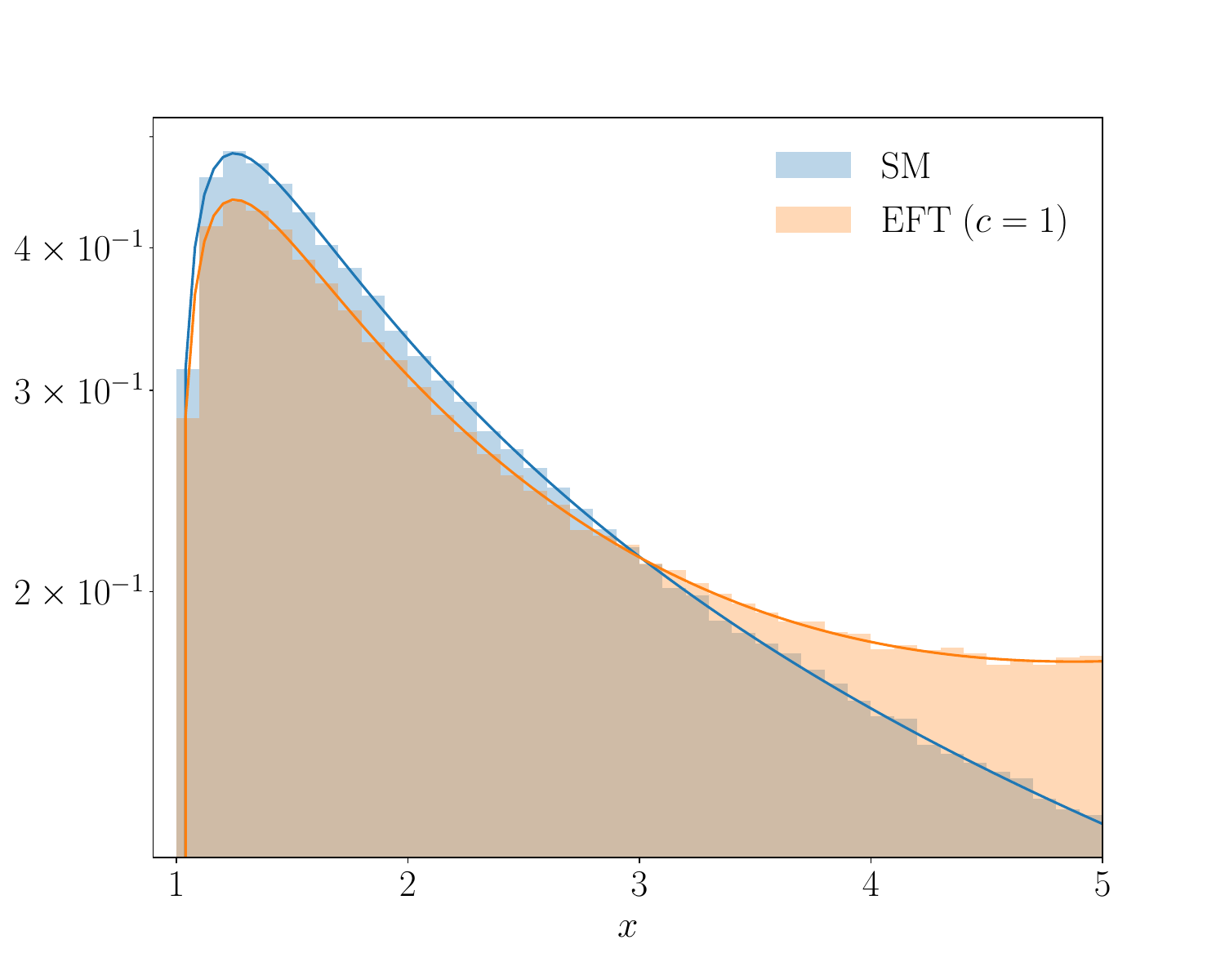}
  \caption{}
  \label{fig:eft_toy_dist}
\end{subfigure}%
\begin{subfigure}{.5\textwidth}
  \centering
  \includegraphics[width=\linewidth]{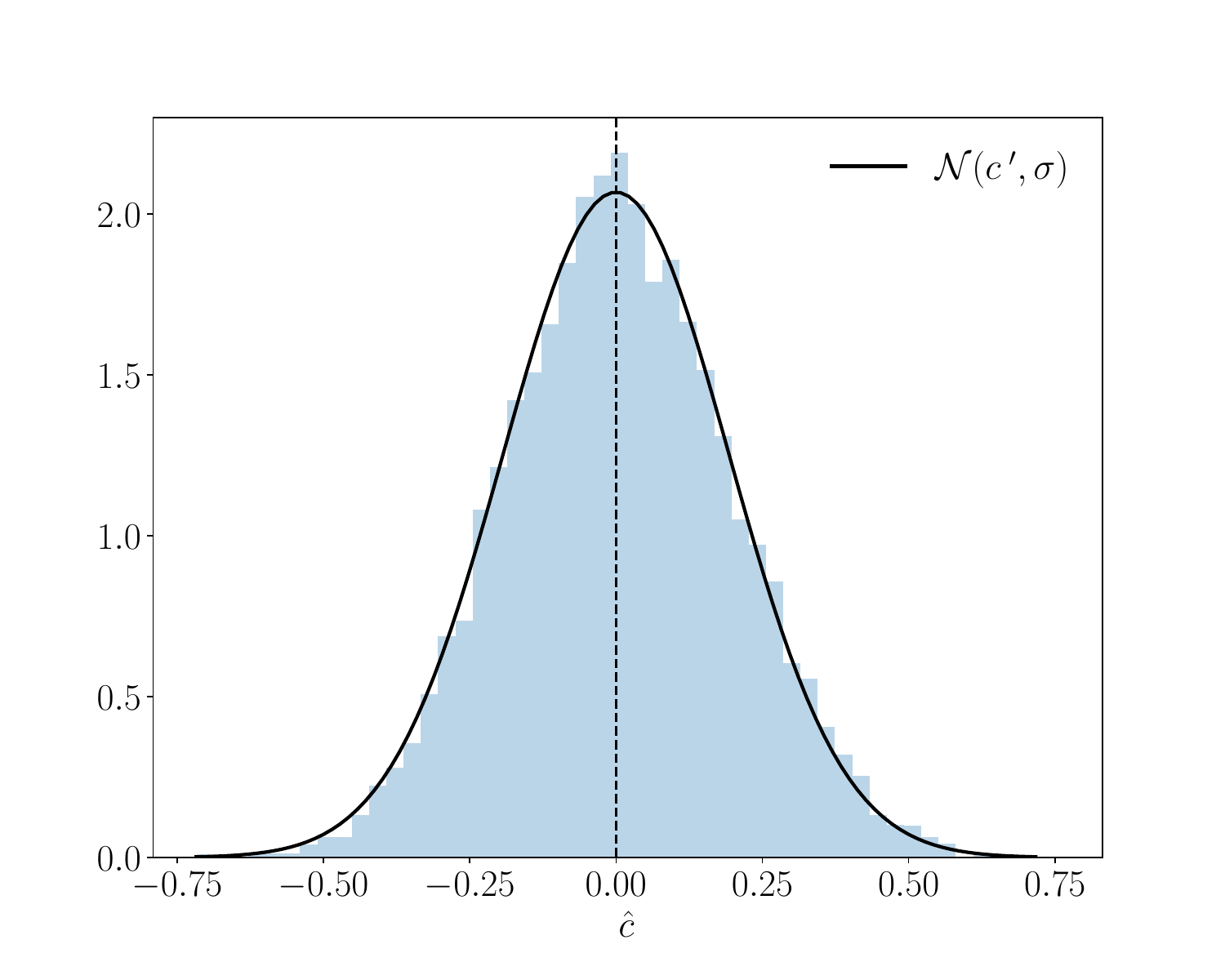}
  \caption{}
  \label{fig:dist_mle_c}
\end{subfigure}
	\caption{Left: the probability distribution $f_\sigma(x,c)$ from Eq.~(\ref{eq:eft_toy_dist}) at $c=1$ and $c=0$, corresponding to the EFT and SM respectively. Right: the MLE $\hat{c}$ under the SM follows a Gaussian distribution centered around $c'$ with standard deviation $\sigma$.}
\label{fig:test}
\end{figure}

In Fig.~\ref{fig:eft_toy_dist} we show the distribution $f_\sigma(x, c)$ at $c=0$ and $c=1$, corresponding to the SM and the EFT hypotheses respectively. One can see a clear energy growing effect in the tail of the EFT distribution. The histograms have been constructed using accept/reject sampling and contain 1M samples from each hypothesis. These will serve as our parent data sets from which we will take subsequent samples in the following. 
Now, taking the EFT at $c=1$ as our null hypothesis, the goal is to compute the expected significance at which we can either accept or reject the null assuming data generated under the SM (the alternative hypothesis). 
We adopt the PLR from Eq.~(\ref{eq:plr}) as our test statistic,
\be
q_c = -2\left[\nu(\hat{c}) - \nu(c) + \sum_{i=1}^{N_{\mathrm{ev}}}\log\frac{\nu(c)f_\sigma(x,c)}{\nu(\hat{c})f_\sigma(x,\hat{c})}\right],
\ee
whose distribution under the SM and the EFT we denote by $\mathrm{pdf}(q_c|0)$ and $\mathrm{pdf}(q_c|c)$ respectively.
One can imagine repeatedly drawing toy datasets from the SM parent data set, computing $q_c$, and evaluating its $p$-value under the null.
The median p-value then serves as a measure of expected sensitivity. Moreover, since the $p$-value is monotonic with $q_c$, it is equivalent to report the $p$-value associated with the median $q_c$ under the SM, denoted $\mathrm{Med}[q_c|0]$,
\be
p_{c, \mathrm{med}} = \int_{q_{c, \mathrm{med}}}^\infty\mathrm{pdf}(q_c|c)dq_c\qquad\text{with}\qquad q_{c, \mathrm{med}}\equiv\mathrm{Med}[q_c|0].
\ee
From the above discussion it becomes clear one not only needs $\mathrm{pdf}(q_c|c)$, which follows a $\chi^2$-distribution as discussed in Sec.~\ref{sec:binned_likelihoods}, but also $\mathrm{pdf}(q_c|c')$ with $c'\neq c$. At this point we employ Wald's theorem \cite{10.2307/1990256}, which states that for a data set of size $N$ generated under $f_\sigma(x, c')$, the PLR $q_c$ can be approximated by
\be 
q_c = \frac{(c-\hat{c})^2}{\sigma^2} + \mathcal{O}\left(1/\sqrt{N}\right),
\label{eq:walds_theorem}
\ee 
where $\hat{c}$ follows a normal distribution with mean $c'$ and a variance $\sigma^2$, i.e. $\hat{c}\sim \mathcal{N}(c', \sigma)$. 
Not only does the approximation provided by Wald's theorem depend on a large sample size $N$~\cite{10.2307/1990256}, but 
the approximation improves as $c\rightarrow \hat{c}\simeq c'$.
This can be understood by Taylor expanding Eq.~(\ref{eq:plr}) about $\hat{c}$. Defining $\Delta \equiv (c-\hat{c})$, we find
\begin{align}
    \nonumber q_c = -2 \log\frac{\mathcal{L}(c)}{\mathcal{L}(\hat{c})}
    &= -2 \log\left[1 + \frac{1}{2\mathcal{L}(\hat{c})}\left.\frac{\partial^2\mathcal{L}}{\partial c^2}\right|_{c=\hat{c}}\Delta^2 + \mathcal{O}\left(\Delta^3\right)\right]\\
    \nonumber&=-\frac{1}{\mathcal{L}(\hat{c})}\left.\frac{\partial^2\mathcal{L}}{\partial c^2}\right|_{c=\hat{c}}\Delta^2 + \mathcal{O}\left(\Delta^3\right) = \frac{\Delta^2}{\sigma^2} + \mathcal{O}\left(\Delta^3\right),
\end{align}
where on the last line we have used that
\be 
\frac{1}{\sigma^2} = -\left.\frac{\partial ^2\log\mathcal{L}}{\partial c^2}\right|_{c=\hat{c}}.
\ee 
Let us define $q_c^{\mathrm{wald}}$ as the leading order approximation of Eq.~(\ref{eq:walds_theorem}),
\be 
q_c^{\mathrm{wald}} \equiv \frac{(c-\hat{c})^2}{\sigma^2}.
\label{eq:qc_wald}
\ee
Recalling that $\hat{c}\sim \mathcal{N}(c', \sigma)$, we note that $q_c^{\mathrm{wald}}$ corresponds to a squared Gaussian random variable with unit variance and non-zero mean. This is known to follow a non-central $\chi^2$ distribution \cite{Cowan:2010js},
\be
\label{eq:ftmulambda}
\mathrm{pdf}(q_c^{\mathrm{wald}}|c') = \frac{1}{2 \sqrt{q_{c}}} \frac{1}{\sqrt{2 \pi}}
\left[ \exp \left( - \frac{1}{2}
\left( \sqrt{q_{c}} + \sqrt{\Lambda} \right)^2 \right) +
\exp \left( - \frac{1}{2} \left( \sqrt{q_{c}} - \sqrt{\Lambda} \right)^2
\right) \right] \;,
\ee
with the non-centrality parameter $\Lambda$ given by
\be 
\Lambda = \frac{(c-c')^2}{\sigma^2}.
\label{eq:lambda_definition}
\ee
Note how Eq.~(\ref{eq:ftmulambda}) reduces to a central $\chi^2$ distribution under the null hypothesis, i.e. when $c' = c$, as dictated by Wilk's theorem. 

The problem of obtaining $\mathrm{pdf}(q_c|c')$ with $c'\neq c$ is now reduced to the question of determining the non-centrality parameter $\Lambda$ to be used in Eq.~(\ref{eq:ftmulambda}).
We can construct this using MC toys: for example, in Fig.~\ref{fig:dist_mle_c} we show the distribution of $\hat{c}$ under the SM, obtained using $10$K toy data sets containing $n_{\rm ev}$ samples each, where $n_{\rm ev}$ is drawn from $\mathrm{Pois}(\nu(0))$ each time a new experiment is run. 
The distribution in Fig.~\ref{fig:dist_mle_c} gives $\sigma=0.19$, resulting in $\Lambda_{\mathrm{MC}} = 27.12$. 
However, although constructing $\mathrm{pdf}(\hat{c}|c')$ using MC toys gives us the info we need to extract $\Lambda$, it is computationally highly inefficient. 
This is especially so in case one needs to perform scans along different $c'$, e.g. to find expected exclusion limits. 

This is where the Asimov data set enters as an efficient tool to obtain $\Lambda$  \cite{Cowan:2010js}. The Asimov data set is defined such that one recovers the true value of $c$ during maximum likelihood estimation, i.e. $\hat{c} = c$. In the case of a Poissonian likelihood $\mathcal{L}_{\mathrm{pois}}$ defined as in Eq.~(\ref{eq:likelihood_binned_poisson}),  
the MLE $\hat{c}$ satisfies
\be
    \left.\frac{d\log\mathcal{L}_{\mathrm{pois}}}{dc}\right|_{c=\hat{c}} = \left.\frac{d}{dc}\left\{-2\sum_{i=1}^{n_{\mathrm{bins}}}\left[n_i\log\nu_i(c)-\nu_i(c) \right]\right\}\right|_{c=\hat{c}} = 0 \, .\\
\label{eq:asimov_dataset_cond}
\ee
The Asimov data set is then constructed such that Eq.~(\ref{eq:asimov_dataset_cond}) implies $\hat{c}=c$, which can only be true if $n_i=\nu_i(c)$. Therefore, we define the Asimov data set under the hypothesis $c$ as
\be 
\mathcal{D}_{\mathrm{Asimov}}\equiv \{n_{i,A}\} = \{\nu_i(c)\}.
\ee 
The non-centrality parameter $\Lambda$ can then be found by evaluating $q_c$ on the Asimov data set corresponding to the alternative hypothesis $c'$, i.e. $n_{i,A} = \nu_i'$,
\be
\Lambda_{\mathrm{A,binned}} = q_{c}(\{n_{i,A}\}) = -2\sum_{i=1}^{n_\mathrm{bins}}\left[\nu_i'\log\left(\frac{\nu_i}{\nu_i'}\right)-\nu_i + \nu_i'\right],
\label{eq:nc_discrete}
\ee
where we have used the fact that $\hat{c} = c'$ and therefore $\hat{\nu_i} = \nu_i'$ by construction.  The relationship between $\Lambda$ and $q_c$ can be understood by comparing Eqs.~(\ref{eq:qc_wald}) and~(\ref{eq:lambda_definition}) and noting that they coincide when $\hat{c}=c'$.

Now, one could wonder how Eq.~(\ref{eq:nc_discrete}) generalizes to the case of unbinned data. An intuitive starting point is to study the behavior of $\Lambda_{\mathrm{A,binned}}$ as one increases the number of bin insertions,
and by doing so we find that after $\sim 10$ bins, $\Lambda_{\mathrm{A,binned}}$ converges to a stable value. Encouraged by this convergence, we take the infinitely narrow bin limit of Eq.~(\ref{eq:nc_discrete}),
\begin{align} 
	\nonumber\Lambda_{\mathrm{A,binned}} &= -2\sum_{i=1}^{n_\mathrm{bins}}\left[\nu_i'\log\left(\frac{\nu_i}{\nu_i'}\right)-\nu_i + \nu_i'\right]\\
	&\rightarrow -2 \int_{-\infty}^{\infty}\left[\nu(c')f_{\sigma}(x, c')\log\left(\frac{\nu(c)f_{\sigma}(x, c)}{\nu(c')f_{\sigma}(x, c')}\right)-\nu(c)f_{\sigma}(x, c) + \nu(c')f_{\sigma}(x, c')\right]dx \equiv \Lambda_{\mathrm{A,unbinned}},
\label{eq:nc_continuous}
\end{align}
where we have sent $\nu_i(c) \rightarrow \nu(c)f_{\sigma}(x, c)dx$. Evaluating the integral in Eq.~(\ref{eq:nc_continuous}) gives $\Lambda_{\mathrm{A,unbinned}} = 22.11$.

\begin{figure}
\centering
\begin{subfigure}{.5\textwidth}
  \centering
  \includegraphics[width=\linewidth]{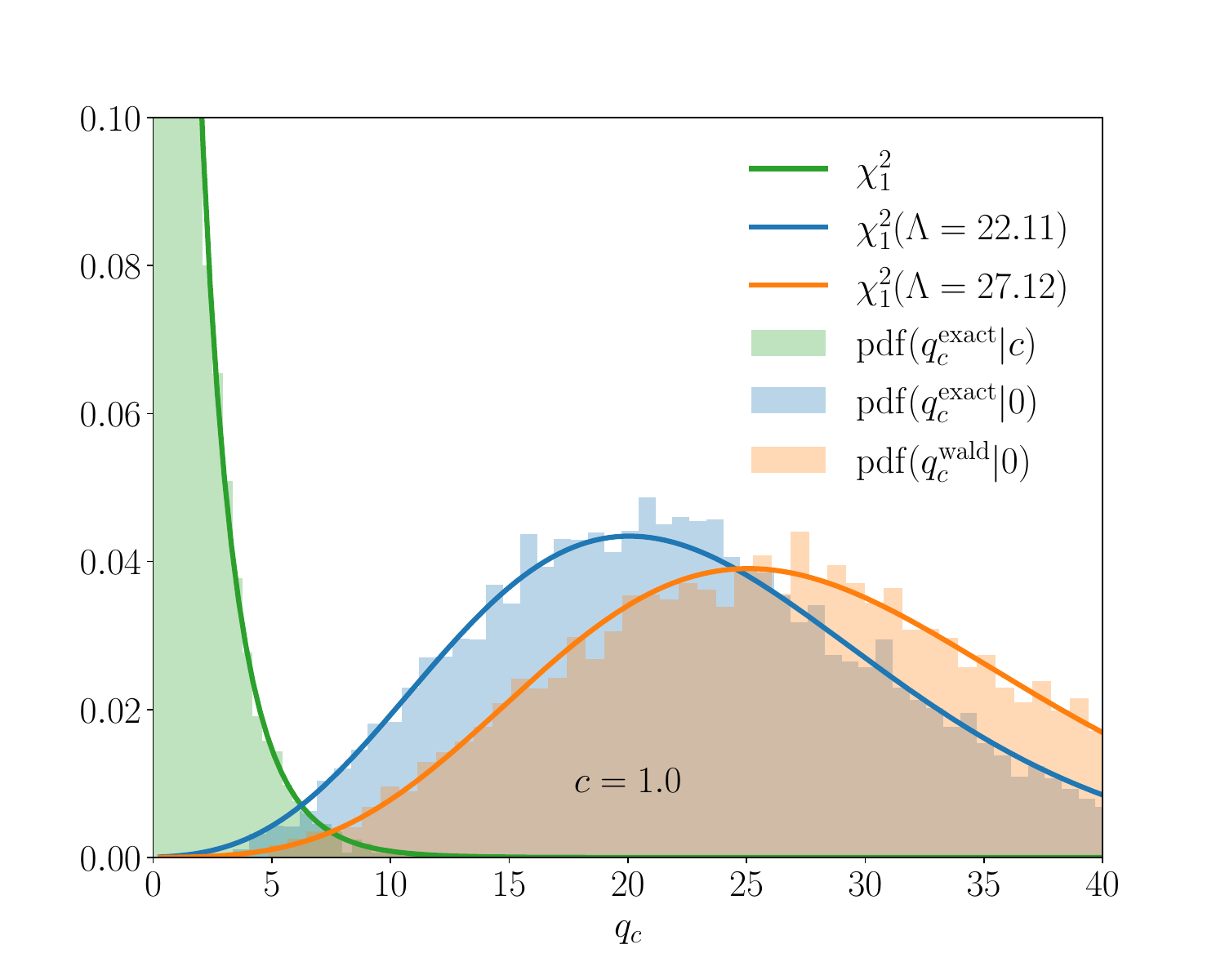}
  \caption{}
  \label{fig:qc_dist_toy_10e-1}
\end{subfigure}%
\begin{subfigure}{.5\textwidth}
  \centering
  \includegraphics[width=\linewidth]{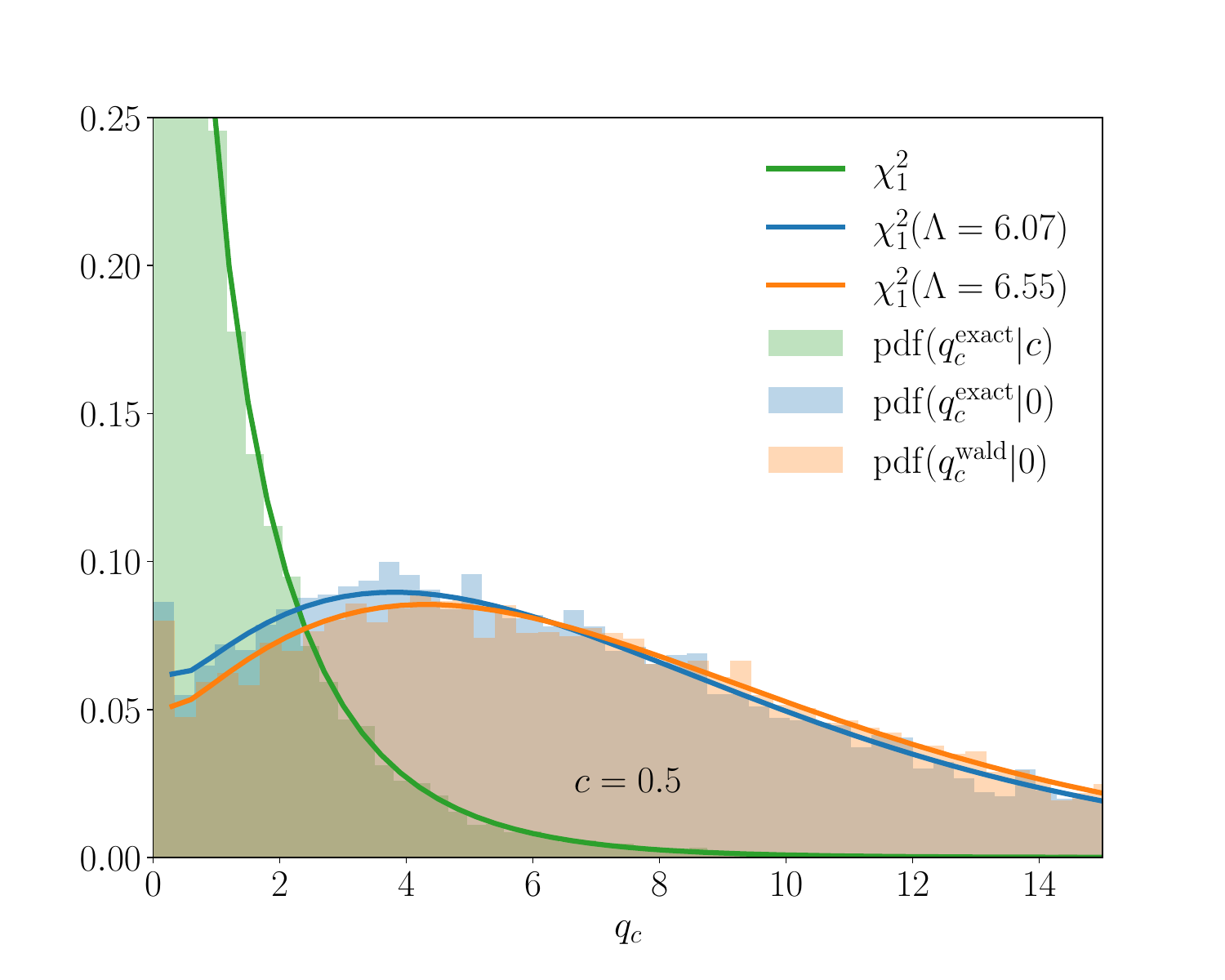}
  \caption{}
  \label{fig:qc_dist_toy_5e-1}
\end{subfigure}
	\caption{The distributions of $q_c$ under the EFT and the SM hypotheses for $c=1.0$ (a) and $c=0.5$ (b). The blue curve agrees well with the distribution $\mathrm{pdf}(q_{c}^{\mathrm{exact}} | c)$, indicating that the distribution of $q_{c}$ is well described by the non-central $\chi^{2}$ evaluated with non-centrality parameter $\Lambda_{\mathrm{A,unbinned}}$.  We notice a disagreement between the distribution of $q_c^{\mathrm{exact}}$ and $q_c^{\mathrm{wald}}$, meaning that neglecting the higher order corrections in Wald's theorem leads to sizable differences. On the right (b) it can be seen that this difference gets smaller when $c \rightarrow c'$, i.e. $\Delta \rightarrow 0$. }
\label{fig:test2}
\end{figure}

In Fig.~\ref{fig:test2} we select three benchmark points at which to visualize the distribution $\mathrm{pdf}(q_{c} | c')$: the SM ($c=0$) and the EFT at $c=0.5$, $c=1$.
Firstly, the green histogram shows the distribution of $q_{c}$ evaluated on data following the EFT hypothesis.
The distribution is obtained by repeatedly drawing samples of size
$n_{ev} \sim \mathrm{Pois}(\nu(c))$ from the parent EFT dataset and calculating $q_{c}$ as given by the PLR in Eq.~(\ref{eq:plr}).  
The solid green curve indicates the distribution of a $\chi^{2}$ with 1 degree of freedom.
We find excellent agreement between the green curve and histogram,
as expected from Wilk's theorem.

A similar procedure is used to obtain the blue histogram, this time corresponding to the SM hypothesis.
We repeatedly draw samples of size
$n_{ev} \sim \mathrm{Pois}(\nu(0))$ from the parent SM dataset and calculate $q_{c}$ as given by the PLR in Eq.~(\ref{eq:plr}). 
The resulting distribution is denoted by $\mathrm{pdf}(q_{c}^{\mathrm{exact}} | c)$.
In contrast, the orange histogram is produced using the approximation given by Wald's theorem in Eq.~(\ref{eq:qc_wald}).  
For each sample drawn, we evaluate $q_{c}^{\mathrm{wald}}$ rather than the full expression for $q_{c}$,
resulting in the distribution $\mathrm{pdf}(q_{c}^{\mathrm{wald}} | c)$.
We notice a non-negligible difference in the distributions of $q_c$ before and after employing Wald's theorem. 
As expected, this is due to the fact that the neglected $\mathcal{O}(\Delta^3)$ term is in fact quite 
sizable, being on the order of a few percent. In Fig.~\ref{fig:qc_dist_toy_5e-1} we show that the difference 
between $\mathrm{pdf}(q_c^{\mathrm{wald}}|c)$ and $\mathrm{pdf}(q_c^{\mathrm{exact}}|c)$ gets smaller when we move to a smaller $\Delta$, corresponding to $c=0.5$.

Finally, we turn to the solid blue and orange curves.  In orange, we plot the distribution of the non-central $\chi^{2}$ as
given by Eq.~(\ref{eq:ftmulambda}), setting the non-centrality parameter $\Lambda$ to
$\Lambda_{\mathrm{MC}} = 27.12$. 
We find good agreement between the orange curve and the orange histogram,
as expected from the fact that both employ Wald's theorem.
The blue curve, on the other hand, indicates the non-central $\chi^{2}$ distribution evaluated at $\Lambda_{\mathrm{A,unbinned}} = 22.11$ 
as calculated using Eq.~(\ref{eq:nc_continuous}).  In this case the agreement between the blue curve and histogram is non-trivial: 
it indicates that the non-centrality parameter evaluated on the unbinned Asimov dataset provides an excellent description
of the distribution of the PLR $q_{c}$ evaluated on unbinned data.  This is the case both at $c=1$ and $c=0.5$,
and provides a validation of the expression for $\Lambda_{\mathrm{A,unbinned}}$ determined in Eq.~(\ref{eq:nc_continuous}).
This agreement suggests  Eq.~(\ref{eq:nc_continuous}) includes some higher order contributions that are otherwise omitted if one adopts instead Eq.~(\ref{eq:lambda_definition}). This was also conjectured in~\cite{Cowan:2010js}.

Finally, Eq.~(\ref{eq:nc_continuous})  generalizes straightforwardly to the case of $n_{\rm p}$ model parameters $\boldsymbol{c}$, and to the case of a multi-dimensional vector $\boldsymbol{x}$ of kinematic variables
\be 
\Lambda_{\mathrm{A,unbinned}} = \int_{-\infty}^{\infty}\left[\nu(\boldsymbol{c}')f_{\sigma}(\boldsymbol{x}, \boldsymbol{c}')\log\left(\frac{\nu(\boldsymbol{c})f_{\sigma}(\boldsymbol{x}, \boldsymbol{c})}{\nu(\boldsymbol{c}')f_{\sigma}(\boldsymbol{x}, \boldsymbol{c}')}\right)-\nu(\boldsymbol{c})f_{\sigma}(\boldsymbol{x}, \boldsymbol{c}) + \nu(\boldsymbol{c}')f_{\sigma}(\boldsymbol{x}, \boldsymbol{c}')\right]d\boldsymbol{x}.
\label{eq:nc_continuous_mult_dim_2}
\ee

\begin{thebibliography}{100}

\bibitem{ATLAS:2022vkf}
{\bfseries ATLAS} Collaboration, ``{A detailed map of Higgs boson interactions
  by the ATLAS experiment ten years after the discovery},''
  \href{http://dx.doi.org/10.1038/s41586-022-04893-w}{{\em Nature} {\bfseries
  607} no.~7917, (2022) 52--59},
  \href{http://arxiv.org/abs/2207.00092}{{\ttfamily arXiv:2207.00092
  [hep-ex]}}.

\bibitem{CMS:2022dwd}
{\bfseries CMS} Collaboration, ``{A portrait of the Higgs boson by the CMS
  experiment ten years after the discovery},''
  \href{http://dx.doi.org/10.1038/s41586-022-04892-x}{{\em Nature} {\bfseries
  607} no.~7917, (2022) 60--68},
  \href{http://arxiv.org/abs/2207.00043}{{\ttfamily arXiv:2207.00043
  [hep-ex]}}.

\bibitem{Dawson:2018dcd}
S.~Dawson, C.~Englert, and T.~Plehn, ``{Higgs Physics: It ain't over till it's
  over},'' \href{http://dx.doi.org/10.1016/j.physrep.2019.05.001}{{\em Phys.
  Rept.} {\bfseries 816} (2019) 1--85},
  \href{http://arxiv.org/abs/1808.01324}{{\ttfamily arXiv:1808.01324
  [hep-ph]}}.

\bibitem{CMS:2012qbp}
{\bfseries CMS} Collaboration, S.~Chatrchyan {\em et~al.}, ``{Observation of a
  New Boson at a Mass of 125 GeV with the CMS Experiment at the LHC},''
  \href{http://dx.doi.org/10.1016/j.physletb.2012.08.021}{{\em Phys. Lett. B}
  {\bfseries 716} (2012) 30--61},
  \href{http://arxiv.org/abs/1207.7235}{{\ttfamily arXiv:1207.7235 [hep-ex]}}.

\bibitem{ATLAS:2012yve}
{\bfseries ATLAS} Collaboration, G.~Aad {\em et~al.}, ``{Observation of a new
  particle in the search for the Standard Model Higgs boson with the ATLAS
  detector at the LHC},''
  \href{http://dx.doi.org/10.1016/j.physletb.2012.08.020}{{\em Phys. Lett. B}
  {\bfseries 716} (2012) 1--29},
  \href{http://arxiv.org/abs/1207.7214}{{\ttfamily arXiv:1207.7214 [hep-ex]}}.

\bibitem{Weinberg:1979sa}
S.~Weinberg, ``{Baryon and Lepton Nonconserving Processes},''
  \href{http://dx.doi.org/10.1103/PhysRevLett.43.1566}{{\em Phys. Rev. Lett.}
  {\bfseries 43} (1979) 1566--1570}.

\bibitem{Buchmuller:1985jz}
W.~Buchmuller and D.~Wyler, ``{Effective Lagrangian Analysis of New
  Interactions and Flavor Conservation},''
  \href{http://dx.doi.org/10.1016/0550-3213(86)90262-2}{{\em Nucl. Phys. B}
  {\bfseries 268} (1986) 621--653}.

\bibitem{Grzadkowski:2010es}
B.~Grzadkowski, M.~Iskrzynski, M.~Misiak, and J.~Rosiek, ``{Dimension-Six Terms
  in the Standard Model Lagrangian},''
  \href{http://dx.doi.org/10.1007/JHEP10(2010)085}{{\em JHEP} {\bfseries 10}
  (2010) 085}, \href{http://arxiv.org/abs/1008.4884}{{\ttfamily arXiv:1008.4884
  [hep-ph]}}.

\bibitem{Manohar:2018aog}
A.~V. Manohar, ``{Introduction to Effective Field Theories},''
  \href{http://arxiv.org/abs/1804.05863}{{\ttfamily arXiv:1804.05863
  [hep-ph]}}.

\bibitem{Alonso:2013hga}
R.~Alonso, E.~E. Jenkins, A.~V. Manohar, and M.~Trott, ``{Renormalization Group
  Evolution of the Standard Model Dimension Six Operators III: Gauge Coupling
  Dependence and Phenomenology},''
  \href{http://dx.doi.org/10.1007/JHEP04(2014)159}{{\em JHEP} {\bfseries 04}
  (2014) 159}, \href{http://arxiv.org/abs/1312.2014}{{\ttfamily arXiv:1312.2014
  [hep-ph]}}.

\bibitem{Boggia:2017hyq}
M.~Boggia {\em et~al.}, ``{The HiggsTools handbook: a beginners guide to
  decoding the Higgs sector},''
  \href{http://dx.doi.org/10.1088/1361-6471/aab812}{{\em J. Phys. G} {\bfseries
  45} no.~6, (2018) 065004}, \href{http://arxiv.org/abs/1711.09875}{{\ttfamily
  arXiv:1711.09875 [hep-ph]}}.

\bibitem{Brivio:2017vri}
I.~Brivio and M.~Trott, ``{The Standard Model as an Effective Field Theory},''
  \href{http://dx.doi.org/10.1016/j.physrep.2018.11.002}{{\em Phys. Rept.}
  {\bfseries 793} (2019) 1--98},
  \href{http://arxiv.org/abs/1706.08945}{{\ttfamily arXiv:1706.08945
  [hep-ph]}}.

\bibitem{Biekoetter:2018ypq}
A.~Biekoetter, T.~Corbett, and T.~Plehn, ``{The Gauge-Higgs Legacy of the LHC
  Run II},'' \href{http://dx.doi.org/10.21468/SciPostPhys.6.6.064}{{\em SciPost
  Phys.} {\bfseries 6} no.~6, (2019) 064},
  \href{http://arxiv.org/abs/1812.07587}{{\ttfamily arXiv:1812.07587
  [hep-ph]}}.

\bibitem{DeBlas:2019ehy}
J.~De~Blas {\em et~al.}, ``{$\texttt{HEPfit}$: a code for the combination of
  indirect and direct constraints on high energy physics models},''
  \href{http://dx.doi.org/10.1140/epjc/s10052-020-7904-z}{{\em Eur. Phys. J. C}
  {\bfseries 80} no.~5, (2020) 456},
  \href{http://arxiv.org/abs/1910.14012}{{\ttfamily arXiv:1910.14012
  [hep-ph]}}.

\bibitem{Ethier:2021bye}
{\bfseries SMEFiT} Collaboration, J.~J. Ethier, G.~Magni, F.~Maltoni,
  L.~Mantani, E.~R. Nocera, J.~Rojo, E.~Slade, E.~Vryonidou, and C.~Zhang,
  ``{Combined SMEFT interpretation of Higgs, diboson, and top quark data from
  the LHC},'' \href{http://dx.doi.org/10.1007/JHEP11(2021)089}{{\em JHEP}
  {\bfseries 11} (2021) 089}, \href{http://arxiv.org/abs/2105.00006}{{\ttfamily
  arXiv:2105.00006 [hep-ph]}}.

\bibitem{Ethier:2021ydt}
J.~J. Ethier, R.~Gomez-Ambrosio, G.~Magni, and J.~Rojo, ``{SMEFT analysis of
  vector boson scattering and diboson data from the LHC Run II},''
  \href{http://dx.doi.org/10.1140/epjc/s10052-021-09347-7}{{\em Eur. Phys. J.
  C} {\bfseries 81} no.~6, (2021) 560},
  \href{http://arxiv.org/abs/2101.03180}{{\ttfamily arXiv:2101.03180
  [hep-ph]}}.

\bibitem{Ellis:2020unq}
J.~Ellis, M.~Madigan, K.~Mimasu, V.~Sanz, and T.~You, ``{Top, Higgs, Diboson
  and Electroweak Fit to the Standard Model Effective Field Theory},''
  \href{http://dx.doi.org/10.1007/JHEP04(2021)279}{{\em JHEP} {\bfseries 04}
  (2021) 279}, \href{http://arxiv.org/abs/2012.02779}{{\ttfamily
  arXiv:2012.02779 [hep-ph]}}.

\bibitem{Bissmann:2020mfi}
S.~Bi\ss{}mann, C.~Grunwald, G.~Hiller, and K.~Kr\"oninger, ``{Top and Beauty
  synergies in SMEFT-fits at present and future colliders},''
  \href{http://dx.doi.org/10.1007/JHEP06(2021)010}{{\em JHEP} {\bfseries 06}
  (2021) 010}, \href{http://arxiv.org/abs/2012.10456}{{\ttfamily
  arXiv:2012.10456 [hep-ph]}}.

\bibitem{Bruggisser:2021duo}
S.~Bruggisser, R.~Sch\"afer, D.~van Dyk, and S.~Westhoff, ``{The Flavor of UV
  Physics},'' \href{http://dx.doi.org/10.1007/JHEP05(2021)257}{{\em JHEP}
  {\bfseries 05} (2021) 257}, \href{http://arxiv.org/abs/2101.07273}{{\ttfamily
  arXiv:2101.07273 [hep-ph]}}.

\bibitem{deBlas:2021wap}
J.~de~Blas, M.~Ciuchini, E.~Franco, A.~Goncalves, S.~Mishima, M.~Pierini,
  L.~Reina, and L.~Silvestrini, ``{Global analysis of electroweak data in the
  Standard Model},'' \href{http://dx.doi.org/10.1103/PhysRevD.106.033003}{{\em
  Phys. Rev. D} {\bfseries 106} no.~3, (2022) 033003},
  \href{http://arxiv.org/abs/2112.07274}{{\ttfamily arXiv:2112.07274
  [hep-ph]}}.

\bibitem{DELPHI:2010ykq}
{\bfseries DELPHI} Collaboration, J.~Abdallah {\em et~al.}, ``{Measurements of
  CP-conserving Trilinear Gauge Boson Couplings WWV (V = gamma,Z) in e+e-
  Collisions at LEP2},''
  \href{http://dx.doi.org/10.1140/epjc/s10052-010-1254-1}{{\em Eur. Phys. J. C}
  {\bfseries 66} (2010) 35--56},
  \href{http://arxiv.org/abs/1002.0752}{{\ttfamily arXiv:1002.0752 [hep-ex]}}.

\bibitem{Diehl:1993br}
M.~Diehl and O.~Nachtmann, ``{Optimal observables for the measurement of three
  gauge boson couplings in e+ e- ---\ensuremath{>} W+ W-},''
  \href{http://dx.doi.org/10.1007/BF01555899}{{\em Z. Phys. C} {\bfseries 62}
  (1994) 397--412}.

\bibitem{Durieux:2018tev}
G.~Durieux, M.~Perell\'o, M.~Vos, and C.~Zhang, ``{Global and optimal probes
  for the top-quark effective field theory at future lepton colliders},''
  \href{http://dx.doi.org/10.1007/JHEP10(2018)168}{{\em JHEP} {\bfseries 10}
  (2018) 168}, \href{http://arxiv.org/abs/1807.02121}{{\ttfamily
  arXiv:1807.02121 [hep-ph]}}.

\bibitem{Cranmer:2021urp}
K.~Cranmer {\em et~al.}, ``{Publishing statistical models: Getting the most out
  of particle physics experiments},''
  \href{http://dx.doi.org/10.21468/SciPostPhys.12.1.037}{{\em SciPost Phys.}
  {\bfseries 12} no.~1, (2022) 037},
  \href{http://arxiv.org/abs/2109.04981}{{\ttfamily arXiv:2109.04981
  [hep-ph]}}.

\bibitem{Arratia:2021otl}
M.~Arratia {\em et~al.}, ``{Publishing unbinned differential cross section
  results},'' \href{http://dx.doi.org/10.1088/1748-0221/17/01/P01024}{{\em
  JINST} {\bfseries 17} no.~01, (2022) P01024},
  \href{http://arxiv.org/abs/2109.13243}{{\ttfamily arXiv:2109.13243
  [hep-ph]}}.

\bibitem{CMS:2022hjj}
{\bfseries CMS} Collaboration, ``{Search for new physics using effective field
  theory in 13 TeV pp collision events that contain a top quark pair and a
  boosted Z or Higgs boson},''
  \href{http://arxiv.org/abs/2208.12837}{{\ttfamily arXiv:2208.12837
  [hep-ex]}}.

\bibitem{CMS:2021aly}
{\bfseries CMS} Collaboration, K.~Lee {\em et~al.}, ``{Probing effective field
  theory operators in the associated production of top quarks with a Z boson in
  multilepton final states at $ \sqrt{s} $ = 13 TeV},''
  \href{http://dx.doi.org/10.1007/JHEP12(2021)083}{{\em JHEP} {\bfseries 12}
  (2021) 083}, \href{http://arxiv.org/abs/2107.13896}{{\ttfamily
  arXiv:2107.13896 [hep-ex]}}.

\bibitem{Campbell:2012cz}
J.~M. Campbell, W.~T. Giele, and C.~Williams, ``{The Matrix Element Method at
  Next-to-Leading Order},''
  \href{http://dx.doi.org/10.1007/JHEP11(2012)043}{{\em JHEP} {\bfseries 11}
  (2012) 043}, \href{http://arxiv.org/abs/1204.4424}{{\ttfamily arXiv:1204.4424
  [hep-ph]}}.

\bibitem{Artoisenet:2010cn}
P.~Artoisenet, V.~Lemaitre, F.~Maltoni, and O.~Mattelaer, ``{Automation of the
  matrix element reweighting method},''
  \href{http://dx.doi.org/10.1007/JHEP12(2010)068}{{\em JHEP} {\bfseries 12}
  (2010) 068}, \href{http://arxiv.org/abs/1007.3300}{{\ttfamily arXiv:1007.3300
  [hep-ph]}}.

\bibitem{Gainer:2013iya}
J.~S. Gainer, J.~Lykken, K.~T. Matchev, S.~Mrenna, and M.~Park, ``{The Matrix
  Element Method: Past, Present, and Future},'' in {\em {Community Summer Study
  2013}: {Snowmass on the Mississippi}}.
\newblock 7, 2013.
\newblock \href{http://arxiv.org/abs/1307.3546}{{\ttfamily arXiv:1307.3546
  [hep-ph]}}.

\bibitem{Fiedler:2010sg}
F.~Fiedler, A.~Grohsjean, P.~Haefner, and P.~Schieferdecker, ``{The Matrix
  Element Method and its Application in Measurements of the Top Quark Mass},''
  \href{http://dx.doi.org/10.1016/j.nima.2010.09.024}{{\em Nucl. Instrum. Meth.
  A} {\bfseries 624} (2010) 203--218},
  \href{http://arxiv.org/abs/1003.1316}{{\ttfamily arXiv:1003.1316 [hep-ex]}}.

\bibitem{Martini:2015fsa}
T.~Martini and P.~Uwer, ``{Extending the Matrix Element Method beyond the Born
  approximation: Calculating event weights at next-to-leading order
  accuracy},'' \href{http://dx.doi.org/10.1007/JHEP09(2015)083}{{\em JHEP}
  {\bfseries 09} (2015) 083}, \href{http://arxiv.org/abs/1506.08798}{{\ttfamily
  arXiv:1506.08798 [hep-ph]}}.

\bibitem{Chen:2020mev}
S.~Chen, A.~Glioti, G.~Panico, and A.~Wulzer, ``{Parametrized classifiers for
  optimal EFT sensitivity},''
  \href{http://dx.doi.org/10.1007/JHEP05(2021)247}{{\em JHEP} {\bfseries 05}
  (2021) 247}, \href{http://arxiv.org/abs/2007.10356}{{\ttfamily
  arXiv:2007.10356 [hep-ph]}}.

\bibitem{DAgnolo:2019vbw}
R.~T. D'Agnolo, G.~Grosso, M.~Pierini, A.~Wulzer, and M.~Zanetti, ``{Learning
  multivariate new physics},''
  \href{http://dx.doi.org/10.1140/epjc/s10052-021-08853-y}{{\em Eur. Phys. J.
  C} {\bfseries 81} no.~1, (2021) 89},
  \href{http://arxiv.org/abs/1912.12155}{{\ttfamily arXiv:1912.12155
  [hep-ph]}}.

\bibitem{DAgnolo:2018cun}
R.~T. D'Agnolo and A.~Wulzer, ``{Learning New Physics from a Machine},''
  \href{http://dx.doi.org/10.1103/PhysRevD.99.015014}{{\em Phys. Rev. D}
  {\bfseries 99} no.~1, (2019) 015014},
  \href{http://arxiv.org/abs/1806.02350}{{\ttfamily arXiv:1806.02350
  [hep-ph]}}.

\bibitem{Brehmer:2019xox}
J.~Brehmer, F.~Kling, I.~Espejo, and K.~Cranmer, ``{MadMiner: Machine
  learning-based inference for particle physics},''
  \href{http://dx.doi.org/10.1007/s41781-020-0035-2}{{\em Comput. Softw. Big
  Sci.} {\bfseries 4} no.~1, (2020) 3},
  \href{http://arxiv.org/abs/1907.10621}{{\ttfamily arXiv:1907.10621
  [hep-ph]}}.

\bibitem{Brehmer:2018kdj}
J.~Brehmer, K.~Cranmer, G.~Louppe, and J.~Pavez, ``{Constraining Effective
  Field Theories with Machine Learning},''
  \href{http://dx.doi.org/10.1103/PhysRevLett.121.111801}{{\em Phys. Rev.
  Lett.} {\bfseries 121} no.~11, (2018) 111801},
  \href{http://arxiv.org/abs/1805.00013}{{\ttfamily arXiv:1805.00013
  [hep-ph]}}.

\bibitem{Brehmer:2018eca}
J.~Brehmer, K.~Cranmer, G.~Louppe, and J.~Pavez, ``{A Guide to Constraining
  Effective Field Theories with Machine Learning},''
  \href{http://dx.doi.org/10.1103/PhysRevD.98.052004}{{\em Phys. Rev. D}
  {\bfseries 98} no.~5, (2018) 052004},
  \href{http://arxiv.org/abs/1805.00020}{{\ttfamily arXiv:1805.00020
  [hep-ph]}}.

\bibitem{Letizia:2022xbe}
M.~Letizia, G.~Losapio, M.~Rando, G.~Grosso, A.~Wulzer, M.~Pierini, M.~Zanetti,
  and L.~Rosasco, ``{Learning new physics efficiently with nonparametric
  methods},'' \href{http://dx.doi.org/10.1140/epjc/s10052-022-10830-y}{{\em
  Eur. Phys. J. C} {\bfseries 82} no.~10, (2022) 879},
  \href{http://arxiv.org/abs/2204.02317}{{\ttfamily arXiv:2204.02317
  [hep-ph]}}.

\bibitem{Chatterjee:2022oco}
S.~Chatterjee, S.~Rohshap, R.~Sch\"ofbeck, and D.~Schwarz, ``{Learning the EFT
  likelihood with tree boosting},''
  \href{http://arxiv.org/abs/2205.12976}{{\ttfamily arXiv:2205.12976
  [hep-ph]}}.

\bibitem{Chatterjee:2021nms}
S.~Chatterjee, N.~Frohner, L.~Lechner, R.~Sch\"ofbeck, and D.~Schwarz, ``{Tree
  boosting for learning EFT parameters},''
  \href{http://dx.doi.org/10.1016/j.cpc.2022.108385}{{\em Comput. Phys.
  Commun.} {\bfseries 277} (2022) 108385},
  \href{http://arxiv.org/abs/2107.10859}{{\ttfamily arXiv:2107.10859
  [hep-ph]}}.

\bibitem{Brehmer:2019gmn}
J.~Brehmer, S.~Dawson, S.~Homiller, F.~Kling, and T.~Plehn, ``{Benchmarking
  simplified template cross sections in $WH$ production},''
  \href{http://dx.doi.org/10.1007/JHEP11(2019)034}{{\em JHEP} {\bfseries 11}
  (2019) 034}, \href{http://arxiv.org/abs/1908.06980}{{\ttfamily
  arXiv:1908.06980 [hep-ph]}}.

\bibitem{Bortolato:2020zcg}
B.~Bortolato, J.~F. Kamenik, N.~Ko\v{s}nik, and A.~Smolkovi\v{c}, ``{Optimized
  probes of $CP$ -odd effects in the $t \bar t h$ process at hadron
  colliders},'' \href{http://dx.doi.org/10.1016/j.nuclphysb.2021.115328}{{\em
  Nucl. Phys. B} {\bfseries 964} (2021) 115328},
  \href{http://arxiv.org/abs/2006.13110}{{\ttfamily arXiv:2006.13110
  [hep-ph]}}.

\bibitem{Butter:2021rvz}
A.~Butter, T.~Plehn, N.~Soybelman, and J.~Brehmer, ``{Back to the Formula --
  LHC Edition},'' \href{http://arxiv.org/abs/2109.10414}{{\ttfamily
  arXiv:2109.10414 [hep-ph]}}.

\bibitem{Arganda:2022qzy}
E.~Arganda, X.~Marcano, V.~M. Lozano, A.~D. Medina, A.~D. Perez, M.~Szewc, and
  A.~Szynkman, ``{A method for approximating optimal statistical significances
  with machine-learned likelihoods},''
  \href{http://dx.doi.org/10.1140/epjc/s10052-022-10944-3}{{\em Eur. Phys. J.
  C} {\bfseries 82} no.~11, (2022) 993},
  \href{http://arxiv.org/abs/2205.05952}{{\ttfamily arXiv:2205.05952
  [hep-ph]}}.

\bibitem{Arganda:2022cl}
E.~Arganda, M.~de~los Rios, A.~D. Perez, and R.~M. Sandá~Seoane, ``{Imposing
  exclusion limits on new physics with machine-learned likelihoods},''
  \href{http://dx.doi.org/10.22323/1.414.1226}{{\em PoS} {\bfseries ICHEP2022}
  (2022) 1226}.

\bibitem{Arganda:2022zbs}
E.~Arganda, A.~D. Perez, M.~d.~l. Rios, and R.~M. Sand\'a~Seoane,
  ``{Machine-Learned Exclusion Limits without Binning},''
  \href{http://arxiv.org/abs/2211.04806}{{\ttfamily arXiv:2211.04806
  [hep-ph]}}.

\bibitem{Gritsan:2020pib}
A.~V. Gritsan, J.~Roskes, U.~Sarica, M.~Schulze, M.~Xiao, and Y.~Zhou, ``{New
  features in the JHU generator framework: constraining Higgs boson properties
  from on-shell and off-shell production},''
  \href{http://dx.doi.org/10.1103/PhysRevD.102.056022}{{\em Phys. Rev. D}
  {\bfseries 102} no.~5, (2020) 056022},
  \href{http://arxiv.org/abs/2002.09888}{{\ttfamily arXiv:2002.09888
  [hep-ph]}}.

\bibitem{DeCastro:2018psv}
P.~De~Castro and T.~Dorigo, ``{INFERNO: Inference-Aware Neural Optimisation},''
  \href{http://dx.doi.org/10.1016/j.cpc.2019.06.007}{{\em Comput. Phys.
  Commun.} {\bfseries 244} (2019) 170--179},
  \href{http://arxiv.org/abs/1806.04743}{{\ttfamily arXiv:1806.04743
  [stat.ML]}}.

\bibitem{Brehmer:2018hga}
J.~Brehmer, G.~Louppe, J.~Pavez, and K.~Cranmer, ``{Mining gold from implicit
  models to improve likelihood-free inference},''
  \href{http://dx.doi.org/10.1073/pnas.1915980117}{{\em Proc. Nat. Acad. Sci.}
  {\bfseries 117} no.~10, (2020) 5242--5249},
  \href{http://arxiv.org/abs/1805.12244}{{\ttfamily arXiv:1805.12244
  [stat.ML]}}.

\bibitem{Wunsch:2020iuh}
S.~Wunsch, S.~J\"orger, R.~Wolf, and G.~Quast, ``{Optimal Statistical Inference
  in the Presence of Systematic Uncertainties Using Neural Network Optimization
  Based on Binned Poisson Likelihoods with Nuisance Parameters},''
  \href{http://dx.doi.org/10.1007/s41781-020-00049-5}{{\em Comput. Softw. Big
  Sci.} {\bfseries 5} no.~1, (2021) 4},
  \href{http://arxiv.org/abs/2003.07186}{{\ttfamily arXiv:2003.07186
  [physics.data-an]}}.

\bibitem{dAgnolo:2021aun}
R.~T. d'Agnolo, G.~Grosso, M.~Pierini, A.~Wulzer, and M.~Zanetti, ``{Learning
  new physics from an imperfect machine},''
  \href{http://dx.doi.org/10.1140/epjc/s10052-022-10226-y}{{\em Eur. Phys. J.
  C} {\bfseries 82} no.~3, (2022) 275},
  \href{http://arxiv.org/abs/2111.13633}{{\ttfamily arXiv:2111.13633
  [hep-ph]}}.

\bibitem{Coccaro:2019lgs}
A.~Coccaro, M.~Pierini, L.~Silvestrini, and R.~Torre, ``{The DNNLikelihood:
  enhancing likelihood distribution with Deep Learning},''
  \href{http://dx.doi.org/10.1140/epjc/s10052-020-8230-1}{{\em Eur. Phys. J. C}
  {\bfseries 80} no.~7, (2020) 664},
  \href{http://arxiv.org/abs/1911.03305}{{\ttfamily arXiv:1911.03305
  [hep-ph]}}.

\bibitem{Gao:2017yyd}
J.~Gao, L.~Harland-Lang, and J.~Rojo, ``{The Structure of the Proton in the LHC
  Precision Era},'' \href{http://dx.doi.org/10.1016/j.physrep.2018.03.002}{{\em
  Phys. Rept.} {\bfseries 742} (2018) 1--121},
  \href{http://arxiv.org/abs/1709.04922}{{\ttfamily arXiv:1709.04922
  [hep-ph]}}.

\bibitem{Kovarik:2019xvh}
K.~Kova\v{r}\'\i{}k, P.~M. Nadolsky, and D.~E. Soper, ``{Hadronic structure in
  high-energy collisions},''
  \href{http://dx.doi.org/10.1103/RevModPhys.92.045003}{{\em Rev. Mod. Phys.}
  {\bfseries 92} no.~4, (2020) 045003},
  \href{http://arxiv.org/abs/1905.06957}{{\ttfamily arXiv:1905.06957
  [hep-ph]}}.

\bibitem{Aggarwal:2022cki}
R.~Aggarwal, M.~Botje, A.~Caldwell, F.~Capel, and O.~Schulz, ``{New constraints
  on the up-quark valence distribution in the proton},''
  \href{http://arxiv.org/abs/2209.06571}{{\ttfamily arXiv:2209.06571
  [hep-ph]}}.

\bibitem{10.1214/aoms/1177732360}
S.~S. Wilks, ``{The Large-Sample Distribution of the Likelihood Ratio for
  Testing Composite Hypotheses},''
  \href{http://dx.doi.org/10.1214/aoms/1177732360}{{\em Annals Math. Statist.}
  {\bfseries 9} no.~1, (1938) 60--62}.

\bibitem{Degrande:2020evl}
C.~Degrande, G.~Durieux, F.~Maltoni, K.~Mimasu, E.~Vryonidou, and C.~Zhang,
  ``{Automated one-loop computations in the standard model effective field
  theory},'' \href{http://dx.doi.org/10.1103/PhysRevD.103.096024}{{\em Phys.
  Rev. D} {\bfseries 103} no.~9, (2021) 096024},
  \href{http://arxiv.org/abs/2008.11743}{{\ttfamily arXiv:2008.11743
  [hep-ph]}}.

\bibitem{NNPDF:2021uiq}
{\bfseries NNPDF} Collaboration, R.~D. Ball {\em et~al.}, ``{An open-source
  machine learning framework for global analyses of parton distributions},''
  \href{http://dx.doi.org/10.1140/epjc/s10052-021-09747-9}{{\em Eur. Phys. J.
  C} {\bfseries 81} no.~10, (2021) 958},
  \href{http://arxiv.org/abs/2109.02671}{{\ttfamily arXiv:2109.02671
  [hep-ph]}}.

\bibitem{NNPDF:2021njg}
{\bfseries NNPDF} Collaboration, R.~D. Ball {\em et~al.}, ``{The path to proton
  structure at 1\% accuracy},''
  \href{http://dx.doi.org/10.1140/epjc/s10052-022-10328-7}{{\em Eur. Phys. J.
  C} {\bfseries 82} no.~5, (2022) 428},
  \href{http://arxiv.org/abs/2109.02653}{{\ttfamily arXiv:2109.02653
  [hep-ph]}}.

\bibitem{Carrazza:2019mzf}
S.~Carrazza and J.~Cruz-Martinez, ``{Towards a new generation of parton
  densities with deep learning models},''
  \href{http://dx.doi.org/10.1140/epjc/s10052-019-7197-2}{{\em Eur. Phys. J. C}
  {\bfseries 79} no.~8, (2019) 676},
  \href{http://arxiv.org/abs/1907.05075}{{\ttfamily arXiv:1907.05075
  [hep-ph]}}.

\bibitem{Loshchilov2017}
I.~Loshchilov and F.~Hutter, ``Decoupled weight decay regularization,''.
  \url{https://arxiv.org/abs/1711.05101}.

\bibitem{Paszke2019}
A.~Paszke, S.~Gross, F.~Massa, A.~Lerer, J.~Bradbury, G.~Chanan, T.~Killeen,
  Z.~Lin, N.~Gimelshein, L.~Antiga, A.~Desmaison, A.~Köpf, E.~Yang, Z.~DeVito,
  M.~Raison, A.~Tejani, S.~Chilamkurthy, B.~Steiner, L.~Fang, J.~Bai, and
  S.~Chintala, ``Pytorch: An imperative style, high-performance deep learning
  library,''. \url{https://arxiv.org/abs/1912.01703}.

\bibitem{Carrazza:2021yrg}
S.~Carrazza, J.~M. Cruz-Martinez, and R.~Stegeman, ``{A data-based
  parametrization of parton distribution functions},''
  \href{http://dx.doi.org/10.1140/epjc/s10052-022-10136-z}{{\em Eur. Phys. J.
  C} {\bfseries 82} no.~2, (2022) 163},
  \href{http://arxiv.org/abs/2111.02954}{{\ttfamily arXiv:2111.02954
  [hep-ph]}}.

\bibitem{NNPDF:2014otw}
{\bfseries NNPDF} Collaboration, R.~D. Ball {\em et~al.}, ``{Parton
  distributions for the LHC Run II},''
  \href{http://dx.doi.org/10.1007/JHEP04(2015)040}{{\em JHEP} {\bfseries 04}
  (2015) 040}, \href{http://arxiv.org/abs/1410.8849}{{\ttfamily arXiv:1410.8849
  [hep-ph]}}.

\bibitem{Giele:2001mr}
W.~T. Giele, S.~A. Keller, and D.~A. Kosower, ``{Parton Distribution Function
  Uncertainties},'' \href{http://arxiv.org/abs/hep-ph/0104052}{{\ttfamily
  arXiv:hep-ph/0104052}}.

\bibitem{Giele:1998gw}
W.~T. Giele and S.~Keller, ``{Implications of hadron collider observables on
  parton distribution function uncertainties},''
  \href{http://dx.doi.org/10.1103/PhysRevD.58.094023}{{\em Phys. Rev. D}
  {\bfseries 58} (1998) 094023},
  \href{http://arxiv.org/abs/hep-ph/9803393}{{\ttfamily arXiv:hep-ph/9803393}}.

\bibitem{Ball:2008by}
{\bfseries NNPDF} Collaboration, R.~D. Ball, L.~Del~Debbio, S.~Forte,
  A.~Guffanti, J.~I. Latorre, A.~Piccione, J.~Rojo, and M.~Ubiali, ``{A
  Determination of parton distributions with faithful uncertainty
  estimation},'' \href{http://dx.doi.org/10.1016/j.nuclphysb.2008.09.037}{{\em
  Nucl. Phys. B} {\bfseries 809} (2009) 1--63},
  \href{http://arxiv.org/abs/0808.1231}{{\ttfamily arXiv:0808.1231 [hep-ph]}}.
  [Erratum: Nucl.Phys.B 816, 293 (2009)].

\bibitem{DelDebbio:2004xtd}
{\bfseries NNPDF} Collaboration, L.~Del~Debbio, S.~Forte, J.~I. Latorre,
  A.~Piccione, and J.~Rojo, ``{Unbiased determination of the proton structure
  function F(2)**p with faithful uncertainty estimation},''
  \href{http://dx.doi.org/10.1088/1126-6708/2005/03/080}{{\em JHEP} {\bfseries
  03} (2005) 080}, \href{http://arxiv.org/abs/hep-ph/0501067}{{\ttfamily
  arXiv:hep-ph/0501067}}.

\bibitem{Cowan:2010js}
G.~Cowan, K.~Cranmer, E.~Gross, and O.~Vitells, ``{Asymptotic formulae for
  likelihood-based tests of new physics},''
  \href{http://dx.doi.org/10.1140/epjc/s10052-011-1554-0}{{\em Eur. Phys. J. C}
  {\bfseries 71} (2011) 1554}, \href{http://arxiv.org/abs/1007.1727}{{\ttfamily
  arXiv:1007.1727 [physics.data-an]}}. [Erratum: Eur.Phys.J.C 73, 2501 (2013)].

\bibitem{Alwall:2014hca}
J.~Alwall, R.~Frederix, S.~Frixione, V.~Hirschi, F.~Maltoni, O.~Mattelaer,
  H.~S. Shao, T.~Stelzer, P.~Torrielli, and M.~Zaro, ``{The automated
  computation of tree-level and next-to-leading order differential cross
  sections, and their matching to parton shower simulations},''
  \href{http://dx.doi.org/10.1007/JHEP07(2014)079}{{\em JHEP} {\bfseries 07}
  (2014) 079}, \href{http://arxiv.org/abs/1405.0301}{{\ttfamily arXiv:1405.0301
  [hep-ph]}}.

\bibitem{Brivio:2017btx}
I.~Brivio, Y.~Jiang, and M.~Trott, ``{The SMEFTsim package, theory and
  tools},'' \href{http://dx.doi.org/10.1007/JHEP12(2017)070}{{\em JHEP}
  {\bfseries 12} (2017) 070}, \href{http://arxiv.org/abs/1709.06492}{{\ttfamily
  arXiv:1709.06492 [hep-ph]}}.

\bibitem{Brivio:2020onw}
I.~Brivio, ``{SMEFTsim 3.0 \textemdash{} a practical guide},''
  \href{http://dx.doi.org/10.1007/JHEP04(2021)073}{{\em JHEP} {\bfseries 04}
  (2021) 073}, \href{http://arxiv.org/abs/2012.11343}{{\ttfamily
  arXiv:2012.11343 [hep-ph]}}.

\bibitem{Ball:2012cx}
R.~D. Ball {\em et~al.}, ``{Parton distributions with LHC data},''
  \href{http://dx.doi.org/10.1016/j.nuclphysb.2012.10.003}{{\em Nucl. Phys. B}
  {\bfseries 867} (2013) 244--289},
  \href{http://arxiv.org/abs/1207.1303}{{\ttfamily arXiv:1207.1303 [hep-ph]}}.

\bibitem{Hahn:2000kx}
T.~Hahn, ``{Generating Feynman diagrams and amplitudes with FeynArts 3},''
  \href{http://dx.doi.org/10.1016/S0010-4655(01)00290-9}{{\em Comput. Phys.
  Commun.} {\bfseries 140} (2001) 418--431},
  \href{http://arxiv.org/abs/hep-ph/0012260}{{\ttfamily arXiv:hep-ph/0012260}}.

\bibitem{Hahn:1998yk}
T.~Hahn and M.~Perez-Victoria, ``{Automatized one loop calculations in
  four-dimensions and D-dimensions},''
  \href{http://dx.doi.org/10.1016/S0010-4655(98)00173-8}{{\em Comput. Phys.
  Commun.} {\bfseries 118} (1999) 153--165},
  \href{http://arxiv.org/abs/hep-ph/9807565}{{\ttfamily arXiv:hep-ph/9807565}}.

\bibitem{Hartland:2019bjb}
N.~P. Hartland, F.~Maltoni, E.~R. Nocera, J.~Rojo, E.~Slade, E.~Vryonidou, and
  C.~Zhang, ``{A Monte Carlo global analysis of the Standard Model Effective
  Field Theory: the top quark sector},''
  \href{http://dx.doi.org/10.1007/JHEP04(2019)100}{{\em JHEP} {\bfseries 04}
  (2019) 100}, \href{http://arxiv.org/abs/1901.05965}{{\ttfamily
  arXiv:1901.05965 [hep-ph]}}.

\bibitem{CMS:2018adi}
{\bfseries CMS} Collaboration, A.~M. Sirunyan {\em et~al.}, ``{Measurements of
  $\mathrm{t\overline{t}}$ differential cross sections in proton-proton
  collisions at $\sqrt{s}=$ 13 TeV using events containing two leptons},''
  \href{http://dx.doi.org/10.1007/JHEP02(2019)149}{{\em JHEP} {\bfseries 02}
  (2019) 149}, \href{http://arxiv.org/abs/1811.06625}{{\ttfamily
  arXiv:1811.06625 [hep-ex]}}.

\bibitem{ATLAS:2017cen}
{\bfseries ATLAS} Collaboration, M.~Aaboud {\em et~al.}, ``{Evidence for the $
  H\to b\overline{b} $ decay with the ATLAS detector},''
  \href{http://dx.doi.org/10.1007/JHEP12(2017)024}{{\em JHEP} {\bfseries 12}
  (2017) 024}, \href{http://arxiv.org/abs/1708.03299}{{\ttfamily
  arXiv:1708.03299 [hep-ex]}}.

\bibitem{Feroz:2013hea}
F.~Feroz, M.~P. Hobson, E.~Cameron, and A.~N. Pettitt, ``{Importance Nested
  Sampling and the MultiNest Algorithm},''
  \href{http://dx.doi.org/10.21105/astro.1306.2144}{{\em Open J. Astrophys.}
  {\bfseries 2} no.~1, (2019) 10},
  \href{http://arxiv.org/abs/1306.2144}{{\ttfamily arXiv:1306.2144
  [astro-ph.IM]}}.

\bibitem{Carrazza:2015hva}
S.~Carrazza, J.~I. Latorre, J.~Rojo, and G.~Watt, ``{A compression algorithm
  for the combination of PDF sets},''
  \href{http://dx.doi.org/10.1140/epjc/s10052-015-3703-3}{{\em Eur. Phys. J. C}
  {\bfseries 75} (2015) 474}, \href{http://arxiv.org/abs/1504.06469}{{\ttfamily
  arXiv:1504.06469 [hep-ph]}}.

\bibitem{Butterworth:2015oua}
J.~Butterworth {\em et~al.}, ``{PDF4LHC recommendations for LHC Run II},''
  \href{http://dx.doi.org/10.1088/0954-3899/43/2/023001}{{\em J. Phys. G}
  {\bfseries 43} (2016) 023001},
  \href{http://arxiv.org/abs/1510.03865}{{\ttfamily arXiv:1510.03865
  [hep-ph]}}.

\bibitem{Brivio:2019ius}
I.~Brivio, S.~Bruggisser, F.~Maltoni, R.~Moutafis, T.~Plehn, E.~Vryonidou,
  S.~Westhoff, and C.~Zhang, ``{O new physics, where art thou? A global search
  in the top sector},'' \href{http://dx.doi.org/10.1007/JHEP02(2020)131}{{\em
  JHEP} {\bfseries 02} (2020) 131},
  \href{http://arxiv.org/abs/1910.03606}{{\ttfamily arXiv:1910.03606
  [hep-ph]}}.

\bibitem{Berger:2019wnu}
N.~Berger {\em et~al.}, ``{Simplified Template Cross Sections - Stage 1.1}''
  \href{http://arxiv.org/abs/1906.02754}{{\ttfamily arXiv:1906.02754
  [hep-ph]}}.

\bibitem{Buckley:2014ana}
A.~Buckley, J.~Ferrando, S.~Lloyd, K.~Nordstr\"om, B.~Page, M.~R\"ufenacht,
  M.~Sch\"onherr, and G.~Watt, ``{LHAPDF6: parton density access in the LHC
  precision era},''
  \href{http://dx.doi.org/10.1140/epjc/s10052-015-3318-8}{{\em Eur. Phys. J. C}
  {\bfseries 75} (2015) 132}, \href{http://arxiv.org/abs/1412.7420}{{\ttfamily
  arXiv:1412.7420 [hep-ph]}}.

\bibitem{Bertone:2014zva}
V.~Bertone, R.~Frederix, S.~Frixione, J.~Rojo, and M.~Sutton, ``{aMCfast:
  automation of fast NLO computations for PDF fits},''
  \href{http://dx.doi.org/10.1007/JHEP08(2014)166}{{\em JHEP} {\bfseries 08}
  (2014) 166}, \href{http://arxiv.org/abs/1406.7693}{{\ttfamily arXiv:1406.7693
  [hep-ph]}}.

\bibitem{Carrazza:2020gss}
S.~Carrazza, E.~R. Nocera, C.~Schwan, and M.~Zaro, ``{PineAPPL: combining EW
  and QCD corrections for fast evaluation of LHC processes},''
  \href{http://dx.doi.org/10.1007/JHEP12(2020)108}{{\em JHEP} {\bfseries 12}
  (2020) 108}, \href{http://arxiv.org/abs/2008.12789}{{\ttfamily
  arXiv:2008.12789 [hep-ph]}}.

\bibitem{Carli:2010rw}
T.~Carli, D.~Clements, A.~Cooper-Sarkar, C.~Gwenlan, G.~P. Salam, F.~Siegert,
  P.~Starovoitov, and M.~Sutton, ``{A posteriori inclusion of parton density
  functions in NLO QCD final-state calculations at hadron colliders: The
  APPLGRID Project},''
  \href{http://dx.doi.org/10.1140/epjc/s10052-010-1255-0}{{\em Eur. Phys. J. C}
  {\bfseries 66} (2010) 503--524},
  \href{http://arxiv.org/abs/0911.2985}{{\ttfamily arXiv:0911.2985 [hep-ph]}}.

\bibitem{ArjonaMartinez:2018eah}
J.~Arjona~Mart\'\i{}nez, O.~Cerri, M.~Pierini, M.~Spiropulu, and J.-R. Vlimant,
  ``{Pileup mitigation at the Large Hadron Collider with graph neural
  networks},'' \href{http://dx.doi.org/10.1140/epjp/i2019-12710-3}{{\em Eur.
  Phys. J. Plus} {\bfseries 134} no.~7, (2019) 333},
  \href{http://arxiv.org/abs/1810.07988}{{\ttfamily arXiv:1810.07988
  [hep-ph]}}.

\bibitem{Abdughani:2018wrw}
M.~Abdughani, J.~Ren, L.~Wu, and J.~M. Yang, ``{Probing stop pair production at
  the LHC with graph neural networks},''
  \href{http://dx.doi.org/10.1007/JHEP08(2019)055}{{\em JHEP} {\bfseries 08}
  (2019) 055}, \href{http://arxiv.org/abs/1807.09088}{{\ttfamily
  arXiv:1807.09088 [hep-ph]}}.

\bibitem{Haisch:2022nwz}
U.~Haisch, D.~J. Scott, M.~Wiesemann, G.~Zanderighi, and S.~Zanoli, ``{NNLO
  event generation for $ pp\to Zh\to
  {\mathrm{\ell}}^{+}{\mathrm{\ell}}^{-}b\overline{b} $ production in the SM
  effective field theory},''
  \href{http://dx.doi.org/10.1007/JHEP07(2022)054}{{\em JHEP} {\bfseries 07}
  (2022) 054}, \href{http://arxiv.org/abs/2204.00663}{{\ttfamily
  arXiv:2204.00663 [hep-ph]}}.

\bibitem{Mazzitelli:2021mmm}
J.~Mazzitelli, P.~F. Monni, P.~Nason, E.~Re, M.~Wiesemann, and G.~Zanderighi,
  ``{Top-pair production at the LHC with MINNLO$_{PS}$},''
  \href{http://dx.doi.org/10.1007/JHEP04(2022)079}{{\em JHEP} {\bfseries 04}
  (2022) 079}, \href{http://arxiv.org/abs/2112.12135}{{\ttfamily
  arXiv:2112.12135 [hep-ph]}}.

\bibitem{Zanoli:2021iyp}
S.~Zanoli, M.~Chiesa, E.~Re, M.~Wiesemann, and G.~Zanderighi,
  ``{Next-to-next-to-leading order event generation for VH production with H
  \textrightarrow{}$ b\overline{b} $ decay},''
  \href{http://dx.doi.org/10.1007/JHEP07(2022)008}{{\em JHEP} {\bfseries 07}
  (2022) 008}, \href{http://arxiv.org/abs/2112.04168}{{\ttfamily
  arXiv:2112.04168 [hep-ph]}}.

\bibitem{CLICPhysicsWorkingGroup:2004qvu}
{\bfseries CLIC Physics Working Group} Collaboration, E.~Accomando {\em
  et~al.}, \href{http://dx.doi.org/10.5170/CERN-2004-005}{``{Physics at the
  CLIC multi-TeV linear collider},''} in {\em {11th International Conference on
  Hadron Spectroscopy}}, CERN Yellow Reports: Monographs.
\newblock 6, 2004.
\newblock \href{http://arxiv.org/abs/hep-ph/0412251}{{\ttfamily
  arXiv:hep-ph/0412251}}.

\bibitem{Linssen:2012hp}
``{Physics and Detectors at CLIC: CLIC Conceptual Design Report},''
  \href{http://arxiv.org/abs/1202.5940}{{\ttfamily arXiv:1202.5940
  [physics.ins-det]}}.

\bibitem{MuonCollider:2022xlm}
{\bfseries Muon Collider} Collaboration, J.~de~Blas {\em et~al.}, ``{The
  physics case of a 3 TeV muon collider stage},''
  \href{http://arxiv.org/abs/2203.07261}{{\ttfamily arXiv:2203.07261
  [hep-ph]}}.

\bibitem{Chen:2022msz}
S.~Chen, A.~Glioti, R.~Rattazzi, L.~Ricci, and A.~Wulzer, ``{Learning from
  radiation at a very high energy lepton collider},''
  \href{http://dx.doi.org/10.1007/JHEP05(2022)180}{{\em JHEP} {\bfseries 05}
  (2022) 180}, \href{http://arxiv.org/abs/2202.10509}{{\ttfamily
  arXiv:2202.10509 [hep-ph]}}.

\bibitem{Buttazzo:2020uzc}
D.~Buttazzo, R.~Franceschini, and A.~Wulzer, ``{Two Paths Towards Precision at
  a Very High Energy Lepton Collider},''
  \href{http://dx.doi.org/10.1007/JHEP05(2021)219}{{\em JHEP} {\bfseries 05}
  (2021) 219}, \href{http://arxiv.org/abs/2012.11555}{{\ttfamily
  arXiv:2012.11555 [hep-ph]}}.

\bibitem{Beenakker:2015rna}
W.~Beenakker, C.~Borschensky, M.~Kr\"amer, A.~Kulesza, E.~Laenen, S.~Marzani,
  and J.~Rojo, ``{NLO+NLL squark and gluino production cross-sections with
  threshold-improved parton distributions},''
  \href{http://dx.doi.org/10.1140/epjc/s10052-016-3892-4}{{\em Eur. Phys. J. C}
  {\bfseries 76} no.~2, (2016) 53},
  \href{http://arxiv.org/abs/1510.00375}{{\ttfamily arXiv:1510.00375
  [hep-ph]}}.

\bibitem{afb}
R.~D. Ball, A.~Candido, S.~Forte, F.~Hekhorn, E.~R. Nocera, J.~Rojo, and
  C.~Schwan, ``{Parton Distributions and New Physics Searches: the Drell-Yan
  Forward-Backward Asymmetry as a Case Study},''
  \href{http://arxiv.org/abs/2209.08115}{{\ttfamily arXiv:2209.08115
  [hep-ph]}}.

\bibitem{Greljo:2021kvv}
A.~Greljo, S.~Iranipour, Z.~Kassabov, M.~Madigan, J.~Moore, J.~Rojo, M.~Ubiali,
  and C.~Voisey, ``{Parton distributions in the SMEFT from high-energy
  Drell-Yan tails},'' \href{http://dx.doi.org/10.1007/JHEP07(2021)122}{{\em
  JHEP} {\bfseries 07} (2021) 122},
  \href{http://arxiv.org/abs/2104.02723}{{\ttfamily arXiv:2104.02723
  [hep-ph]}}.

\bibitem{Carrazza:2019sec}
S.~Carrazza, C.~Degrande, S.~Iranipour, J.~Rojo, and M.~Ubiali, ``{Can New
  Physics hide inside the proton?},''
  \href{http://dx.doi.org/10.1103/PhysRevLett.123.132001}{{\em Phys. Rev.
  Lett.} {\bfseries 123} no.~13, (2019) 132001},
  \href{http://arxiv.org/abs/1905.05215}{{\ttfamily arXiv:1905.05215
  [hep-ph]}}.

\bibitem{Iranipour:2022iak}
S.~Iranipour and M.~Ubiali, ``{A new generation of simultaneous fits to LHC
  data using deep learning},''
  \href{http://dx.doi.org/10.1007/JHEP05(2022)032}{{\em JHEP} {\bfseries 05}
  (2022) 032}, \href{http://arxiv.org/abs/2201.07240}{{\ttfamily
  arXiv:2201.07240 [hep-ph]}}.

\bibitem{Liu:2022plj}
D.~Liu, C.~Sun, and J.~Gao, ``{Machine learning of log-likelihood functions in
  global analysis of parton distributions},''
  \href{http://dx.doi.org/10.1007/JHEP08(2022)088}{{\em JHEP} {\bfseries 08}
  (2022) 088}, \href{http://arxiv.org/abs/2201.06586}{{\ttfamily
  arXiv:2201.06586 [hep-ph]}}.

\bibitem{Gao:2022srd}
J.~Gao, M.~Gao, T.~J. Hobbs, D.~Liu, and X.~Shen, ``{Simultaneous CTEQ-TEA
  extraction of PDFs and SMEFT parameters from jet and $t{\bar t}$ data},''
  \href{http://arxiv.org/abs/2211.01094}{{\ttfamily arXiv:2211.01094
  [hep-ph]}}.

\bibitem{10.2307/1990256}
A.~Wald, ``Tests of statistical hypotheses concerning several parameters when
  the number of observations is large,'' {\em Transactions of the American
  Mathematical Society} {\bfseries 54} no.~3, (1943) 426--482.
  \url{http://www.jstor.org/stable/1990256}.

\end{thebibliography}
\providecommand{\href}[2]{#2}\begingroup\raggedright\endgroup

\end{document}